\documentclass[aps,prb,reprint,twocolumn,citeautoscript,superscriptaddress,showkeys,showpacs]{revtex4-1}
\usepackage{xr}
\usepackage[fleqn,tbtags]{amsmath}
\usepackage{amsfonts, amssymb,amsxtra,textcomp}
\usepackage[]{graphicx}
\usepackage{grffile}
\usepackage{color}                                      
\definecolor{d_red}{cmyk}{0.00, 0.81, 1.00, 0.27}
\definecolor{d_orange}{cmyk}{0.00, 0.33, 1.00, 0.00}
\definecolor{d_blue}{cmyk}{0.78, 0.47, 0.00, 0.20}
\definecolor{d_lgreen}{cmyk}{0.07, 0.00, 0.79, 0.29}
\definecolor{d_green}{cmyk}{0.66, 0.00, 0.71, 0.56}
\definecolor{d_blue}{cmyk}{0.78, 0.47, 0.00, 0.20}
\definecolor{d_dblue}{cmyk}{0.91, 0.79, 0.00, 0.22}
\definecolor{d_pink}{cmyk}{0.0, 0.79, 0.37, 0.29}
\definecolor{d_purple}{cmyk}{0.16, 0.54, 0.00, 0.70}
\definecolor{d_paleblue}{cmyk}{0.669, 0.338, 0.00, 0.373}
\definecolor{d_dpaleblue}{cmyk}{0.441, 0.290, 0.00, 0.580}
\definecolor{d_brown}{cmyk}{0.0, 0.490, 0.930, 0.350}
\definecolor{d_turquoise}{cmyk}{0.630, 0.04, 0.0, 0.440}

\newcommand{\av}[1]{\langle #1 \rangle}

\newcommand{\bfss}{{\boldsymbol{S}}}
\newcommand{\bfrr}{{\boldsymbol{R}}}
\newcommand{\bfqq}{{\boldsymbol{Q}}}

\newcommand{\bfll}{{\boldsymbol{L}}}
\newcommand{\bfxx}{{\boldsymbol{X}}}
\newcommand{\bfq}{{\boldsymbol{q}}}
\newcommand{\bfp}{{\boldsymbol{p}}}
\newcommand{\bfk}{{\boldsymbol{k}}}
\newcommand{\bfn}{{\boldsymbol{n}}}

\newcommand{\bfb}{{\boldsymbol{b}}}
\newcommand{\bfa}{{\boldsymbol{a}}}
\newcommand{\bfm}{{\boldsymbol{m}}}
\newcommand{\bfh}{{\boldsymbol{h}}}
\newcommand{\bft}{{\boldsymbol{t}}}
\newcommand{\bfr}{{\boldsymbol{r}}}
\newcommand{\bfx}{{\boldsymbol{x}}}
\newcommand{\bfy}{{\boldsymbol{y}}}

\def\bmx{\begin{pmatrix}}
\def\emx{\end{pmatrix}}

\usepackage{hyperref}
\usepackage[figure,table]{hypcap}               
\hypersetup{
  pdftitle = {},
  pdfsubject = {},
  pdfauthor = {},
  pdfkeywords = {},
  pdfcreator = {},
  pdfproducer = {LaTeX with hyperref},
  colorlinks = true,
  linkcolor = blue, 
  anchorcolor = blue,
  citecolor = blue, 
  filecolor = red, 
  menucolor = red, 
  pagecolor = red, 
  urlcolor  = blue, 
  breaklinks = true,
  pdfstartview = FitV,
  pdfhighlight = /I,
  pdfpagelayout = OneColumn,
  hypertexnames=true 
}

\begin{document}
\title{Emergent Criticality and Friedan Scaling in a 2D Frustrated Heisenberg Antiferromagnet}
\author{Peter P. Orth}
\affiliation{Institute for Theory of Condensed Matter, Karlsruhe Institute of Technology (KIT), 76131 Karlsruhe, Germany}
\author{Premala Chandra}
\affiliation{Center for Materials Theory, Rutgers University, Piscataway, New Jersey 08854, USA}
\author{Piers Coleman}
\affiliation{Center for Materials Theory, Rutgers University, Piscataway, New Jersey 08854, USA}
\affiliation{Hubbard Theory Consortium and Department of Physics, Royal Holloway, University of London, Egham, Surrey TW20 0EX, UK}
\author{J\"org Schmalian}
\affiliation{Institute for Theory of Condensed Matter, Karlsruhe Institute of Technology (KIT), 76131 Karlsruhe, Germany}
\affiliation{Institute for Solid State Research, Karlsruhe Institute of Technology (KIT), 76131 Karlsruhe, Germany} 
\date{\today}

\begin{abstract}
 We study a two-dimensional frustrated Heisenberg antiferromagnet on the windmill lattice consisting of triangular and dual honeycomb lattice sites. In the classical ground state the spins on different sublattices are decoupled, but quantum and thermal fluctuations drive the system into a coplanar state via an ``order from disorder'' mechanism. We obtain the finite temperature phase diagram using renormalization group approaches. In the coplanar regime, the relative U$(1)$ phase between the spins on the two sublattices decouples from the remaining degrees of freedom, and is described by a six-state clock model with an emergent critical phase. At lower temperatures the system enters a $\mathbb{Z}_6$ broken phase with long-range phase correlations. We derive these results by two distinct renormalization group approaches to two-dimensional magnetism: by Wilson-Polyakov scaling and by Friedan's geometric approach to nonlinear sigma models where the scaling of the spin-stiffnesses is governed by the Ricci flow of a 4D metric tensor. 
\end{abstract}

\pacs{75.10.-b, 75.10.Jm}

\maketitle

\section{Introduction}
\label{sec:intr-heis-model}
Two-dimensional systems with continuous symmetry and short-range interactions obey the Hohenberg-Mermin-Wagner (HMW) theorem\cite{PhysRev.158.383,PhysRevLett.17.1307} and thus exhibit true long-range order only at strictly zero temperature. Nevertheless it is now known that (geometrically) frustrated two-dimensional (2D) Heisenberg spin systems can circumvent this theorem such that long-range \emph{discrete order} occurs at finite temperatures~\cite{PhysRevLett.64.88,PhysRevLett.91.177202,PhysRevLett.92.157202,PhysRevB.81.214419,PhysRevB.78.094423,annurev-conmatphys-070909-104138,PhysRevLett.110.077201}. This ``order from disorder'' is driven by short-wavelength quantum and thermal spin fluctuations,\cite{villain-JPhysFrance-1977,Villain80,Shender82,Gukasov88,PhysRevLett.62.2056,ChandraColeman-LesHouches,PhysRevLett.68.855,PhysRevB.58.12049}. The emergent order parameter is defined as the relative orientation of spins. Remarkably, long-range order exists despite a finite magnetic correlation length of the underlying Heisenberg system. This can lead to finite temperature phase transitions such as an $\mathbb{Z}_2$ Ising or $\mathbb{Z}_3$ Potts phase transition~\cite{PhysRevLett.64.88,PhysRevLett.92.157202,PhysRevLett.91.177202,PhysRevB.81.214419,PhysRevB.78.094423,annurev-conmatphys-070909-104138,PhysRevLett.110.077201}. This phenomenon is well-established in the $J_1$-$J_2$ Heisenberg model on the square lattice,\cite{PhysRevLett.64.88,ChandraColeman-LesHouches,PhysRevLett.91.177202,PhysRevLett.92.157202} and has recently found unexpected application in the physics of iron-based superconductors\cite{PhysRevB.77.224509, PhysRevB.78.020501, PhysRevLett.105.157003,Fernandes2012,PhysRevB.85.024534}, where it induces a nematic structural phase transition of the lattice in the absence of long-range magnetic order. Emergent discrete order occurs in a range of strongly correlated materials~\cite{annurev-conmatphys-070909-104138,PhysRevB.86.115443,PhysRevLett.106.207202,BorziMackenzie-Science-2007}.
\begin{figure}[t!]
  \centering
  \includegraphics[width=\linewidth]{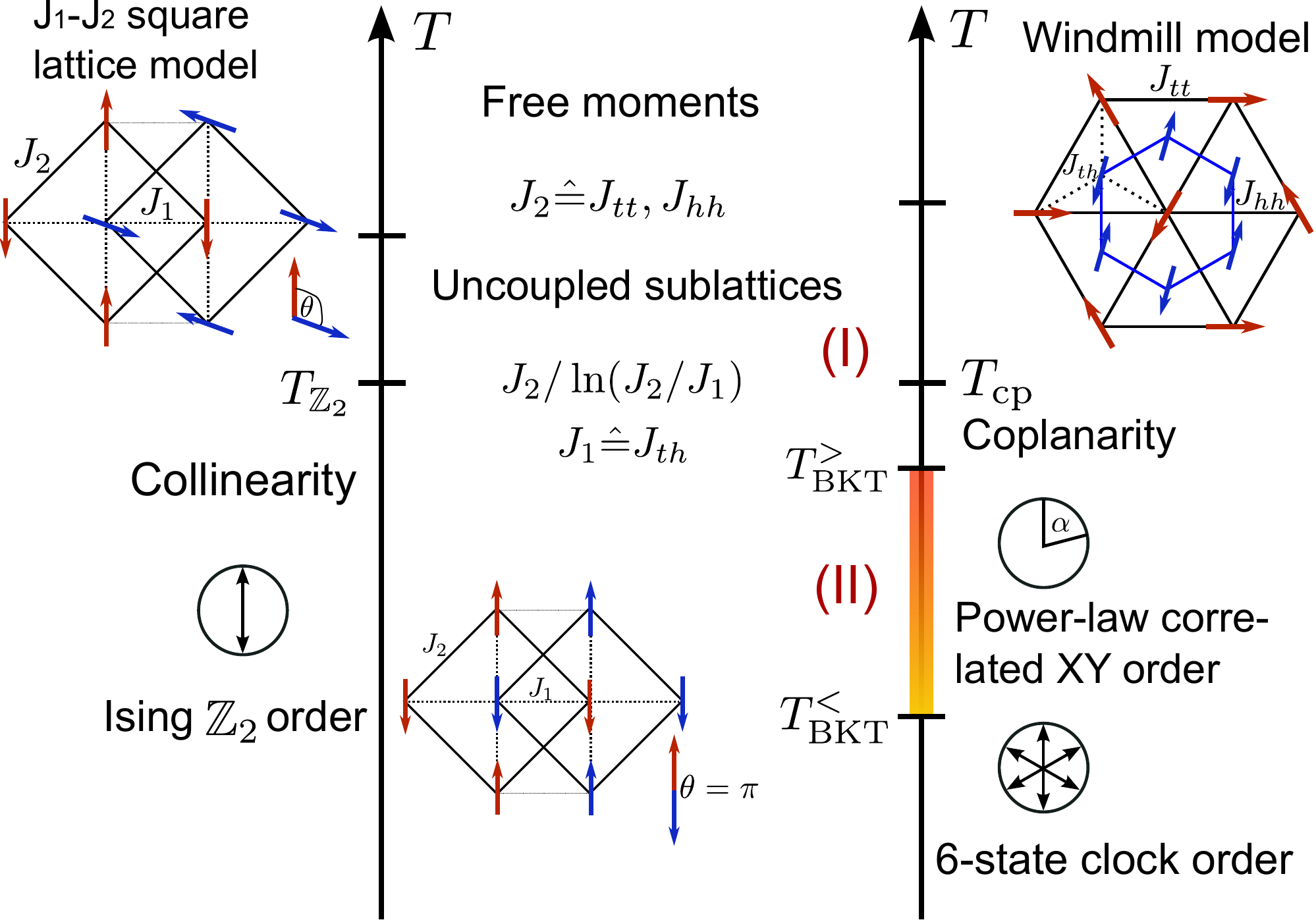}
  \caption{(Color online). Schematic phase diagram summarizing the main results of our study of the ``windmill'' Heisenberg model of interpenetrating triangular and hexagonal lattices; the phase behavior of its square-lattice counterpart is also shown (on the left) for reference where in each case $J_1$ and $J_2$ refer to the inter- and intra-lattice couplings respectively. We note that the development of fluctuation-induced collinearity is a transition in the square-lattice problem whereas its analogue in the windmill model, the development of coplanarity, is a crossover.  (I) and (II) refers to the development of coplanarity and criticality in the windmill model and are discussed extensively in the main text and in the Appendices.} 
  \label{fig:1}
\end{figure}

Knowing about this escape clause of the HMW theorem, here we ask whether one may also find a critical phase with algebraic order and an associated Berezinskii-Kosterlitz-Thouless (BKT) phase transition~\cite{Thouless98,ChaikinLubensky-Book} in an isotropic Heisenberg spin system in two dimensions. In order to construct such a Hamiltonian, we exploit the fact that discrete $\mathbb{Z}_p$ clock models host a critical phase for $p \geq 5$~\cite{PhysRevB.16.1217,Ortiz2012780}. In this article we study a frustrated 2D Heisenberg model with an emergent $\mathbb{Z}_6$ order parameter. The order parameter describes the relative orientation of spins on different sublattices. Using a renormalization group (RG) analysis, we show that these emergent discrete degrees of freedom are described by a $\mathbb{Z}_6$ clock model that admits a critical phase bracketed by two Berezinskii-Kosterlitz-Thouless (BKT) transitions at finite temperature~\cite{PhysRevB.16.1217,Ortiz2012780}. In addition to discussing details of this 
work reported briefly elsewhere,\cite{PhysRevLett.109.237205} we include a self-contained presentation of a ``Ricci flow'' methodology to study classical 2D magnetism based on Friedan's geometric approach to nonlinear sigma models~\cite{PhysRevLett.45.1057,Friedan-AnnPhys-1985}; at each stage all results are compared with those obtained by Wilson-Polyakov scaling~\cite{polyakov_nlsm_75, RevModPhys.47.773,RevModPhys.55.583,Polyakov_GaugeFieldsAndStrings_Book}.

Generalizing previous work on the $J_1$-$J_2$ Heisenberg model on the square lattice~\cite{PhysRevLett.64.88,PhysRevLett.91.177202,PhysRevLett.92.157202}, here we study a $J_1$-$J_2$ Heisenberg Hamiltonian on interpenetrating triangular and honeycomb lattices that we call the ``windmill'' lattice Heisenberg model. Both models consider coupling of spins on a given lattice to spins on the corresponding dual lattice. Exchange couplings exist between all nearest-neighbor pairs within both sublattices and between the sublattices. The couplings within each triangular and honeycomb sublattice $J_{tt}$ and $J_{hh}$ play the role of $J_2$, while the coupling between different sublattices $J_{th}$ corresponds to $J_1$. In Fig.~\ref{fig:1}, we display the main results of this article as a schematic phase diagram using the
square $J_1$-$J_2$ model as a reference.  At high temperatures $T \gg J_2$ both spin systems display free moment behavior, and then at $T \sim J_2$ they each become two decoupled lattices where the local exchange field of one of the sublattice on the spins on the other sublattice is identically zero. In the simpler square lattice case, a renormalization group analysis indicates that at low temperatures short-wavelength thermal and quantum fluctuations break the $\mathbb{Z}_4$ lattice symmetry down to $\mathbb{Z}_2$ and select two collinear states from the 
ground-state manifold leading to long-range discrete ($\mathbb{Z}_2$) order. A finite $\mathbb{Z}_2$ phase transition occurs at  $T \sim \frac{J_2}{\ln(\frac{J_2}{J_1})}$ when the domain wall thickness separating the two states is less than the Heisenberg spin correlation length~\cite{PhysRevLett.64.88,ChandraColeman-LesHouches,PhysRevLett.91.177202,PhysRevLett.92.157202}.

The corresponding physics in the windmill lattice model occurs in two distinct stages, as indicated schematically in Fig.~\ref{fig:1}. At $T \sim J_2$ the two sublattices are decoupled leading to a $\text{SO}(3) \times \text{O}(3)/\text{O}(2)$ order parameter. Its low-energy description, derived from its microscopic Heisenberg Hamiltonian, takes the form of a nonlinear sigma model (NLSM), that contains two additional potential terms arising from short-wavelength quantum and thermal spin-wave fluctuations. One of these potential terms forces the spins on both sublattices to be coplanar (I in Fig.~\ref{fig:1}) at a crossover temperature $T_{cp}\sim \frac{J_2}{\ln(\frac{J_2}{J_1})}$ with $\text{SO}(3) \times \text{U}(1)$ order where no symmetry is explicitly broken; the other potential term sets a six-fold potential in the plane. Using an RG analyis, we explicitly show that in the coplanar state the $\text{U}(1)$ degrees of freedom decouple to form an $\text{XY}$ model with a six-fold potential. Following the well-known RG program of this BKT problem~\cite{PhysRevB.16.1217,Thouless98,ChaikinLubensky-Book,Ortiz2012780}, we find that the vortex-unbinding transition temperature to enter the critical phase is of the same order as that of the coplanar crossover. Ultimately at low temperatures the six-fold potential term becomes relevant, and the system enters a $\mathbb{Z}_6$ broken phase; the two transitions bracketing the critical phase are both in the BKT universality class.  To our knowledge, this is the first identification and characterization of a 2D isotropic Heisenberg spin system with a finite temperature power-law correlated phase and the associated BKT transitions.  We do note that
such a scenario was previously found on a Kitaev-Heisenberg model resulting from a conceptually different mechanism~\cite{PhysRevLett.109.187201,PhysRevB.88.024410}, and also for discrete spins on the triangular lattice~\cite{PhysRevB.29.5250,PhysRevB.68.104409}.

A novel feature of our work is that we apply Friedan's gravitational scaling approach~\cite{PhysRevLett.45.1057} to 2D classical magnetism. Here the configurations of the 2D spin system described by four Euler angles correspond to the worldsheet of a string evolving in four dimensions where the metric is determined by the spin stiffnesses. Using Friedan's coordinate-independent approach to nonlinear sigma models~\cite{PhysRevLett.45.1057}, we then identify the renormalization of the spin stiffnesses with the Ricci flow of the corresponding metric tensor; all results in this article are presented using both the Wilson-Polyakov renormalization group~\cite{polyakov_nlsm_75, RevModPhys.47.773,RevModPhys.55.583,Polyakov_GaugeFieldsAndStrings_Book} and Friedan's coordinate-independent approach~\cite{PhysRevLett.45.1057} with technical details in the Appendices. Using this analogy, the decoupling of the $\text{U}(1)$ phase in our system can be viewed as a toy model for compactification of a four-dimensional string theory; we note that this nontrivial decoupling of the $\text{U}(1)$ phase is essential for the occurance of the emergent critical phase.    

We now describe the modular structure of this article. In Sec.~\ref{sec:long-wavelength-nlsm} we introduce the microscopic Heisenberg Hamiltonian of the windmill model and compute its spinwave spectrum.  We also derive its long-wavelength action that takes the form of a coupled SO(3) $\times$ O(3)/O(2) NLSM. In Sec.~\ref{sec:outline-rg-program} we outline the renormalization group (RG) program that we use to determine the system's phase diagram, discussing key features of the Wilson-Polyakov and the Friedan approaches to scaling and presenting the main results of the subsequent analysis obtained with these two distinct methods. High-temperature behavior, where the two sublattices are approximately uncoupled, is studied in Sec.~\ref{sec:renorm-group-equat}; we derive and analyze the corresponding RG scaling equations of the spin stiffnesses and the potential terms coupling the two sublattices. ``Order from disorder'' soon drives the system into a coplanar state, where spins on the honeycomb and the triangular lattice are lying in the same plane in spin space. In Sec.~\ref{sec:rg-analysis-coplanar}, we derive and analyze the scaling of the spin stiffnesses in the coplanar regime where the system is described by a coupled SO(3) $\times$ U(1) NLSM. We show that the U(1) relative in-plane angle of triangular and honeycomb spins decouples, and analyze the resulting low-energy action of this emergent U(1) degree of freedom in Sec.~\ref{sec:phase-deco-effect}; it takes the form of a $\mathbb{Z}_6$ clock model where the six-fold potential results from the discrete lattice environment. We adapt a BKT RG analysis to our specific situation and show that the system exhibits two consecutive BKT phase transitions which frame a critical phase with power-law correlations in the relative U(1) angle. At low temperatures the six-fold potential is RG relevant and leads to a spontanous breaking of the $\mathbb{Z}_6$ symmetry and long-range discrete order. We summarize our results, discuss experimental realizations and open questions for future research in Sec.~\ref{sec:summary-outlook}. We present predominantly results in the main text; technical details of the calculations, using both the Wilson-Polyakov RG and the Friedan coordinate-independent approaches are provided in several Appendices. We also provide electronic Supplemental Material in the form of a \emph{Mathematica} file that includes the calculation of the RG equations using the Friedan approach~\cite{SupplMat-FriedanScaling}.

\section{Windmill lattice Heisenberg antiferromagnet}
\label{sec:long-wavelength-nlsm}
\begin{figure}[t]
  \centering
  \includegraphics[width=\linewidth]{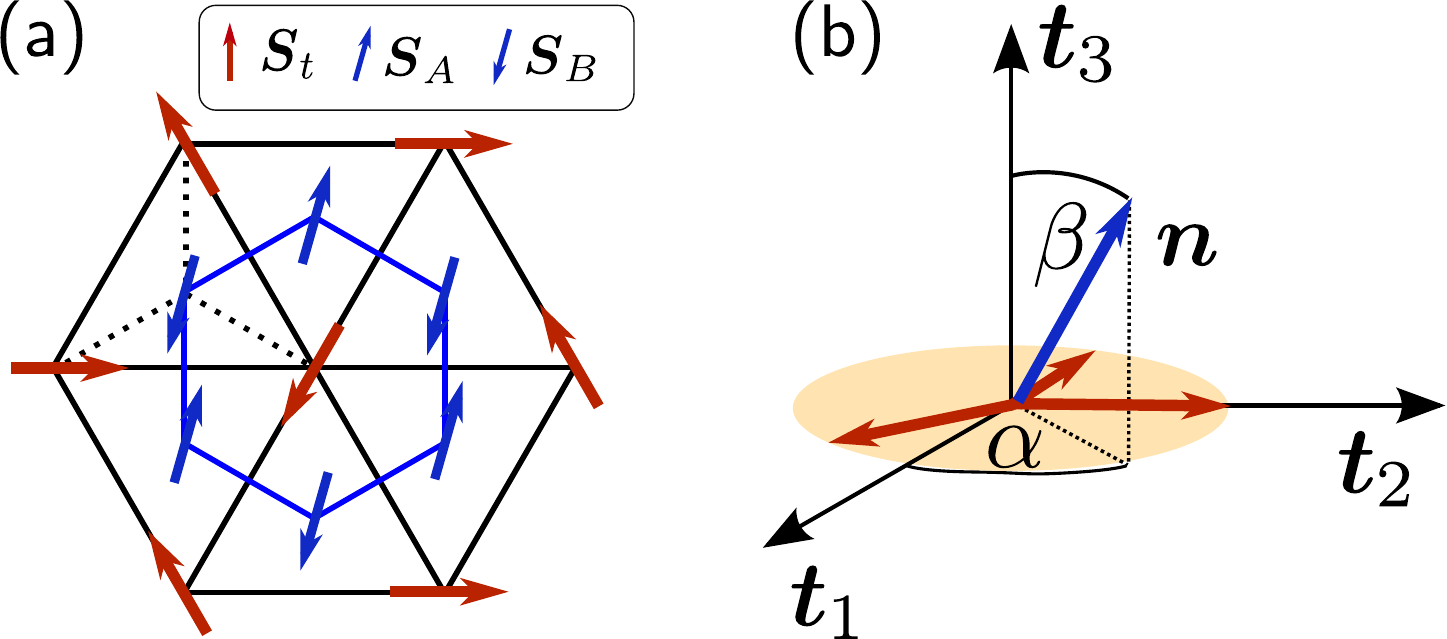}
  \caption{(Color online). (a) Windmill lattice Heisenberg model consisting of spins $S_a$ on sites of both triangular ($t$) and honeycomb ($A,B$) lattice. Exchange interaction $J_{ab}$ exists between all nearest-neighbor spins with $a, b \in \{t, A, B\}$. Interaction between spins on different sublattices $J_{tA}=J_{tB}$ (dashed links, for clarity only shown in one plaquette) is assumed to be weaker than between same sublattice spins $J_{tt}, J_{AB}$ (solid links). (b) Definition of angles $\alpha$ and $\beta$ that describe telative orientation of magnetic order parameter $\bfn$ for O(3)/O(2) Neel order on the honeycomb lattice and tripod $\{\bft_1, \bft_2, \bft_3\}$ for the SO(3) order on the triangular lattice. Note that $\beta = \pi/2$ corresponds to coplanar order with honeybomb (blue) and triangular spins (red) sharing a common plane. }
  \label{fig:2}
\end{figure}

Here we introduce the ``windmill'' model, an antiferromagnetic Heisenberg model on interpenetrating two-dimensional triangular and 
honeycomb lattices, shown in Fig.~\ref{fig:2}(a), that we study in detail in this article. The underlying Bravais lattice is triangular with primitive lattice vectors $\bfa_1 = \frac{a_0}{2} (1, \sqrt{3})$ and $\bfa_2 = \frac{a_0}{2} (-1, \sqrt{3})$. It contains three basis sites per unit cell at positions $\bfb_t = a_0 (0, 2/\sqrt{3})$, $\bfb_A = (0,0)$ and $\bfb_B = a_0 (0, 1/\sqrt{3})$, where $t$ refers to the triangular and $A,B$ to the two honeycomb basis sites. In the following we set the lattice constant $a_0 = 1$. The Hamiltonian consists of nearest-neighbor coupling terms on the same sublattice as well as between the two sublattices, and is given by
\begin{align}
  \label{eq:1}
  H = H_{tt} + H_{AB} + H_{tA} + H_{tB}
\end{align}
with 
\begin{align}
  \label{eq:2}
  H_{ab} &= J_{ab} \sum_{m=1}^{N_L} \sum_{\{\boldsymbol{\delta}_{ab}\}} \bfss_a(\bfr_m) \cdot \bfss_b(\bfr_m + \boldsymbol{\delta}_{ab}) \,.
\end{align}
Here, $\bfss_a(\bfr_m)$ denote spin operators at Bravais lattice site $\bfr_m$ and basis site $a \in \{t, A,B\}$ and $N_L$ is the number of Bravais lattice sites. Antiferromagnetic Heisenberg exchange coupling constants $J_{ab} > 0$ act between pairs of nearest-neighbor spins on sublattices $a$ and $b$. The vectors $\{\boldsymbol{\delta}_{ab}\}$ point between nearest neighbors on sublattices $a$ and $b$. Explicitly, they are given by $\{\boldsymbol{\delta}_{tt}\} = \{ \pm \bfa_1, \pm \bfa_2, \pm (\bfa_1 - \bfa_2) \}$, $\{\boldsymbol{\delta}_{hh}\} = \{(0,0), - \bfa_1, - \bfa_2\}$, $\{\boldsymbol{\delta}_{tA}\} = \{\bfa_1, \bfa_2, \bfa_1 + \bfa_2\}$ and $\{\boldsymbol{\delta}_{tB}\} = \{ (0,0), \bfa_1, \bfa_2\}$. 

In this article, we focus on the regime where the Heisenberg exchange couplings $J_{th}$ between spins on different sublattices are 
smaller than the couplings within the two sublattices
\begin{align}
  \label{eq:3}
  J_{th} < J_{tt}, J_{hh} \,.
\end{align}
This situation is realized, for example, in a system of two layers with weak interlayer couplings; we will discuss possible experimental realizations in Sec.~\ref{sec:summary-outlook}. A good starting point for our analysis is therefore the ground state of individual honeycomb and triangular sublattices, and in the following sections we derive the low-energy action around the classical ground state. 

\subsection{Order Parameter Symmetry and Long-Wavelength Gradient Action}
\label{sec:order-param-symm}
Let us start from the ground state of decoupled sublattices, \emph{i.e} considering $J_{th} = 0$. This state will turn out to be stable up to some critical coupling $J_{th} > 0$. In agreement with the Hohenberg-Mermin-Wagner (HMW) theorem~\cite{PhysRev.158.383,PhysRevLett.17.1307}, magnetic order only occurs at strictly zero temperature. At $T=0$, the honeycomb lattice exhibits uniaxial N\'eel order since it is a bipartite lattice. The magnetic order is described by a normalized vector $\bfn = (n_x, n_y, n_z)$ that points along the magnetization on the $A$ sites. The magnetization on the $B$ sites points along $(-\bfn)$. The symmetry of the honeycomb order parameter is therefore $\bfn \in \text{O}(3)/\text{O}(2)$. The magnetic ground state of the triangular lattice, on the other hand, is non-collinear. Neighboring spins on a plaquette arrange in a $120^\circ$ configuration with respect to each other (see Fig.~\ref{fig:2}). The order is described by three orthonormal vectors $\{\bft_1, \bft_2, \bft_3\}$, where we take $\bft_{1}$ and $\bft_2$ to span the plane of the triangular magnetization. The chirality of the magnetic order is encoded in the direction of the third vector $\bft_3 = \bft_1 \times \bft_2$ [or $\bft_3 = - (\bft_1 \times \bft_2)$]. We may group the vectors into an orthogonal matrix $t = (\bft_1, \bft_2, \bft_3)$, and the chirality is thus determined by the sign of $\det(t) = \pm 1$. Since for smooth spin configurations, which we restrict ourselves to, the sign of $\det(t)$ cannot change by continuity, the order parameter manifold reads $t \in \text{SO(3)}$. 

At finite temperatures $T>0$, magnetic correlations decay exponentially on both sublattices over finite correlation length-scales, $\xi_h$ and $\xi_t$ for the honeycomb and
the triangular lattices respectively. The order parameters $\bfn(x)$ and $t(x)$ are now spatially fluctuating. We assume that the magnetic correlation length is larger than the lattice spacing $\xi_{h}, \xi_t \gg a_0$, which is the case for temperatures $T < J_{tt}, J_{hh}$. 

The gradient part of the long-wavelength action takes the form of a O(3)/O(2) $\times$ SO(3) NLSM. As we derive in Appendix~\ref{sec:deriv-long-wavel}, it reads 
\begin{align}
  \label{eq:4}
  S_0 = \int d^2 x \Bigl( \frac{K}{2} (\partial_\mu \bfn)^2 + \sum_{j=1}^3 \frac{K_j}{2} (\partial_\mu \bft_j)^2 \Bigr)\,.
\end{align}
This equation describes the elastic energy cost of long-wavelength spatial
spin-wave fluctuations of the order parameter fields. The
dimensionless elastic energy scale is set by the spin stiffnesses
$\{K, K_j\}$, which are determined microscopically by the ratio of
Heisenberg exchange couplings $J_{ab}$ to temperature $T$.
In a $1/S$ expansion, where $S$ is the length of the spins, we show in Appendix~\ref{sec:deriv-long-wavel} that the spin stiffnesses are given by~\cite{PhysRevLett.50.1153,PhysRevB.39.6797,PhysRevB.39.2344}
\begin{align}
  \label{eq:5}
  K &= \frac{J_{hh} S^2}{\sqrt{3} T} \\
  K_1 &= K_2 = \frac{\sqrt{3} J_{tt} S^2}{4 T} \\
  K_3 &= 0 \,.
\end{align}
Since the coupling constant $K_3$ will be generated during the RG flow, it is included already in the beginning. 

In contrast to the $J_1$-$J_2$ square lattice
case~\cite{PhysRevLett.64.88}, in the ``windmill model'' there are no
gradient terms coupling the different sublattices and $S_0$ is
independent of $J_{th}$ (see Appendix~\ref{sec:coupling-term}). In the $J_{1}$-$J_{2}$ square lattice model, 
the long-wavelength action includes 
a gradient coupling between the two-antiferromagnetic
sublattices of the form~\cite{PhysRevLett.64.88}
\begin{equation}
\label{eq:213}
S_{\text{sq.;coupling}}\sim \int d^{2}x\left(\partial_{x} {\bfn_{1}}\cdot
\partial_{y} {\bfn_{2}} - 
\partial_{y} {\bfn_{1}}\cdot
\partial_{x} {\bfn_{2}}  
 \right)
\end{equation}
where $\bfn_{1}$ and $\bfn_{2}$ are the sublattice magnetizations of
the two interpenetrating antiferromagnets. This term is invariant under time-reversal and the point-group symmetries of the lattice.  One might expect a similar coupling of the form
\begin{equation}
\label{eq:223}
S_{{c1}}\sim \int d^{2}x \;
\kappa_{\alpha \beta }\left(\partial_{\alpha } {\bft^{1,2}}\cdot
\partial_{\beta } {\bfn} 
 \right)
\end{equation}
between $\bfn $ and the ``inplane'' components of the SO(3) order parameter $\bft^{1}$ and $\bft^{2}$, or alternatively, 
\begin{equation}
\label{eq:225}
S_{{c2}}\sim \int d^{2}x
\; \kappa_{\alpha \beta }\left(\partial_{\alpha } {\bft^{3}}\cdot
\partial_{\beta } {\bfn} 
 \right)
\end{equation}
between the third component of the SO(3) order parameter and $\bfn $. Here, $\kappa_{\alpha \beta}$ refers to the coupling between different sublattices. 
However, Eq.~\eqref{eq:223} is not invariant under 60$^{\circ }$ lattice
rotations and Eq.~\eqref{eq:225} is not invariant under time-reversal;
this is because $\bfn $ reverses under time-reversal whereas 
$\bft^{3}$, a pseudo-vector, does not. Therefore coupling terms like $S_{c1}$ and 
$S_{c2}$ are not permitted by symmetry.  In this way we can qualitatively 
eliminate the possibility of gradient couplings between the two sublattices, and a  
rigorous analysis is presented in Appendix~\ref{sec:coupling-term}.

\subsection{Potential terms in the long-wavelength action}
\label{sec:potent-terms-acti}
In the absence of fluctuations, \emph{i.e.}, for classical spins at zero temperature, one easily sees that at each site the exchange fields from all neighboring spins exactly cancel each other, both for triangular and honeycomb spins. Since apart from global O(3)/O(2) $\times$ SO(3) transformations the ground state is non-degenerate, we can conclude by continuity that this state remains the classical ground state of the system for a range of small non-zero couplings $J_{th}$. 
Quantum and thermal fluctations, on the other hand, will induce a coupling of the magnetic order parameters on different sublattices. This is the well-known ``order-from-disorder'' mechanism. It is a general principle that spins tend to align themselves perpendicular to the fluctuating Weiss field of the surrounding spins on the other sublattice~\cite{PhysRevLett.62.2056}, thereby maximizing the coupling of their respective fluctuating exchange fields. Since the fluctuating Weiss field of a given spin points perpendicular to the direction of this spin, it follows that spins on different sublattices prefer a ``maximally aligned'' relative configuration. Below we will find this from an explicit calculation. 

In addition to the gradient terms $S_0$, the long-wavelength action contains potential terms arising from those short-wavelength spin fluctuations~\cite{PhysRevLett.64.88}. They probe the local environment of the spins, and favor a certain \emph{relative orientation} of the two order parameters $\bfn(x)$ and $t(x)$. 
Below, we derive the potential terms in a $1/S$ expansion and find
\begin{align}
  \label{eq:6}
  S_c &= \frac12 \int d^2 x \bigl( \gamma \cos^2 (\beta) + \lambda \sin^6 (\beta) \sin^2 (3 \alpha) \bigr)
\end{align}
with $\gamma > 0$ and $\lambda > 0$. The azimuth $\alpha$ and polar angle $\beta$ describe the relative orientation of spins on different sublattices as defined in Fig.~\ref{fig:2}(b). In terms of the local order parameter triads the two potential terms read 
\begin{align}
  \label{eq:7}
  \gamma \cos ^2 (\beta) = \gamma (\bfn \cdot \bft_3)^2 
\end{align}
and 
\begin{align}
  \label{eq:8}
  \lambda \sin^6 (\beta) \sin^2 (3 \alpha) = \lambda \bigl[ ( \bfn \cdot \bft_2)^3 - 3 (\bfn \cdot \bft_2) (\bfn \cdot \bft_1)^2 \bigr]^2 \,.
\end{align}
The amplitude $\gamma$ describes the tendency towards a coplanar spin configuration where the honecomb spins lie everywhere in the plane of the spatially varying triangular magnetization $\bfn(x) \perp \bft_3(x)$. The six-fold potential term $\lambda$ energetically favors a configuration where the honeycomb spins point along one of the six equivalent directions parallel or anti-parallel to one of the three neighboring triangular spins on a plaquette. 

\begin{figure}[t]
  \centering
  \includegraphics[width=\linewidth]{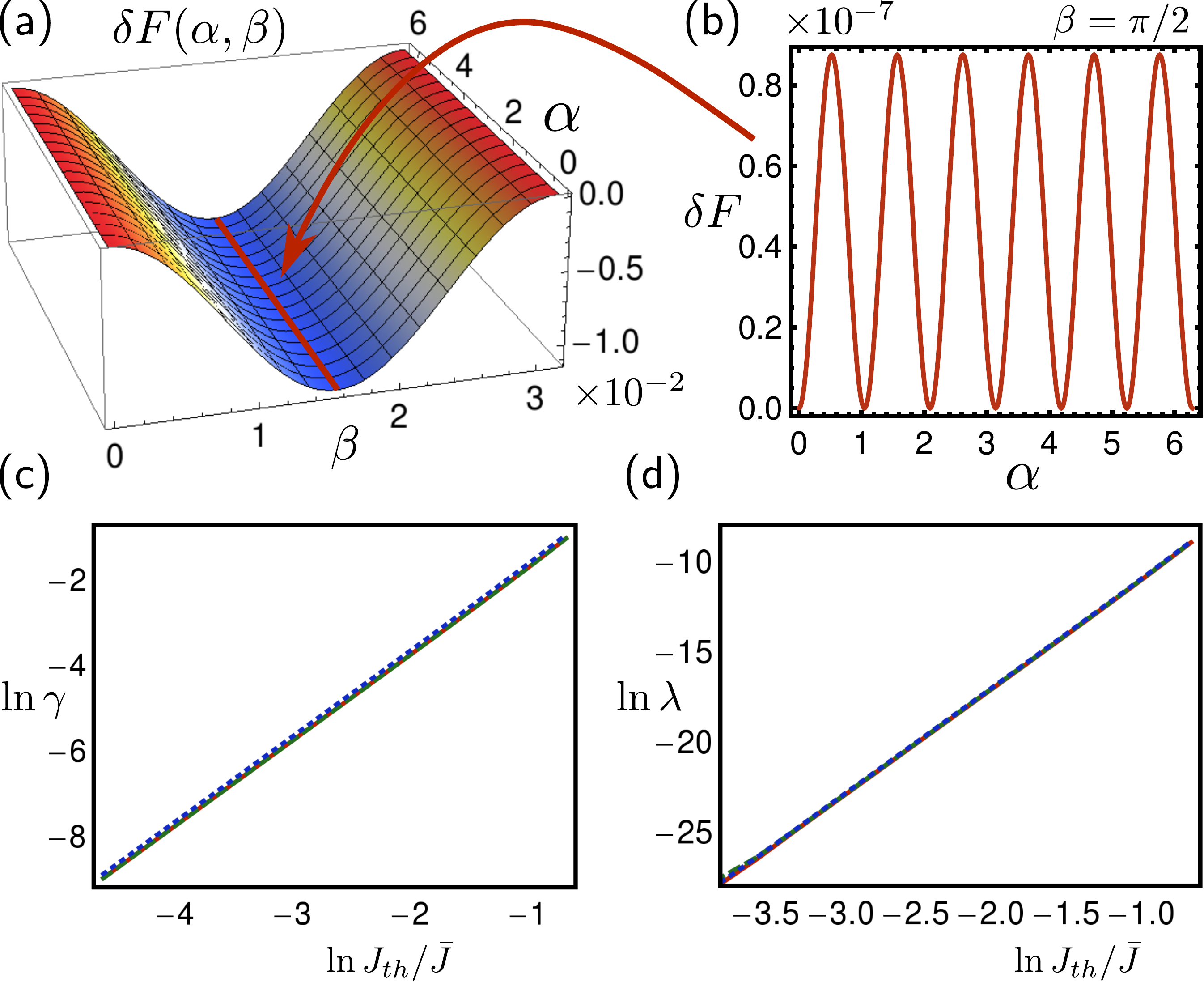}
  \caption{(Color online). (a) Fluctuation free energy $\delta F(\alpha, \beta)$ for $J_{tt} = J_{hh} = 1$, $J_{th} = 0.2 J_{tt}$ and $T = 0.5 J_{tt}$. (b) Fluctuation free energy $\delta F(\alpha, \pi/2)$ exhibits six-fold symmetry as function of in-plane angle $\alpha$. (c) Coplanar amplitude $\gamma$ as function of $J_{th}/\bar{J}$ exhibits $\gamma \sim (J_{th}/\bar{J})^2$ scaling. Plot is for $T = 0.5 \bar{J}$ and includes three different values of $(J_{tt}, J_{hh}) = \{(2,0.5), (1,1), (1,4) \}$ (red, green dashed, blue dotted). The dependence on the ratio $J_{tt}/J_{hh}$ is weak. Insets zoom into certain region of the plot. (d) Six-fold potential $\lambda$ as function of $J_{th}/\bar{J}$ exhibits $\gamma \sim (J_{th}/\bar{J})^6$ scaling. Parameters are identical to panel (c).}
  \label{fig:3}
\end{figure}

The potential terms in Eq.~\eqref{eq:6} are derived by calculating corrections to the free energy due to spin fluctuations. We perform a Holstein-Primakov spin-wave expansion around the classical ground state in Fig.~\ref{fig:2}, which takes both quantum and thermal fluctuations into account. Details can be found in Appendix~\ref{sec:spin-wave-expansion}, where we show that the fluctuation correction to the free energy $\delta F = F(J_{th}) - F(J_{th} = 0)$ as a function of angles $\alpha$ and $\beta$ takes the form 
\begin{align}
  \label{eq:9}
  \delta F(\alpha, \beta) &= T \sum_{\bfp \in \text{MBZ}} \sum_{i} \ln \biggl( \frac{\sinh\bigl[ E_{i,\bfp}(J_{th})/2T \bigr] }{ \sinh \bigl[ E_{i, \bfp}(0)/2T \bigr] } \biggr) \,.
\end{align}
Here, $\bfp$ is taken from the magnetic Brillouin zone (MBZ) and $E_{i,\bfp}(J_{th}, \alpha, \beta)$ is the spin-wave energy of the $i$th band, which is numerically known exactly. We present $\delta F(\alpha, \beta)$ for fixed values of $J_{ab}$ and $T$ in Fig.~\ref{fig:3}a and~b. 
From the free energy $\delta F$ we can identify the coupling action $S_c = \delta F /T$ with bare potential strengths
\begin{align}
  \label{eq:10}
  \gamma &= (J_{th}/\bar{J})^2 \; A_\gamma(J_{tt}/J_{hh}, \bar{J}/T) \\
\label{eq:11}
  \lambda &= (J_{th}/\bar{J})^6 \; A_\lambda(J_{tt}/J_{hh}, \bar{J}/T)\,.
\end{align}
We have defined $\bar{J} = \sqrt{J_{tt} J_{hh}}$ and the dimensionless functions $A_\gamma$ and $A_\lambda$ depend only weakly on the ratio $J_{tt}/J_{hh}$ (see Fig.~\ref{fig:3}c and~d. While the coplanar term $\propto \gamma$ appears already at second order perturbation theory in $J_{th}$, the six-fold potential term $\propto \lambda$ appears only at sixth order. It involves interaction of a honeycomb spin with all its three neighboring triangular spins. 

The sign of $\gamma$ determines whether the magnetization of the honeycomb lattice tends to lie perpendicular to the plane of triangular magnetization ($ \gamma < 0$) or coplanar ($\gamma > 0$). We find $\gamma > 0$ favoring coplanarity (see Fig.~\ref{fig:3}a), which is in agreement with the ``order-from-disorder'' principle of ``maximal relative alignment'' mentioned above. The six-fold symmetric potential $\lambda$, which is only relevant for $\gamma > 0$, requires zooming into Fig.~\ref{fig:3}a as $\lambda/\gamma \sim \mathcal{O}(J_{th}^4/\bar{J}^4) \ll 1$. This is shown in Fig.~\ref{fig:3}(b) for the coplanar configuration $\beta = \pi/2$.

The functions $A_\gamma$ and $A_\lambda$ can be exactly calculated numerically and we show in Fig.~\ref{fig:3}(c) and (d) for different ratios of $J_{tt}/J_{hh}$ that $A_\gamma$ and $A_\lambda$ are only very weakly dependent on the ratio $J_{tt}/J_{hh}$. Explicit analytical expressions are obtained by combining an expansion at high and at low temperatures compared to the bandwidth of the spin-wave spectrum, where one finds
\begin{align}
  \label{eq:12}
  A_\gamma &=  f_T(J_{tt}/J_{hh}) \mathcal{G}_T + f_Q(J_{tt}/J_{hh})\mathcal{G}_Q \frac{\bar{J} S }{T}  \\
\label{eq:13}
  A_\lambda &= f_T(J_{tt}/J_{hh}) \mathcal{H}_T + f_Q(J_{tt}/J_{hh}) \mathcal{H}_Q \frac{\bar{J} S }{T} 
\end{align}
with $f_T(x) \approx \frac{0.015}{\sqrt{x}} + 0.98 + 0.005 \sqrt{x} $, $\mathcal{G}_T = 0.95$, $\mathcal{G}_Q = 0.09$ and $f_Q(x) \approx -\frac{0.23}{\sqrt{x}} + 1.37 - 0.19 \sqrt{x}$ $\mathcal{H}_T = 5 \times 10^{-3}$, $\mathcal{H}_Q = 2 \times 10^{-4}$. The form of the functions $f_{T}$ and $f_Q$, which fulfill $f_T(1) = f_Q(1) = 1$, are obtained from a simple fit of the exact numerical result. 

\subsection{Complete long-wavelength action}
\label{sec:compl-long-wavel-1}
We arrive at the full long-wavelength action $S = S_0 + S_c$ by combining the gradient term $S_0$ in Eq.~\eqref{eq:4} and the potential terms in Eq.~\eqref{eq:6}:
\begin{align}
  \label{eq:14}
  S &= \int d^2 x \Bigl( \frac{K}{2} (\partial_\mu \bfn)^2 + \sum_{j=1}^3 \frac{K_j}{2} (\partial_\mu \bft_j)^2 \Bigr) \nonumber \\ & \quad + \frac12 \int d^2 x \bigl( \gamma \cos^2 (\beta) + \lambda \sin^6 (\beta) \sin^2 (3 \alpha) \bigr) \,.
\end{align}
The $\text{O(3)/O(2)} \times \text{SO(3)}$ gradient terms describe the elastic energy of spatial spin fluctuations and turns out to be independent of $J_{th}$. The potential terms, on the other hand, couples the order parameters $\bfn(x)$ and $t(x)$ of the two sublattices and depends on the relative orientation of the spins on different sublattices. The derivation of the action $S$ assumes a classical ground of the form depicted in Fig.~\ref{fig:2}, which is the ground state of the system for relatively weak inter-sublattice coupling $J_{th} < J_{tt}, J_{hh}$. We also assume that the magnetic correlation lengths on the two sublattices $\xi_t$ and $\xi_h$, respectively, are both larger than the lattice constant  $a_0$, which holds for temperatures $T \lesssim \bar{J}$. 


\section{Wilson-Polyakov and Friedan RG approaches}
\label{sec:outline-rg-program}
The action $S$ in Eq.~\eqref{eq:14} is the starting point for the renormalization group (RG) analysis that we perform to determine the phase diagram of the system. The RG analysis is separated into three temperature regions, going from high to low temperatures, as described briefly in the introduction. In this section we set the stage to perform this RG analysis, by first describing the two distinct scaling procedures that we employ. 

We want to discuss and contrast the conceptual underpinnings of the two scaling procedures, the Wilson-Polyakov~\cite{polyakov_nlsm_75,RevModPhys.47.773,RevModPhys.55.583,Polyakov_GaugeFieldsAndStrings_Book,ChandraColeman-LesHouches} and the Friedan approaches~\cite{PhysRevLett.45.1057,Friedan-AnnPhys-1985}, used in this paper to follow the renormalization group flows of the two-dimensional windmill model. Both methods integrate or "smooth'' out the short-wavelength fluctuations in the magnetization of the spin system, following the resulting flow of its spinwave stiffnesses; however the methodologies are very different but yield the same results. 

In general the local orientation of the axes of an antiferromagnet are determined by a $D$ dimensional vector $\bfxx(\bfx)$ parametrized by coordinates $\bfx$ in $d$ dimensions. In the following we allow for general dimensions $d$ with $d=2$ in case of the windmill model. For example in a simple uniaxial magnet with order parameter symmetry $\text{O(3)/O(2)}$ the vector $\bfxx= (\theta,\phi) $ is a two-dimensional spin magnitude containing the spherical co-ordinates of the magnetization, whereas for a biaxial helical magnet with order parameter symmetry $\text{SO}(3)$, $\bfxx= (\theta,\phi,\psi)$ are the three Euler angles that define the orientation of a local triad of vectors.  The gradient part of the action can then be written as (cf. Eq.~\eqref{eq:4})
\begin{align}
\label{eq:15}
  S_0 &= \frac12 \int d^dx \sum_{i,j=1}^D\sum_{\mu = 1}^d g_{ij}\bigl[ \bfxx(\bfx) \bigr] \; \bigl(\partial_\mu X^i \bigr) \bigl(\partial_\mu X^j\bigr) \,,
\end{align}
where the metric $g_{ij}(\bfxx)$ define the spinwave stiffnesses and the vector $\bfxx(\bfx)$ depends on $\bfx = (x_1,x_2,\ldots, x_d)$, which are the spatial coordinates in $d = 2 + \epsilon$ dimensions. This is an Euclidean version of a Nambu-Goto string theory action~\cite{Polchinski_Book}. Whereas in magnetism $\bfx$ is the physical coordinate and $\bfxx$ is the magnetization, in the context of string theory $\bfxx$ is the string displacement in $D$-dimensional spacetime and $\bfx = (\tau, y_1,...y_{d-1})$ is the parameter space where $\tau$ is time and $\bfy$ is the coordinate along the string ($d$-brane).

The basic philosophy underlying Wilson-Polyakov scaling of two-dimensional spin systems is to divide the spin fluctuations into short- and long-wavelength components, integrating out the fast degrees of freedom while maintaining the spin amplitude fixed, a sort of ``poor man's scaling'' approach to magnetism~\cite{PhysRevLett.23.89,PhysRevB.1.1522}. The magnetization $\bfxx(\bfx)$ is divided into a coarse-grained slow long-wavelength component $\bfxx_<(\bfx)$ and one due to short-wavelength fluctuations $\bfxx_>(\bfx)$, 
\begin{equation}
\bfxx(\bfx) = \bfxx_<(\bfx) + \bfxx_>(\bfx)\,.
\end{equation}
If the Fourier transform of $\bfxx(\bfxx)$ involves wavevectors from $\bfq \in [0,\Lambda]$ then the Fourier transform of $\bfxx_<$ involves wavevectors $\bfq\in [0,\Lambda/b]$, where $b = e^l > 1$ is the dilation factor, while $\bfxx_>$ involves wavevectors in the small sliver $\bfq \in [\Lambda/b, \Lambda]$ of momentum space~\cite{Cardy_Book}. The action is then expanded to Gaussian order in the fast fluctuations,
\begin{equation}
S_0[\bfxx_< + \bfxx_>] = S_0[\bfxx_<] + \frac{\delta S_0}{\delta \bfxx_>} \bfxx_> + \frac{1}{2} \bfxx_> \frac{\delta ^2 S_0}{\delta \bfxx_>} \bfxx_> \,.
\end{equation}
By integrating out the fast Gaussian degrees of freedom $\bfxx_>$ and rescaling $ \bfx \rightarrow \bfx b $, the action is now renormalized; the renormalizations in the stiffnesses are described by a set of $\beta$ functions, 
\begin{equation}
\frac{\partial g_{ij}}{\partial \ln \Lambda} \equiv \frac{\partial g_{ij}}{\partial l } = \beta_{ij} [g]
\end{equation}
with $l = \ln b$ and
\begin{equation}
\beta_{ij} = (d-2)g_{ij} + \mathcal{O}(g^2) \,.
\end{equation}
The first term results from the rescaling of spatial coordinates, and the terms quadratic in $g$ emerge from the Gaussian integral over $\bfxx_>$. 

\begin{figure}[t]
  \centering
  \includegraphics[width=\linewidth]{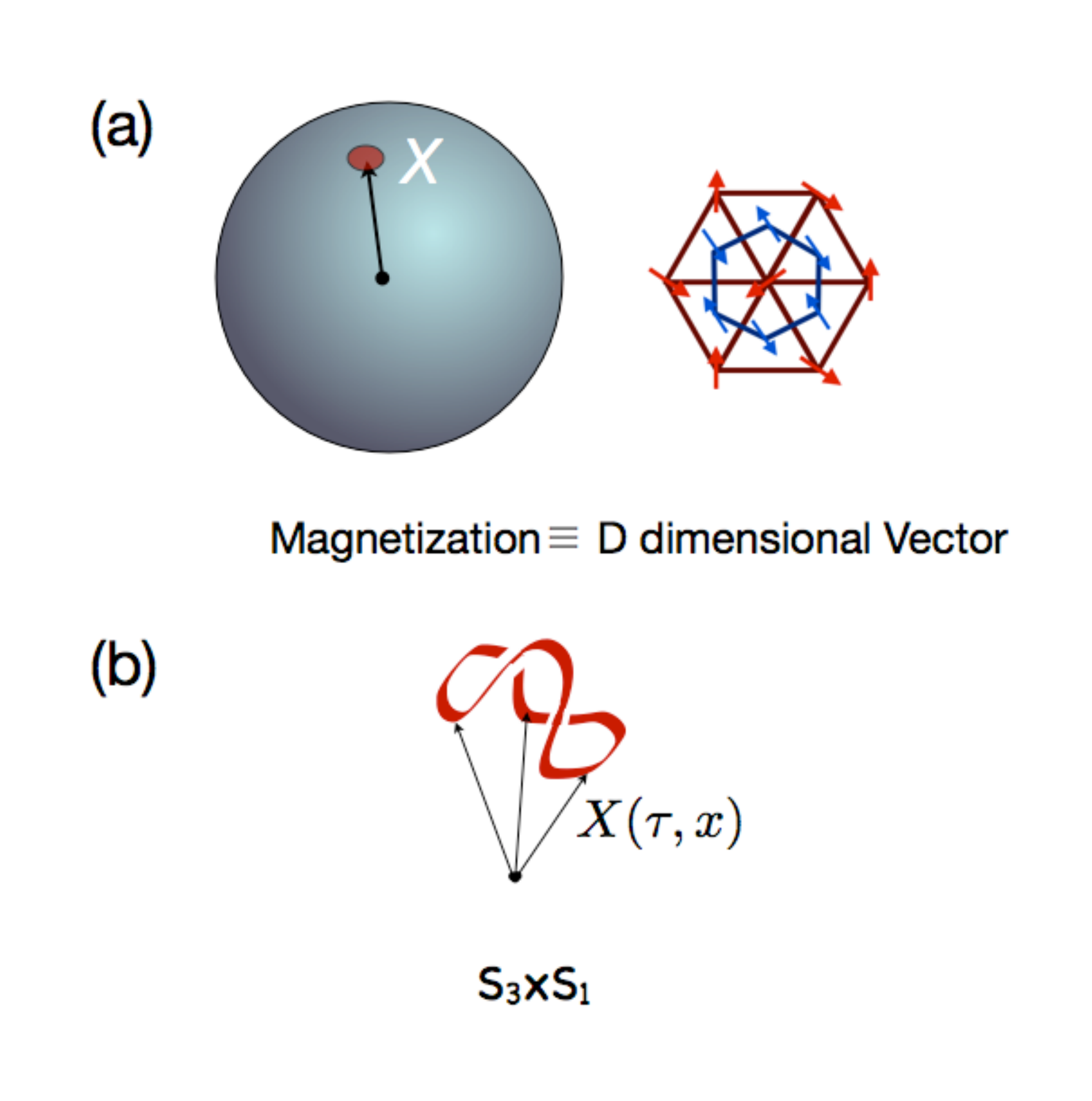}
  \caption{(Color online). Schematic to illustrate the Friedan approach\cite{PhysRevLett.45.1057} to the windmill model. (a) The magnetization is in general a $D$-dimensional vector where $D=4$ for the windmill model. (b) In Friedan's methodology the long-wavelength action of the magnet is treated as a Nambu-Goto action of a string with coordinates $X(\tau, x)$ moving in a $D$-dimensional target space. Here, $\tau$ refers to the time and $x$ to the position along the string. For the coplanar regime of the windmill model the target space is a 4-dimensional manifold $S_3 \times S_1$ associated with the $SO(3) \times U(1)$ symmetry of the action.} 
  \label{fig:4}
\end{figure}

By contrast, in the Friedan approach~\cite{PhysRevLett.45.1057,Friedan-AnnPhys-1985} the action of the $2+\epsilon$-dimensional spin system is treated as a kind of ``mini-string theory'' where the coordinates $\bfxx(\bfx)$ are regarded as the coordinates of a string (or ``$d$-brane'') in a $D$-dimensional target space. In a $d=2$ dimensional coordinate space (note the distinction with the $D=4$ dimensional target space that will be relevant for the windmill model here), we can identify the first component of $\bfx = (x,y)$ as the time coordinate $\tau$, so that $(x,y) \rightarrow (\tau, y)$ and  $\bfxx(\tau, y)$ describes the time evolution of the string coordinate at time $\tau$ and at position $y$ along the string. For the windmill model, as we shall discuss in detail shortly, the magnetization in the coplanar regime is a function of four Euler angles and thus is a $D=4$ vector; in Fig.~\ref{fig:4}, we display a schematic to depict the Friedan approach in this case.

Friedan's essential observation was that the action of the system is covariant under coordinate changes in target space, $\bfxx\rightarrow \bfxx'$, provided that
\begin{equation}
\label{eq:16}
g_{ij}[\bfxx]\rightarrow g_{ij}'[\bfxx'] = \sum_{k,l} g_{kl} \frac{\partial X^{k}}{\partial X^{'i}} \frac{\partial X^{l}}{\partial X^{'j}}  \,.
\end{equation}
This is precisely the covariance of a metric tensor
\begin{equation}
\label{eq:17}
ds^2 = \sum_{i,j} g_{ij} dX^i dX^j
\end{equation}
under the coordinate transformation $\bfxx \rightarrow \bfxx'$. With this identification, Friedan established a mapping between the renormalization group flows of NLSMs and  ``Ricci flow'' describing the slow time evolution of a geometric manifold. Friedan reasoned that since the action $S[\bfxx]$ is covariant, the same is true of the scaling; thus the coefficients of the $\beta$ function must be second-rank tensors with the same transformation properties as the metric tensor $g_{ij}$. Indeed the only tensors available are $g_{ij}$ itself, and two-component contractions of the Riemann tensor $R^k_{lij}$ defined below; this places significant constraints on the form of the $\beta$ function.  For $(2+\epsilon)$-dimensional NLSM, Friedan showed that the renormalization group flow of the spin stiffnesses up to two-loop order is given by the Ricci flow of the metric tensor~\cite{PhysRevLett.45.1057,Friedan-AnnPhys-1985}
\begin{align}
\label{eq:18}
  \frac{d g_{ij}}{dl} &= \epsilon g_{ij} - \frac{1}{2 \pi} R_{ij} - \frac{1}{8 \pi^2} R_i{}^{klm} R_{jklm} \,.
\end{align}
The Riemann tensor $R^k{}_{lij}$ is determined by the Christoffel symbols 
\begin{align}
\label{eq:19}
  \Gamma^i_{jk} &= \frac12 g^{il} \bigl( g_{jl,k} + g_{kl,j} - g_{jk,l} \bigr)
\end{align}
as
\begin{align}
\label{eq:20}
  R^k{}_{lij} &= \Gamma^k_{lj,i} - \Gamma^k_{li,j} + \Gamma^k_{ni} \Gamma^n_{lj} - \Gamma^k_{nj} \Gamma^n_{li}\,.
\end{align}
and here we use the standard notation $g_{ij,k} = \frac{\partial g_{ij}}{\partial X^k}$. 
The leading order loop contribution of the RG flow is determined by the Ricci tensor $R_{ij}$, which is a contraction of the Riemann tensor
\begin{align}
\label{eq:21}
  R_{ij} &= R^k{}_{ikj} \,.
\end{align}

The application of the Friedan approach to two-dimensional magnetism on a lattice provides a beautiful link between the statistical mechanics of $d=2$  magnetism and the geometry of a string theory.  Integrating out the short-wavelength fluctuations of the magnet, we find that its stiffness renormalizes.  In the Friedan mapping this corresponds to integrating out the high-frequency fluctuations of the string.  When these fluctuations are removed, the metric and hence the underlying geometry of space defined by $ds^2=\sum_{i,j} g_{ij}dX^i dX^j$ evolves according to Ricci flow.  $g$ becomes smaller and the size of the ``universe'' decreases; thus the renormalization of the spinwave stiffness in a $d=2$ Heisenberg magnet is linked with the compactification of spacetime in a $D$-dimensional string theory. In the windmill model we will see later that the decoupling of the $\text{U(1)}$ degrees of freedom to form a decoupled $XY$ magnet can be viewed from the string perspective as the formation of a  one-dimensional ``universe'', decoupled from its compactified $D-1=3$ interior dimensions.  

As we demonstrate in this paper, the Wilson-Polyakov and the Friedan scaling approaches yield identical results for the renormalization flows of the spin stiffnesses.  In order to be self-contained and to introduce the interested reader to both methodologies, we have included detailed technical Appendices where all results are derived with both approaches, and as electronic Supplementary Material we also provide a \emph{Mathematica} file that includes the computation of the RG equations via the Ricci flow~\cite{SupplMat-FriedanScaling}. In the main text, however, we focus mainly on the results of these calculations for the frustrated windmill model. 

\section{RG  analysis at high temperatures}
\label{sec:renorm-group-equat}
In this section we investigate the windmill model at high temperatures. The triangular and honeycomb sublattices are then approximately uncoupled, because the bare potential values $\gamma, \lambda \ll 1$ since $J_{th}/\bar{J} \ll 1$. The symmetry of the system is SO(3) $\times$ O(3)/O(2). The RG flow equations are therefore given by those of the uncoupled honeycomb and triangular lattices~\cite{polyakov_nlsm_75,PhysRevLett.64.3175}. In order for the reader to obtain familiarity with the Wilson-Polyakov and Friedan scaling methods, we rederive those equations in Appendix~\ref{sec:derivation-rg-flow}. As electronic Supplementary Material we provide a \emph{Mathematica} file that includes the computation of the RG equations via Friedan scaling~\cite{SupplMat-FriedanScaling}.

The potential terms are both RG relevant, since they contain no derivatives. Thus, they increase exponentially under the RG. As soon as coplanar amplitude becomes of order unity $\gamma(l_\gamma) \simeq 1$, scaling stops and the system undergoes a crossover into a coplanar regime, which is discussed in Sec.~\ref{sec:rg-analysis-coplanar}.

\subsection{Derivation of RG equations}
\label{sec:deriv-rg-equat}
The RG proceeds from the action $S$ in Eq.~\eqref{eq:14} and successively integrates out short-wavelength degrees of freedom to arrive at an effective action $S'$ that only contains slow modes. Those modes dominate the behavior at low temperatures. The effective action $S'$ has the same form as $S$, but contains modified parameters $\{K(l), K_i(l), \gamma(l), \lambda(l)\}$ that depend on the RG flow parameter $l$ that determines the increased lattice constant of the effective action $a(l) = a_0 e^l$. We first bring the action $S$ into a form amenable to the two RG procedures discussed above. We then derive the RG equations in the uncoupled regime. Technical details are given in Appendix~\ref{sec:derivation-rg-flow}. 

To bring the action $S$ into a suitable form to perform the RG calculation, we first rewrite the action~\eqref{eq:14} in terms of matrix fields 
\begin{align}
  \label{eq:22}
 t(x) = \bigl(\bft_1(x), \bft_2(x), \bft_3(x) \bigr)  \in \text{SO(3)}
\end{align}
and
\begin{align}
  \label{eq:23}
  h(x) = \bigl( \bfh_1(x), \bfh_2(x), \bfh_3(x) \bigr) \in \text{SO(3)} \,.
\end{align}
Here, $\bfn(x) = \bfh_1(x)$ denotes the direction of the staggered magnetization on the honeycomb lattice, and $\bfh_2$ and $\bfh_3$ are two orthonormal vectors that complete the local triad describing magnetic order on the honeycomb lattice. In matrix form the action in Eq.~\eqref{eq:14} reads
\begin{align}
  \label{eq:24}
  S &= \frac{1}{4} \int d^2x \; \text{Tr} \Bigl[(\partial_\mu Q_h)^T (\partial_\mu Q_h)  \Bigr] \nonumber \\ & \qquad + \frac12 \int d^2x \; \text{Tr} \Bigl[ K_t ( \partial_\mu t^{-1}) (\partial_\mu t) \Bigr]  + S_c \,,
\end{align}
where we have defined the matrix $Q_h = h K_h h^{-1}$ and the diagonal stiffness matrices
\begin{align}
  \label{eq:25}
  K_h &= \text{diag} (\sqrt{K}, 0,0) \\
  K_t &= \text{diag} (K_1, K_2, K_3) \,.
\end{align}
The first (second) term in Eq.~\eqref{eq:24} describes spins on the honeycomb (triangular) lattice. In general, the triangular coupling matrix $K_t$ contains three independent stiffnesses $\{K_1, K_2, K_3\}$, but in our case it holds initially that $K_1 = K_2$ and this is preserved during the RG flow. 

The first term in Eq.~\eqref{eq:24} defines the $O(3)/O(2)$ NLSM of the honeycomb lattice. Here, two elements $h(x)$ and $h'(x) = h(x) r(x)$ of the coset space are identical, if they only differ by (local) rotation $r(x) \in (O(2)$ around the $\bfh_1$ axis. It is therefore useful to define the NLSM in terms of the matrix $Q_h = h K_h h^{-1}$ since $Q_h$ is constant if $[K_h, h] = 0$. A functional integral over the matrices $Q_h$ thus runs automatically over the coset space $O(3)/O(2)$. Note that a straightforward expansion shows that the action in Eq.~\eqref{eq:24} is identical to Eq.~\eqref{eq:14}.

It will be useful for us to rewrite the action~\eqref{eq:24} in yet another form using angular velocities as 
\begin{align}
  \label{eq:26}
  S &= \frac12 \int_x \biggl\{ K [ (\Omega_\mu^2)^2 + (\Omega_\mu^3)^2 ] + \sum_{a=1}^3 I_a (\tilde{\Omega}_\mu^a)^2 \biggr\} + S_c 
\end{align}
with $\int_x = \int d^2 x$ and $I_a = K_b + K_c$ where $a \neq b \neq c$. Here, we have defined angular velocities for the order parameter on the honeycomb and triangular lattice 
\begin{align}
  \label{eq:27}
  \Omega_\mu &= h^{-1} (\partial_\mu h) = - i \sum_{a=1}^3\Omega_\mu^a \tau_a \\
\label{eq:28}
  \tilde{\Omega}_\mu &= t^{-1} (\partial_\mu t) = - i \sum_{a=1}^3 \tilde{\Omega}_\mu^a \tau_a
\end{align}
The 3$\times$3 matrices $\tau_a$ fulfill the SU(2) algebra $[\tau_a, \tau_b] = i \epsilon_{abc} \tau_c$ and take the adjoint form $(\tau_a)_{bc} = i \epsilon_{bac}$. Different components of the angular velocity are obtained from $\Omega_\mu^a = \frac{i}{2} \text{Tr}(\Omega_\mu \tau_a)$ and $\tilde{\Omega}_\mu^a = \frac{i}{2} \text{Tr}(\tilde{\Omega}_\mu \tau_a)$. Note the analogy of Eq.~\eqref{eq:26} to the action of a spinning top with moments of inertia $I_a$ around the principal axes.

Next, we express the matrix fields $t,h$ in terms of Euler angles, and write
\begin{align}
  \label{eq:29}
  h &= e^{- i \phi_h \tau_2} e^{- i \theta_h \tau_3} e^{- i \psi_h \tau_1 } \\
\label{eq:30}
  t &= e^{- i \phi_t \tau_2} e^{- i \theta_t \tau_3} e^{- i \psi_t \tau_1} \,.
\end{align}
We use a convention of Euler angles such that the angle $\psi_h$ immediately drops out of the action as $[K_h, \tau_1] = 0$ and $Q_h$ is independent of $\psi_h$. In total, five Euler angles are required to describe the local orientation of spins, three angles $\{\phi_t, \theta_t, \psi_t\}$ for the triangular lattice and two angles $\{\phi_h, \theta_h\}$ for the honeycomb lattice, reflecting the SO(3) $\times$ O(3)/O(2) symmetry. 

The action is now in a form useful to derive scaling equations for the spin stiffnesses within both RG schemes; both methods are discussed in detail in Appendix~\ref{sec:derivation-rg-flow}. Here, we focus on Polyakov scaling which proceeds by separating $t$ and $h$ into slow and fast fields, performing an integration over the fast modes which is followed by momentum and field rescaling. First, the matrix fields are expressed as a product of matrices containing only slow and fast components in the Euler angles: $h = h_< h_>$ and $t = t_< t_>$. Here, $h_<, t_<$ are rotation matrices that only contain slowly fluctuating Euler angles 
\begin{align}
  \label{eq:31}
  h_< = e^{- i \phi_h^< \tau_2} e^{- i \theta_h^< \tau_3} e^{- i \psi_h^< \tau_1 }
\end{align}
and $h_>, t_>$ contain only fast fluctuating fields
\begin{align}
  \label{eq:32}
  h_> = e^{- i \phi_h^> \tau_2} e^{- i \theta_h^> \tau_3} e^{- i \psi_h^> \tau_1 } \,.
\end{align}
Corresponding equations exist for $t_<$ and $t_>$. Then, one expands to quadratic order in the fast angles and performs the functional integral over the fast modes. Expanding to quadratic order corresponds to a one-loop approximation, the small parameters being inverse stiffnesses $g_h = 1/K \ll 1$ and $g_t = 1/K_1 \ll 1$. Finally, we rescale momenta and fields to arrive at the renormalized action. The coupling of fast and slow modes leads to a renormalization of spin stiffnesses and potential amplitudes. 

\subsection{Scaling equations and coplanar crossover}
\label{sec:analysis-rg-flow}
Iterating the RG procedure as shown in Appendix~\ref{sec:derivation-rg-flow} one obtains the scaling equations for the spin stiffnesses
\begin{align}
  \label{eq:33}
  \frac{d}{ dl} K &= - \frac{1}{2 \pi} \\
\label{eq:34}
  \frac{d}{dl} K_1 &= - \frac{ ( 1 + \eta)^2}{8 \pi} \\
\label{eq:35}
  \frac{d}{dl} \eta &= - \frac{\eta (1 + \eta)^2}{4 \pi K_1}  \,,
\end{align}
where we have defined the triangular lattice anisotropy 
\begin{align}
  \label{eq:36}
  \eta = \frac{K_1 - K_3}{K_1 + K_3} \,.
\end{align}
and the flow parameter $l$ determines the running cutoff $\Lambda(l) = a_0^{-1} e^{-l}$.
These equations hold in the uncoupled lattice regime at high temperatures and are the known flow equations of individual honeycomb and triangular lattice~\cite{polyakov_nlsm_75, PhysRevLett.64.3175}.
Solving Eqs.~\eqref{eq:33}-\eqref{eq:35} yields 
\begin{align}
  \label{eq:37}
  K(l) &= K(0) - \frac{l}{2 \pi} \\
\label{eq:38}
  K_1(l) &= K_1(0)/\Bigl( \sqrt{3}  \tan\Bigl[ \frac{\pi}{6} + \frac{\sqrt{3}}{8 \pi} \frac{l}{K_1(0)}  \Bigr] \Bigr) \\
\label{eq:39}
  \eta(l) &= \eta(0) [K_1(l)/K_1(0)]^2 \,,
\end{align}
where we have used that initially $K_1(0) = K_2(0)$. The stiffnesses are reduced at longer length-scales, which is in agreement with the Hohenberg-Mermin-Wagner theorem. If it holds initially that $K_1 = K_2$, this is preserved during the RG flow. Importantly, the anisotropy $\eta(l)$ is irrelevant and flows from its initial value of $\eta(0) = 1$ towards zero. The stiffnesses of the triangular lattice approach an isotropic fixed point with all stiffnesses being equal. These equations are derived under the assumption that the potential terms are small $\gamma, \lambda \ll 1$, \emph{i.e.} neglecting $S_c$. The potential amplitudes $\gamma$ and $\lambda$, however, scale as
\begin{align}
  \label{eq:40}
  \frac{d}{dl} \gamma &= 2 \gamma \\
\label{eq:41}
  \frac{d}{dl} \lambda &= 2 \lambda \,,
\end{align}
and thus grow exponentially 
\begin{align}
  \label{eq:42}
  \gamma(l) &= \gamma(0) e^{2l} \\
\label{eq:43}
  \lambda(l) &= \lambda(0) e^{2l} \,.
\end{align}
Scaling therefore stops as soon as $\gamma(l_\gamma) = 1$, which defines the coplanar lengthscale 
\begin{align}
  \label{eq:44}
  a_\gamma = a_0 e^{l_\gamma} \simeq a_0 \bar{J}/J_{th} \,.
\end{align}
This condition marks a crossover to a coplanar regime where the honeycomb spins tend to lie in the plane of the triangular spins. This transition occurs as a crossover rather than a phase transition since no symmetry is being broken. The crossover occurs when $a_\gamma$ is comparable to the shorter of the two magnetic correlation lengths $\xi_t$ and $\xi_h$. In case of $J_{hh} < J_{tt}$ this occurs at the coplanar crossover temperature
\begin{align}
  \label{eq:45}
  T_{cp} \simeq \frac{J_{hh} S^2}{1 + \ln(1/\gamma(0 ))/4 \pi} \,.
\end{align}
In the opposite case of $J_{tt} < J_{hh}$ one obtains an implicit expression for the coplanar temperature
\begin{align}
  \label{eq:46}
    T_{cp} &= \frac{J_{tt} S^2}{4} \cot \biggl[ \frac{1}{8 \pi} \Bigl( \frac{4 \pi^2}{3} + \frac{2 \ln(1/\gamma(0) ) }{J_{tt} S^2} T_{cp} \Bigr) \biggr] \,,
\end{align}
that also approaches zero only logarithmically as $\gamma(0) \rightarrow 0$ [see Eq.~\eqref{eq:10} and Fig.~\ref{fig:3}(c)]. The coplanar temperature is defined as $T_{cp} = \text{min}_{\alpha=1,2} \, T_{cp}^{(\alpha)} $, where $T^{(1)}_{cp}$ is determined by the conditions $ K\bigl(l_\gamma, T_{cp}^{(1)}\bigr) = 1$ and $K_1\bigl(l_\gamma, T_{cp}^{(1)}\bigr) > 1$, while $T^{(2)}_{cp}$ is determined by the conditions $ K\bigl(l_\gamma, T_{cp}^{(2)}\bigr) > 1$ and $K_1\bigl(l_\gamma, T_{cp}^{(2)}\bigr) = 1$.

\section{Coplanar regime at intermediate temperatures}
\label{sec:rg-analysis-coplanar}
For temperatures below $T_{cp}$, spins on different sublattices order coplanar. Once they are coplanar, we can assume that $\bfn \cdot \bft_3 = 0$ since fluctuations of the polar angle around $\beta = \pi/2$ are massive. The azimuth $\alpha$ remains as a soft U(1) degree of freedom. The coplanar system is thus determined by a SO(3) $ \times$ U(1) order parameter defined in terms of three Euler angles $\{\phi, \theta, \psi\}$ and a single relative phase $\alpha$.
In this section, we derive the RG equations in the coplanar regime by enforcing this condition as a hard-core constraint. The final values of the previous flow in the uncoupled regime $\{K(l_\gamma), K_i(l_\gamma), \lambda(l_\gamma)\}$ serve as initial parameters in the coplanar RG equations. 

Solving the RG scaling equations, we prove that the U(1) angle $\alpha$ asymptotically decouples from the underlying SO(3) Euler angles, which exhibit correlations only over finite lengthscales. This decoupling is crucial for the emergence of a critical phase and associated BKT transitions, since otherwise, vortices in the relative angle $\alpha$ would not necessarily interact logarithmically due to screening effects that occur via the coupling to the SO(3) degrees of freedom. Within the Friedan geometric scaling approach, this decoupling of the phase can be regarded as a toy model for the compactification of a four-dimensional string theory. 

\subsection{Action in the coplanar regime}
\label{sec:acti-copl-regime}
To implement the constraint $\bfn \cdot \bft_3 = 0$, we express the triangular matrix field $t = \bigl( \bft_1(x), \bft_2(x), \bft_3(x) \bigr)$ in terms of the honeycomb matrix field $h$ as 
\begin{align}
  \label{eq:47}
  t = h U
\end{align}
where
\begin{align}
  \label{eq:48}
  U &= \exp( - i \alpha \tau_3) \,.
\end{align}
The azimuth $\alpha$ determines the relative in-plane orientation of the spins on the two sublattices (see Fig.~\ref{fig:2}(b)). In the coplanar regime it is convenient to choose a different convention for Euler angles and write
\begin{align}
  \label{eq:49}
  h &= e^{- i \phi_h \tau_3} e^{- i \theta_h \tau_1} e^{- i \psi_h \tau_3} \,.
\end{align}
The physical content of the theory is of course independent of the choice of Euler angles, but with Eq.~\eqref{eq:49} the phase angle $\alpha$ simply shifts the Euler angle $\psi$. Substituting $t = h U$ into the action in Eq.~\eqref{eq:24} yields 
\begin{align}
  \label{eq:50}
  S &= - \frac12 \int_x \text{Tr} \Bigl[ K_t \Bigl\{ \Omega_\mu^2 + u_\mu^2 + 2 u_\mu \Omega_\mu \Bigr\} \Bigr]\\ & \qquad  +\frac{1}{4} \int_x \; \text{Tr} \Bigl[(\partial_\mu Q_h)^T (\partial_\mu Q_h)  \Bigr]  + S_c\Bigl(\beta = \frac{\pi}{2}\Bigr) \nonumber 
\end{align}
with angular velocities $\Omega_\mu = h^{-1} (\partial_\mu h)$ as well as $u_\mu = U^{-1} (\partial_\mu U)$. Repeated indices $\mu = 1,2$ are summed over. We have used that $[U, K_t] = 0$ in case of $K_1 = K_2$. The initial values of the parameters $\{K_j, K\}$ ($j = 1,2,3$) are set by the final values of the flow in the uncoupled regime at $l = l_\gamma$.

If we insert $t = h U$ in Eq.~\eqref{eq:26}, we immediately see that the coplanar action can alo be written in the general form 
\begin{align}
  \label{eq:51}
  S &= \frac12 \int_x \Bigl( I_1 (\Omega_\mu^1)^2 + I_2 (\Omega_\mu^2)^2 + I_3 (\Omega_\mu^3)^2 \nonumber \\ & \qquad + I_\alpha (\partial_\mu \alpha)^2 + \kappa (\partial_\mu \alpha) \Omega_\mu^3 \Bigr) + S_c
\end{align}
with $\Omega_\mu^a = \frac{i}{2} \text{Tr} (\Omega_\mu \tau_a)$ and SO(3) stiffnesses 
\begin{align}
  \label{eq:52}
  I_1 &= K_2(l_\gamma)  + K_3(l_\gamma) \\
\label{eq:53}
  I_2 &= K_1(l_\gamma) + K_3(l_\gamma) + K(l_\gamma) \\
\label{eq:54}
  I_3 &= K_1(l_\gamma) + K_2(l_\gamma) + K(l_\gamma)\,,
\end{align}
where $K_1 (l_\gamma) = K_2(l_\gamma)$. Note that in contrast to the pure triangular case, here it turns out that $I_1 \neq I_2$ due to the coupling of the two sublattices. The U(1) degree of freedom $\alpha$ has an initial stiffness of
\begin{align}
  \label{eq:55}
  I_\alpha &= 2 K_1(l_\gamma) \,.
\end{align}
The coupling constant between the SO(3) and U(1) sectors is given by
\begin{align}
  \label{eq:56}
  \kappa = 2 [ K_1(l_\gamma) + K_2(l_\gamma) ]\,,
\end{align}
which is of the same order as the stiffnesses and thus not small. The sixfold potential 
\begin{align}
  \label{eq:57}
  S_c\Bigl(\beta = \frac{\pi}{2} \Bigr) &= \frac{\lambda}{2} \int_x \sin^2(3 \alpha) 
\end{align}
is a small but relevant perturbation to the gradient part of the action. 

\subsection{Derivation of RG equations}
\label{sec:deriv-rg-equat-1}
To derive the RG flow equations in the coplanar regime both the Wilson-Polyakov as well as the Friedan RG approach may be used, and we present both calculations in Appendix~\ref{sec:derivation-rg-flow-1}. Within the Wilson-Polyakov scheme we perform a one-loop RG by introducing fast and slow modes $h = h_< h_>$, $U = U_< U_>$ and $\alpha = \alpha_< + \alpha_>$, expanding in and integrating over the fast modes and performing the rescaling. This procedure is presented in Appendix~\ref{sec:poly-scal-copl}. Alternatively, we may use the Friedan approach and exploit the analogy between the Ricci flow of a relativistic metric of a string theory and the RG equation of the NLSM. This yields the flow equations up to two-loops. This calculation is presented in detail in Appendix~\ref{sec:ricci-flow-rg}, and in the electronic Supplementary Material~\cite{SupplMat-FriedanScaling}. The scaling equations of the six-fold potential $\lambda$ are derived in Appendix~\ref{sec:flow-six-fold}. 

The main question that we have to answer is whether the U(1) sector decouples from the non-Abelian SO(3) part with a finite stiffness $I_\alpha$. It is more natural to formulate clear decoupling criteria within the Friedan approach, where the gradient part of the action takes the form of Eq.~\eqref{eq:15} with a stiffness metric tensor
\begin{align}
  \label{eq:58}
  g &= \bmx g^{\text{SO(3)}} & \mathcal{K}^T \\ \mathcal{K} & I_\alpha \emx  \,.
\end{align}
It contains a coupling $ \mathcal{K} = \frac{\kappa}{2} \bigl( \cos \theta, 0, 1\bigr)$ between the U(1) part $I_\alpha$ and the SO(3) part that reads
\begin{widetext}
\begin{align}
  \label{eq:59}
  g^{\text{SO(3)}} &= \bmx (I_1 \sin^2 \psi + I_2 \cos^2\psi) \sin^2 \theta + I_3 \cos^2 \theta & (I_1 - I_2) \sin \theta \cos \psi \sin \psi & I_3 \cos \theta \\
  (I_1 - I_2)  \sin \theta \cos \psi \sin \psi & I_1 \cos^2 \psi + I_2 \sin^2 \psi & 0 \\
  I_3 \cos \theta & 0 & I_3 \emx\,,
\end{align}
\end{widetext}
In contrast to the isolated triangular lattice, here, the stiffnesses $I_1 \neq I_2$ (see Eqs.~\eqref{eq:52} and~\eqref{eq:53}).
The coupling term $\mathcal{K}$ can be eliminated by a variable transformation of the Euler angle 
\begin{align}
  \label{eq:60}
  \psi \rightarrow \psi' = \psi + r \alpha 
\end{align}
with shift $r = \kappa/2 I_3$. This yields a metric 
\begin{align}
  \label{eq:61}
  g  = \bmx g^{SO(3)}(\theta, \phi, \psi'(\alpha) ) & 0 \\ 0 &  I'_\alpha \emx
\end{align}
with $\mathcal{K} = 0$ and rescaled U(1) stiffness
\begin{align}
  \label{eq:62}
  I'_\alpha &= I_\alpha - \frac{\kappa^2}{4 I_3} \,.
\end{align}
The coupling between the U(1) and the SO(3) sectors is hidden in the fact that $\psi'$ depends on the U(1) phase $\alpha$. From this gauge transformation to the appropriate center of mass coordinates two clear decoupling criteria emerge: the metric $g^{\text{SO(3)}}$ becomes independent of the angle $\alpha$ if either the system becomes isotropic in the $I_1$-$I_2$-plane
  \begin{align}
    \label{eq:63}
    |I_2 - I_1| \ll \sqrt{I_1 I_2}
  \end{align}
or if the shift of the Euler angle $\psi \rightarrow \psi' $ is small
\begin{align}
  \label{eq:64}
  r \ll 1 \,.
\end{align}
In both cases, the U(1) phase $\alpha$ decouples from the dynamics of the noncollinear magnetic degrees of freedom $\{\theta, \phi, \psi\}$. The first criterion follows from the fact that $g^{\text{SO(3)}}$ is independent of the angle $\psi'$ if $I_1 = I_2$ (see Eq.~\eqref{eq:59}), while the second criterion implies that the shift of the Euler angle $\psi$ is negligible. As we show below, it depends on the ratio $J_{tt}/J_{hh}$ which of the decoupling criteria applies. 
\subsection{Analysis of scaling equations}
\label{sec:analys-rg-equat}
The derivation of the RG flow equations of the variables $I_1$, $I_2$, $I_3$, $I'_\alpha$ and $r$ is presented in Appendix~\ref{sec:derivation-rg-flow-1}. The flow equation for the six-fold potential $\lambda$ is derived in Appendix~\ref{sec:flow-six-fold}. The qualitative results is already fully captured by the one-loop equations, which are given by
\begin{figure}[t]
  \centering
  \includegraphics[width=\linewidth]{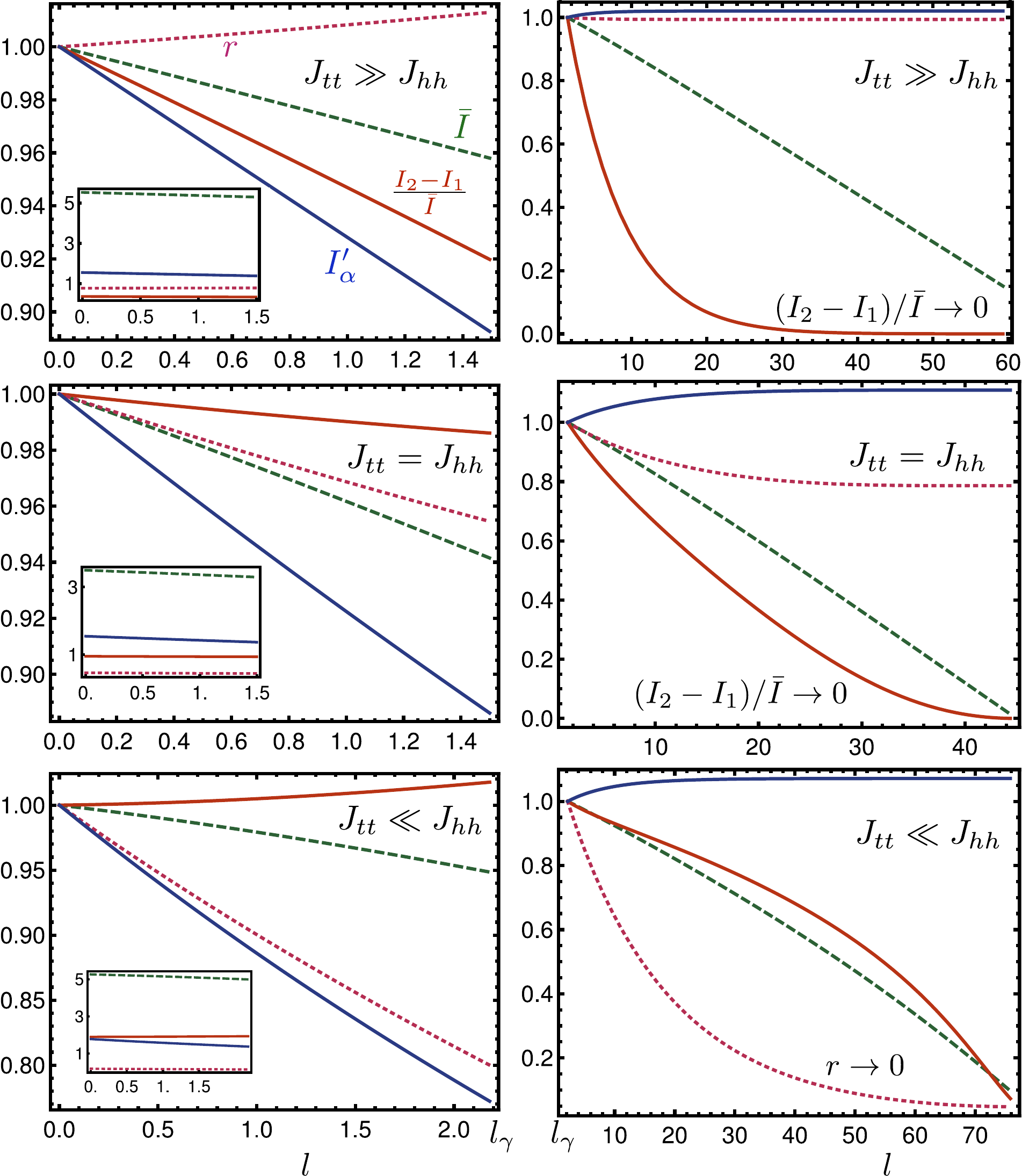}
  \caption{(Color online) Renormalization group flow of spin stiffnesses and coupling constants $\bar{I}=(I_1 I_2 I_3)^{1/3}$ (green dashed), $(I_2 - I_1)/\bar{I}$ (red), $r$ (pink dotted) and $I'_\alpha$ (blue). Left column is flow in the uncoupled lattice regime, where $\gamma(l) \ll 1$ and right column is in the coplanar regime. Curves are normalized to initial values at $l=0$ ($l=l_\gamma$) for uncoupled (coplanar) flow. Inset shows non-normalized results. 
Upper panel is for $J_{tt} \gg J_{hh}$ with $J_{tt} = 2$, $J_{hh} = 0.5$, $J_{th} = 0.2$, $T = 0.25$. Initial values at $l=0$ read $\bar{I} = 5.5$, $(I_2 - I_1)/\bar{I}=0.36$, $r = 0.78$, $I'_\alpha = 1.55$, and initial values at $l = l_\gamma$ are given by $\bar{I} = 5.30$, $(I_2 - I_1)/\bar{I}=0.33$, $r = 0.79$, $I'_\alpha = 1.39$. 
Middle panel is for isotropic system $J_{tt} = J_{hh}$ with $J_{tt} = 1$, $J_{hh} = 1$, $J_{th} = 0.2$, $T = 0.3$. Initial values at $l=0$ read $\bar{I} = 3.5$, $(I_2 - I_1)/\bar{I}=0.95$, $r = 0.46$, $I'_\alpha = 1.55$, and initial values at $l = l_\gamma$ are given by $\bar{I} = 3.30$, $(I_2 - I_1)/\bar{I}=0.94$, $r = 0.44$, $I'_\alpha = 1.38$. 
Lower panel is for $J_{tt} \ll J_{hh}$ with $J_{tt} = 0.5$, $J_{hh} = 2$, $J_{th} = 0.2$, $T = 0.3$. Initial values at $l=0$ read $\bar{I} = 3.5$, $(I_2 - I_1)/\bar{I}=0.95$, $r = 0.46$, $I'_\alpha = 1.55$, and initial values at $l = l_\gamma$ are given by $\bar{I} = 3.30$, $(I_2 - I_1)/\bar{I}=0.94$, $r = 0.44$, $I'_\alpha = 1.38$. 
}
  \label{fig:5}
\end{figure}
\begin{align}
  \label{eq:65}
  \frac{d}{dl} I_1 &= \frac{- I_1^2 + (I_2 - I_3)^2}{4 \pi I_2 I_3} + \frac{(I_2^2 - I_1^2) r^2 }{ 4 \pi I_2 I'_\alpha} \\
\label{eq:66}
  \frac{d}{dl} I_2 &= \frac{ - I_2^2 + (I_1 - I_3)^2}{ 4 \pi I_1 I_3} + \frac{(I_1^2 - I_2^2) r^2}{ 4 \pi I_1 I'_\alpha } \\
\label{eq:67}
  \frac{d}{dl} I_3 &= \frac{ - I_3^2 + (I_1 - I_2)^2}{ 4 \pi I_1 I_2} \\
\label{eq:68}
  \frac{d}{dl} I'_\alpha &= \beta_{I'_\alpha} = \frac{(I_1 - I_2)^2 r^2}{4 \pi I_1 I_2} \\
\label{eq:69}
  \frac{d}{dl} r &= - \frac{(I_1 - I_2)^2 r }{4 \pi I_1 I_2 I_3} \\
 \label{eq:70}
  \frac{d}{dl} \lambda &= \Bigl( 2 - \frac{9}{\pi I'_\alpha} \Bigr) \lambda \,.
\end{align}
The initial values of the flow are given in Eqs.~\eqref{eq:52}-\eqref{eq:56} and $\lambda(l_\gamma) = \lambda(0) e^{2 l_\gamma}$. The shift of the decoupling transformation follows as 
\begin{align}
  \label{eq:71}
  r(l_\gamma) = \Bigl(1 + \frac{K(l_\gamma)}{2 K_1(l_\gamma)}\Bigr)^{-1}
\end{align}
and the rescaled U(1) stiffness as 
\begin{align}
  \label{eq:72}
  I'_\alpha(l_\gamma) = \Bigl(\frac{1}{2K_1(l_\gamma)} + \frac{1}{K(l_\gamma)}\Bigr)^{-1} \,.
\end{align}
In Fig.~\ref{fig:5} we present the coplanar RG flow for different sets of microscopic parameters corresponding to both weak and strong initial anistropies $|I_1 - I_2|/\sqrt{I_1 I_2}$.

Like in the case of the isolated SO(3) magnet, the spin stiffnesses $I_1$, $I_2$, $I_3$ are reduced during the flow towards longer lengthscales and approach an isotropic fixed point with $I_1 = I_2 = I_3$. The initial anisotropy in the $I_1$-$I_2$ plane is given by $|I_2 - I_1|/\sqrt{I_1 I_2} = K/[(K_1 + K_3)(K_1 + K_3 + K)]$. This anisotropy flows to zero faster if the coupling $r$ is large. This follows from the second term on the right hand side of Eqs.~\eqref{eq:65} and~\eqref{eq:66}. For weak initial anisotropies $K \ll K_1$, which is the case for $J_{hh} \ll J_{tt}$, it implies that the coupling $r \approx 1$ is large. Therefore, the decoupling of $\alpha$ emerges rapidly in this case since $|I_1 - I_2|/\sqrt{I_1 I_2} \rightarrow 0$ quickly in this case (see Fig.~\ref{fig:5} upper right). The SO(3) sector becomes isotropic in the $I_1$-$I_2$-plane. Note also that the shift $r$ remains almost constant during the flow for small anistropies, which follows directly from Eq.~\eqref{eq:69}. 

In the opposite regime of strong initial anisotropies $K \gg K_1$, which is the case for $J_{hh} \gg J_{tt}$, $|I_2 - I_1|/\sqrt{I_1 I_2}$ is not small. In this situation, however, we observe that the shift $r \approx 2K_1/K \ll 1$ is small. Moreover, $r$ vanishes faster in the presence of large anistropies, which follows from Eq.~\eqref{eq:69}. For $J_{hh} \gg J_{tt}$ the decoupling of $\alpha$ thus emerges because the shift $r$ of the Euler angle $\psi$ becomes negligible. 

In both cases of large and small initial anisotropy, the phase angle $\alpha$ rapidly emerges as an independent degree of freedom during the flow. The phase stiffness $I'_\alpha$ is determined by the smaller of the stiffnesses $K_1$ and $K$ - just like a reduced mass - and actually increases slightly during the flow. Therefore, $I'_\alpha$ is always finite at the decoupling lengthscale. As expected its $\beta$-function $\beta_{I'_\alpha}$ in Eq.~\eqref{eq:68} approaches zero once either of the two decoupling conditions $|I_1 - I_2| \rightarrow 0$ or $r\rightarrow 0$ is fulfilled. From then on, no further renormalization of the U(1) stiffness $I'_\alpha$ due to spin waves occurs perturbatively. Vortex excitations, on the other hand, lead to a further renormalization of $I'_\alpha$. We take this into account in the next Sec.~\ref{sec:phase-deco-effect}. 

Analyzing the scaling of the six-fold potential, we find that according to Eq.~\eqref{eq:70} the relevance of the six-fold potential $\lambda$ depends on the value of the $I'_\alpha$. The potential term is relevant only at sufficiently low temperatures when $I'_\alpha \geq 9/2 \pi$. At larger temperatures, $\lambda$ is irrelevant and flows to zero. The flow of $\lambda$ depends on the discrete symmetry of the potential term. For a potential with discrete $\mathbb{Z}_p$ symmetry, the flow equation for the coupling strength $\lambda_p$ is given by $\frac{d}{dl} \lambda_p = \bigl( 2 - p^2/4 \pi I'_\alpha) \lambda_p$. Since it holds that $I'_\alpha \geq 1$, the potential can only become irrelevant for $p \geq 5$. For the coplanar term $\gamma$, for example, $p = 2$ and the $p$-dependent correction term is negligible compared to the dominant tree-level scaling part.

\subsection{Geometric interpretation of decoupling}
\label{sec:geom-interpr-deco}
The geometric formulation of the RG flow allows for an intriguing interpretation of the decoupling of the U(1) phase $\alpha$ from the non-Abelian SO(3) sector of the theory. Computing the Ricci scalar $R = g^{ij} R_{ij}$ during the flow, we find
\begin{align}
  \label{eq:73}
  R = R^{\text{SO(3)}} - \frac{1}{2 \pi I'_\alpha} \beta_{I'_\alpha} \,.
\end{align}
It is given by a sum of the Ricci scalar of the SO(3) sector
\begin{align}
  \label{eq:74}
  R^{\text{SO(3)}} &= \sum_{j = 1}^3 \Bigl(I_j^{-1} - \frac{I_j^2}{2 I_1 I_2 I_3} \Bigr)
\end{align}
and a contribution from the coupled U(1) part that is proportional to the $\beta$-function $\beta_{I'_\alpha}$ (see Eq.~\eqref{eq:68}). 

\begin{figure}[t]
  \centering
  \includegraphics[width=.55\linewidth]{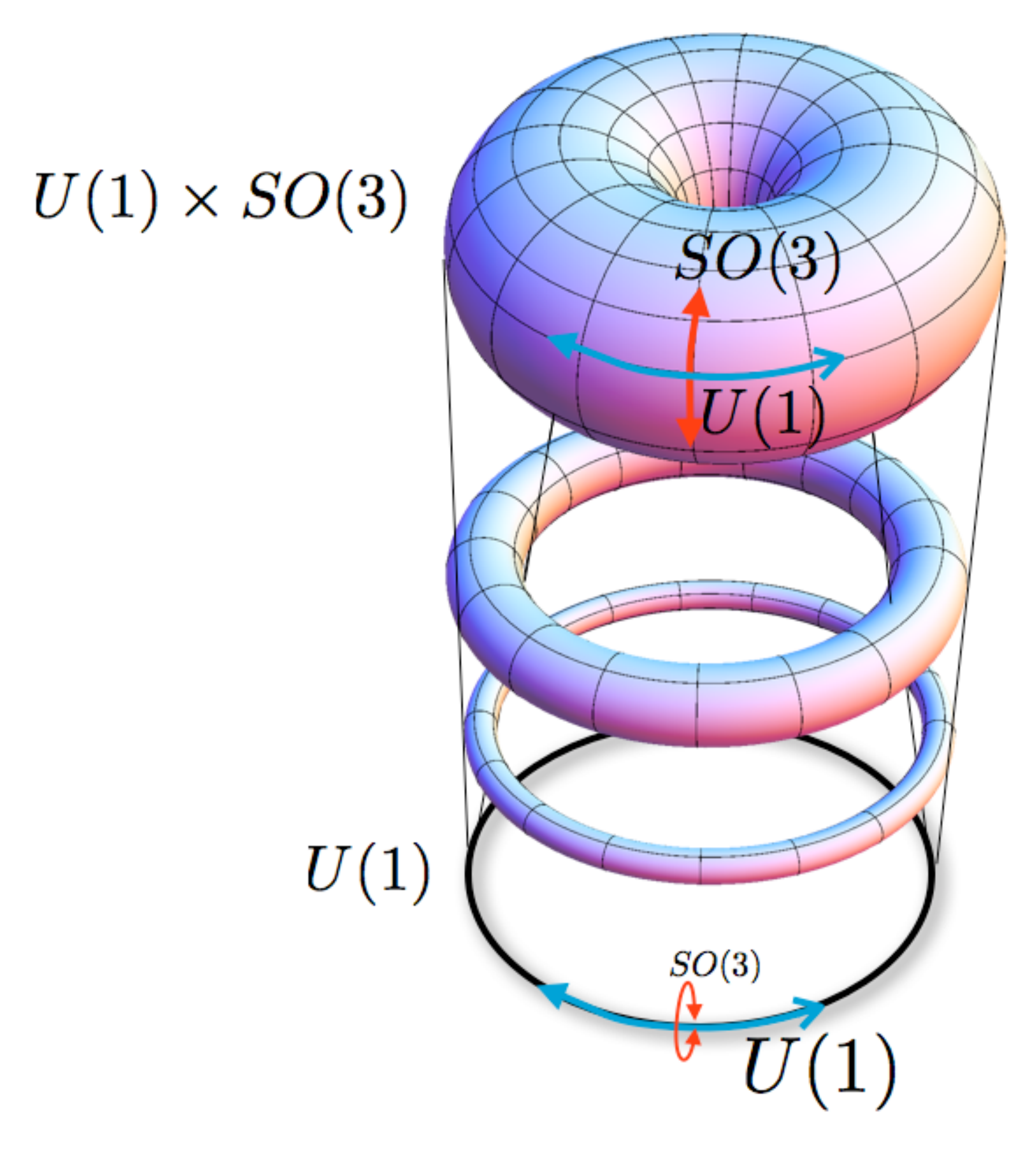}
  \caption{(Color online). Schematic of the decoupling of the flat U(1) sector and the "curling up" of the remaining SO(3) manifold 
under the renormalization group flow towards longer lengthscales in the windmill model. This is analogous to the phenomenon of compactification in string theory. }
  \label{fig:6}
\end{figure}

Once the decoupling occurs $\beta_{I'_\alpha} \rightarrow 0$, the contribution to $R$ from the U(1) sector becomes negligible. On the other hand, $R \rightarrow R^{\text{SO(3)}}$ grows under renormalization since the stiffnesses $I_j$ decrease. As shown schematically in Fig.~\ref{fig:6}, this corresponds to the intriguing situation of a curved four-dimensional manifold at large energies, which separates into a flat one-dimensional U(1) part that is only weakly coupled to the remaining three-dimensional SO(3) manifold at low energies. Towards smaller energies, the curvature of the SO(3) part grows larger and larger such that $R \rightarrow \infty$. This asymptotic decoupling of a subspace and ``curling-up'' of the complementary dimensions is analogous to the phenomenon of compactification in string theory. 

\section{Low temperature regime and phase diagram}
\label{sec:phase-deco-effect}
Once the decoupling of the U(1) phase $\alpha$ has occured the resulting low energy theory of the system is given by $S = S_{\text{SO(3)}} + S_{\mathbb{Z}_6}$ with 
\begin{align}
  \label{eq:75}
  S_{\text{SO(3)}} = \frac12 \int_x \sum_{i,j = 1}^3 g^{\text{SO(3)}}_{ij} (\partial_\mu X^i) (\partial_\mu X^j)
\end{align}
and 
\begin{align}
  \label{eq:76}
  S_{\mathbb{Z}_6} &= \frac12 \int d^2x \bigl[ I'_\alpha (\partial_\mu \alpha)^2 + \lambda \sin^3(3 \alpha) \bigr]
\end{align}
being the familiar action of the six-state clock model~\cite{PhysRevB.16.1217}. It describes a two-dimensional XY model in the presence of an additional six-fold potential $\lambda$. The $\mathbb{Z}_6$ clock model exhibits two consecutive BKT transitions~\cite{PhysRevB.16.1217,Ortiz2012780}. The upper transition temperature $T^>_{\text{BKT}}$ separates a disordered regime at high temperatures from a critical phase at lower temperatures, where correlations $\av{\exp(i[\alpha(x) - \alpha(x')])}$ in the relative phase angle $\alpha(x)$ decay as a power law in the distance $|x - x'|$. At the lower transition temperature $T^<_{\text{BKT}}$ the discrete $\mathbb{Z}_6$ symmetry is spontaneously broken. Below $T^<_{\text{BKT}}$ one observes true long-range order with phase $\alpha = n \pi/3$ $(n \in \{1, \ldots, 6 \})$ being locked into one of the six discrete minima of the potential. 

These conclusions follow from the BKT flow equations for the spin stiffness $I'_\alpha$ and the vortex fugacity $Y$, which we derive below, in combination with the flow equation~\eqref{eq:70} of the six-fold potential $\lambda$. We must rederive the BKT flow for our model to account for the fact that the vortex core size is given by the coplanar lengthscale $a_\gamma$ which is much larger than the microscopic lattice spacing $a_0 \ll a_\gamma$. A similar situation is described in detail in Ref.~\onlinecite{PhysRevLett.109.155703}, where it is shown that an increased vortex core size $a_\gamma/a_0 \gg 1$ leaves the BKT flow equations invariant but leads to an increased initial value of the vortex fugacity $y \rightarrow Y = y (a_\gamma/a_0)^2$. This enhancement of the fugacity can be understood physically by noticing that vortex excitations interact logarithmically only on lengthscales larger that $a_\gamma$, but the entropy associated with those excitations is obtained from counting different centers of the vortex core, which involves the microscopic lattice scale $a_0$~\cite{PhysRevLett.109.155703}. 

To obtain the low energy phase diagram we thus have to analyze the RG flow equations~\cite{Berezinskii-1972,KosterlitzThouless-JPhysC-1973,Kosterlitz-JPhysC-1974,ChaikinLubensky-Book,PhysRevB.16.1217, PhysRevLett.109.155703}
\begin{align}
  \label{eq:77}
  \frac{d}{dl} I'_\alpha{}^{-1} &= 4 \pi^3 Y^2 \\
\label{eq:78}
  \frac{d}{dl} Y &= ( 2- \pi I'_\alpha) Y \\
\label{eq:79}
  \frac{d}{dl} \lambda &= \Bigl( 2 - \frac{9}{\pi I'_\alpha} \Bigr) \lambda \,,
\end{align}
As noted above the initial value of the vortex fugacity is enhanced to $Y = (a_\gamma/a_0)^2 \exp[- S_c(T)]$~\cite{PhysRevLett.109.155703}. Here, $S_c \simeq \pi[ 1 + \text{min}(K, K_1, K_2)]$ denotes the core action of a vortex with core size $a_\gamma$. It contains a contribution from the elastic energy $S_0 = \frac{K}{2} \int_x (\partial_\mu \bfn)^2 + \sum_{j=1}^3 \frac{K_j}{2} \int_x (\partial_\mu \bft_j)^2 $ and one from to the potential energy $S_\gamma = \frac{1}{2 a_0^2} \int_x \gamma \cos^2 \beta$.

\begin{figure}[t]
  \centering
  \includegraphics[width=.55\linewidth]{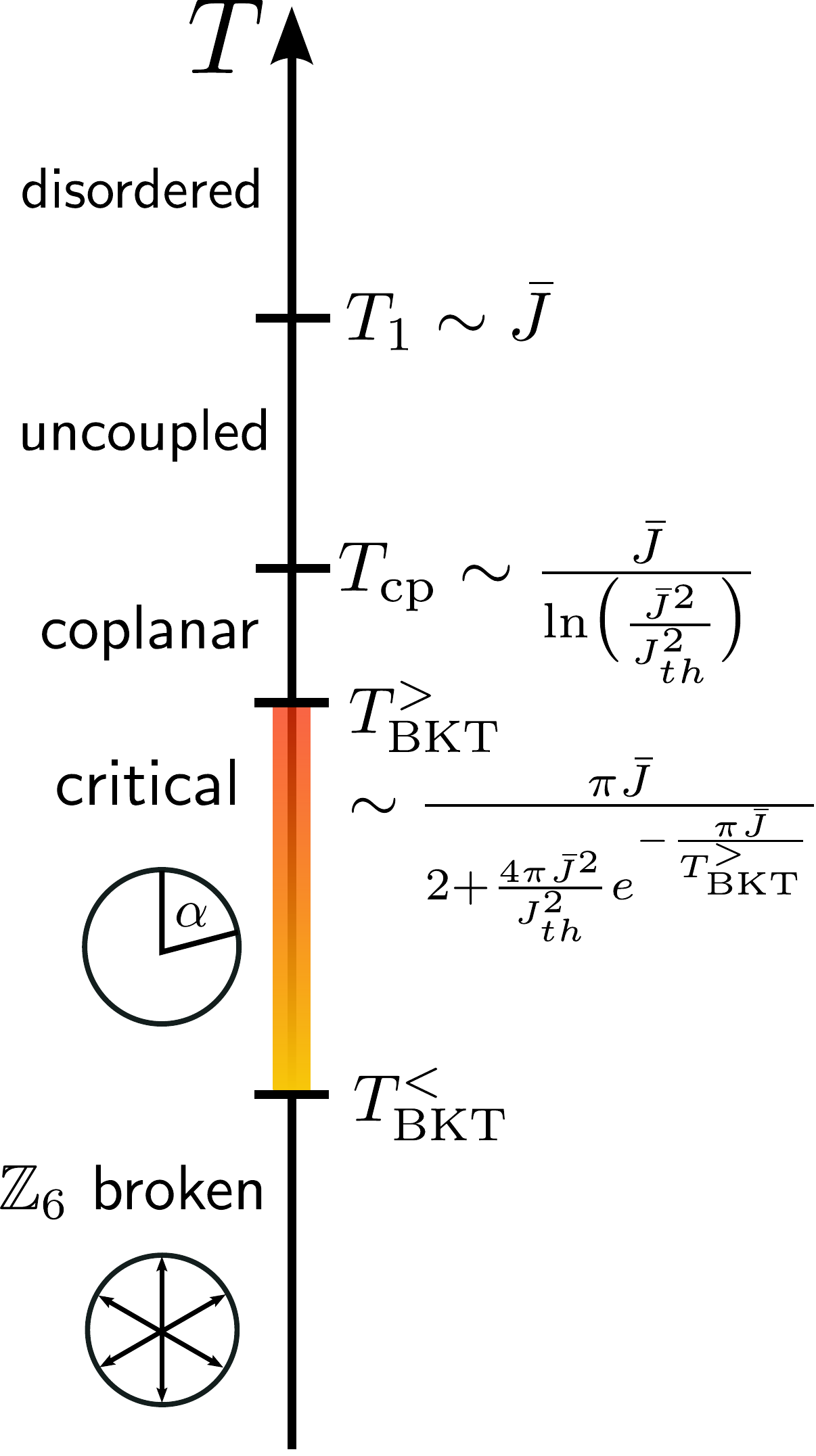}
  \caption{(color online). Schematic phase diagram of the system with approximate transition temperatures. While correlations in the relative U(1) angle decay exponentially $G_\alpha(x,x') = \av{e^{i [\alpha(x) - \alpha(x')]}}\sim e^{-|x-x'|/\xi}$ for $T > T^>_{\text{BKT}}$, the system shows algebraic order in the critical phase with $G_\alpha(x,x') \sim |x-x'|^{-\eta}$. In the $\mathbb{Z}_6$ broken phase below $T^<_{\text{BKT}}$ one observes true long range order with $ \lim_{|x-x'| \rightarrow \infty} G_\alpha(x,x') = \text{const.}$.   }
  \label{fig:7}
\end{figure}

We obtain the phase diagram for temperatures below the coplanar crossover from analyzing Eqs.~\eqref{eq:77}-\eqref{eq:79}. The resulting phase diagram including estimates for the transition temperatures is shown in Fig.~\ref{fig:7}. From Eq.~\eqref{eq:78} it follows that the vortex fugacity $Y$ is only relevant \emph{above} a certain temperature when $I'_\alpha(T) \leq 2/\pi$. On the other hand, the six-fold potential $\lambda$ is only relevant \emph{below} a certain temperature when $I'_\alpha \geq 9/2 \pi$. Since $\frac{9}{2 \pi} > \frac{2}{\pi}$, there is an intermediate temperature regime where both vortex fugacity and six-fold potential are irrelevant. In this regime, the system exhibits a critical phase. 

For a general $\mathbb{Z}_6$ symmetric potential, one finds that it becomes relevant below the temperature $I'_\alpha \geq p^2/8 \pi$. The intermediate temperature regime and critical phase, where both vortex fugacity and six-fold potential scale to zero, thus only exists for $p\geq 5$ (and thus $p^2 > 16$).

At higher temperatures above the critical phase, free vortices proliferate and the system shows only short-range order. At lower temperatures below the critical phase, the six-fold potential is relevant and leads to a locking of the phase into one of the six equivalent minima $\av{\alpha} = n \pi/3$ with $n \in \{ 1, \ldots, 6 \}$. The discrete $\mathbb{Z}_6$ symmetry is then spontaneously broken. 

We find the upper BKT transition temperature $T^>_{\text{BKT}}$, where vortices unbind, from the flow equations \eqref{eq:77}-\eqref{eq:79} in implicit form as 
\begin{align}
  \label{eq:80}
  I'_\alpha(T^>_{\text{BKT}})^{-1} &= \frac{\pi}{2 + 4 \pi Y(T^>_{\text{BKT}})} 
\end{align}
with $Y(T) = (a_\gamma/a_0)^2 e^{-S_c(T)} = e^{-S_c(T)}/\gamma$ and core action $S_c(T) \simeq \pi[ 1 + \text{min}(K, K_1, K_2)]$. As we show below, we conclude from Eq.~\eqref{eq:80} that the upper BKT transition occurs soon after the system becomes coplanar 
\begin{align}
  \label{eq:81}
  T^>_{\text{BKT}} \lesssim T_{\text{cp}} \,.
\end{align}
The BKT transition temperature is only numerically smaller than the coplanar crossover temperature. The system enters the critical phase soon after it becomes coplanar. The enhancement of the fugacity shifts the BKT transition temperature to the coplanar crossover scale. 

Let us now explicitly show this by solving Eq.~\eqref{eq:80} to leading order in small $\gamma$. The stiffness $I'_\alpha=(\frac{1}{K} + \frac{1}{2K_1})^{-1}$ is determined by the smaller of the two stiffnesses $K$ and $2K_1$ at the decoupling lengthscale. Since the scaling of $I'_\alpha$ is negligible in the coplanar regime (see Fig.~\ref{fig:5}) we can equally well use the values of $K$ and $2K_1$ at the coplanar crossover scale $a_\gamma$. Let us assume in the following that $K(l_\gamma) < 2 K_1(l_\gamma)$ and thus $I'_\alpha = K(l_\gamma) \simeq J/T$ (the other case of $2K_1 < K$ is analogous). Here we have neglected the reduction of $K$ during the high temperature flow for simplicity. This can easily be incorporated and does not change our conclusion. 

We then introduce the dimensionless variable $x= \pi J/T^>_{\text{BKT}}$ such that Eq.~\eqref{eq:80} takes the form~\cite{PhysRevLett.109.155703} 
\begin{align}
  \label{eq:82}
   \gamma &= \frac{4 \pi e^{-x}}{x-2} \,.
\end{align}
We are interested in a solution as $\gamma \sim (J_{th}/\bar{J})^2 \rightarrow 0$, which implies that $x\rightarrow \infty$. To leading order we thus find that
\begin{align}
  \label{eq:83}
  T^>_{\text{BKT}} &\sim \frac{J}{\ln(1/\gamma)}
\end{align}
which is of the same order as the coplanar crossover temperature scale $T_{\text{cp}}$ in Eq.~\eqref{eq:45}.

The lower BKT transition temperature $T^<_{\text{BKT}}$ is of the same order as the upper one $T^>_{\text{BKT}}$~\cite{PhysRevB.16.1217, Ortiz2012780}. 
Below $T^<_{\text{BKT}}$ the $\mathbb{Z}_6$ symmetry is spontaneously broken and the system exhibits true long range order with the phase variable $\alpha = n \pi/3$ locked into one of the six minima at $n \in \{1, \ldots, 6\}$. 

\section{Summary and Open Questions}
\label{sec:summary-outlook}
To summarize, we have identified an emergent critical phase at finite temperatures in a 2D isotropic Heisenberg ``windmill'' spin model. Like in the $J_1$-$J_2$ model on the square lattice, the windmill model considers coupling of a lattice with its dual lattice. Using both Wilson-Polyakov RG and Friedan covariant approaches, we have studied its phase diagram in the limit of weak intersublattice coupling $J_{th} < J_{tt}, J_{hh}$. Short-wavelength thermal and quantum fluctuations couple spins on different sublattices into a coplanar state by an ``order from disorder'' mechanism; this crossover occurs at a temperature $T_{\text{cp}} \sim \bar{J}/\ln (\bar{J}^2/J_{th}^2)$. In the coplanar regime the system is described by a coupled SO(3) $\times$ U(1) NLSM. Analyzing the scaling of the coupling strength and the spin stiffnesses, we show that the U(1) sector quickly decouples from the non-Abelian SO(3) degrees of freedom. The emergent U(1) degree of freedom of the relative in-plane angle of honeycomb and triangular spins is described by the action of a six-state clock model. It exhibits two consecutive BKT phase transitions that bracket a critical phase with algebraic correlations in the relative angle. At low temperatures the discrete $\mathbb{Z}_6$ symmetry is spontaneously broken and $\alpha$ exhibits true long-range order. 

Naturally there remain various open questions for further study. On the theoretical side, there is clearly the challenge of investigating this windmill model numerically, particularly in the coupling regimes that are inaccessible to our analytic approach. A purely $1 + 1$ quantum analogue of emergent criticality would also be appealing, particularly as to our knowledge a purely quantum version of fluctuation-selected discrete order has not yet been demonstrated. The experimental realization of this windmill model is another open task. One promising route is to use spin-resolved Scanning Tunnelling Microscopy techniques for the nanofabrication and characterization of stacked triangular and honeycomb monolayers of magnetic atoms like $Cr$ or $Co$~\cite{manoharan-TwoKondoMirage-Nature-2000,Manoharan-ArtificialMolecularGraphene-Nature-2012,PhysRevLett.101.267205,PhysRevB.82.012402}. Other experimental candidates include cold spinful atoms in optical lattices in the limit of large on-site interactions~\cite{PhysRevLett.91.090402,PhysRevLett.107.165301,Trotzky-SuperexchangeColdAtoms-Science-2008,PhysRevLett.107.210405}. An XY version of our model could be realized using ultracold bosons in a similar manner to that recently reported for the triangular lattice.\cite{Struck19082011}  However here an issue would be the competition between the BKT transition of the underlying XY system and those of the fluctuation-selected degrees of freedom, and more theoretical analysis needs to be done to identify the parameter regime where these temperature scales are distinct.

Finally we note that the type of Ricci flow discussed here plays 
a central
role in the Perelman proof~\cite{perelman} of the Poincar\' e conjecture in
four-dimensional space; Perelman's approach involves surgically
removing singularities that develop in the standard Ricci flow in a systematic
fashion. Our work relating 2D classical magnetism and Friedan scaling
suggests that Ricci flow is just the leading term in a 
renormalization group scheme that smooths out the short-wavelength
fluctuations in a manifold. From a statistical mechanics perspective,
the singularities that develop in standard Ricci flow are false Landau
poles in the renormalization flow that result from neglecting higher
order terms in the $\beta $ function.  If this is true, then a proper
implementation of the RG scheme may well eliminate the Landau poles
and thus the need for ``surgery''. 
This line of reasoning then suggests that 
that well-characterized 2D Heisenberg antiferromagnets could be used
to ``simulate'' generalized, surgery-free Ricci flows 
of topological manifolds.

\acknowledgments
We acknowledge useful discussions with S. T. Carr, R. Fernandes, D. Friedan, E. J. K\"onig, D. Nelson, V. Oganesyan, P. Ostrovsky, N. Perkins, J. Reuther, S. Sondhi, O. Sushkov and A. Tsvelik. The Young Investigator Group of P.P.O. received financial support from the ``Concept for the Future'' of the KIT within the framework of the German Excellence Initiative. This work was supported by DOE grant DE-FG02-99ER45790 (P. Coleman) and SEPNET (P.C., P.C. and J.S.). P.C., P.C. and J.S. acknowledge the hospitality of Royal Holloway, University of London where this work was begun, and P.C. and P.C. acknowledge visitors' support from the Institute for the Theory of Condensed Matter, Karlsruhe Instiute for Technology (KIT) where this project was continued. P.C., P.C. and J.S. also thank the Aspen Center for Physics and the National Science Foundation Grant No. PHYS-1066293 for hospitality where this work was further developed and discussed.

\appendix
\section{Derivation of long-wavelength action for windmill lattice Heisenberg antiferromagnet}
\label{sec:deriv-long-wavel}
In this section, we provide details of the derivation of the gradient part $S_0$ of the long-wavelength action on the windmill lattice. We perform a gradient expansion around the classical ground state on the windmill lattice for $J_{th} \ll J_{tt}, J_{hh}$. We consider each term of the Hamiltonian $H = H_{tt} + H_{hh} + H_{th}$ separately in the next sections. We follow the derivation of Ref.~\onlinecite{PhysRevB.39.6797,PhysRevB.38.7181} for the parts $H_{tt}$ and $H_{hh}$. 
\subsection{Triangular lattice}
\label{sec:triangular-lattice}
The part of the Hamiltonian coupling spins on the triangular lattice is given by 
\begin{align}
  \label{eq:158}
  H_{tt} = \frac{J_{tt}}{2} \sum_{\bfr_m} \sum_{\boldsymbol{\delta}_\alpha = \boldsymbol{\delta}_1}^{\boldsymbol{\delta}_6} \bfss(\bfr_m) \bfss(\bfr_m + \boldsymbol{\delta}_\alpha) \,.
\end{align}
Here, $\bfr_m$ denotes a Bravais lattice vector and $\boldsymbol{\delta}_\alpha = \{ \pm \bfa_1, \pm \bfa_2, \pm (\bfa_1 - \bfa_2) \}$ are the vectors to nearest-neighbor sites. We parametrize the three spin directions close to of the $120^\circ$ ground state of the triangular lattice by $(i=1,2,3)$
\begin{align}
  \label{eq:210}
  \bfss_i &= \frac{S R(x) [\bfn_i + a_0 \bfll(x)]}{\sqrt{1 + 2 a_0 \bfn_i \bfll(x) + a_0^2 \bfll(x)^2 }}  \,.
\end{align}
Here, $R(x)$ is a rotation matrix that varies in space slowly and $\bfn_i$ are three fixed directions in spin space, which fulfill $\sum_{i=1}^3 \bfn_i = 0$. We choose $\bfn_1 = (0, 1, 0)$, $\bfn_2 = ( \sqrt{3}/2, -1/2, 0 )$ and $\bfn_3 =(- \sqrt{3}/2 , - 1/2, 0 )$. The vector $\bfll(x)$ defines the tilting of the three triangular spins on one plaquette away from a $120^\circ$ configuration, where we assume that $a_0 |\bfll| \ll 1$. The long-wavelength action arises from an expansion in spatial derivatives and $a_0 |\bfll| \ll 1$. We will keep only terms up to second order in either of the two. To first order in $\bfll$, we find $\bfss_i = S R [ \bfn_i + a_0 \{ \bfll - (\bfn_i \bfll) \bfn_i \} ]$. Therefore, $\bigl( \bfss_1 + \bfss_2 + \bfss_3 \bigl)_\alpha = 3 a_L S R T_{\alpha \gamma} L_\gamma$ with tensor $T_{\alpha \gamma} = \delta_{\alpha \gamma} - \frac13 \sum_{i=1}^3 n_i^\alpha n_i^\gamma$. 

By summing over the three sublattice directions $\bfss_i$ and dividing by three, the Hamiltonian becomes 
\begin{align}
  \label{eq:212}
  H_{tt} &= \frac{J_{tt}}{2} \sum_{\bfr_m} \frac13 \sum_{i=1}^3 \sum_{j\neq i} \sum_{k=1}^3 \bfss_i(\bfr_m) \bfss_j(\bfr_m + \boldsymbol{\delta}_k) \,.
\end{align}
We now perform a gradient expansion $\bfss_j(\bfr_m + \boldsymbol{\delta}_k) = \bfss_j(\bfrr_{\nu}) + (\boldsymbol{\delta}^k_{ij} \cdot \nabla) \bfss_j(\bfr_m) + \frac12 a_0^2 (\boldsymbol{\delta}^k_{ij} \cdot \nabla)^2 \bfss_j(\bfr_m) + \ldots$, where $\boldsymbol{\delta}^k_{ij}$ denote the nearest-neighbor vectors between spins $\bfss_i$ and $\bfss_j$. Those vectors are given by $\{\boldsymbol{\delta}^k_{ij}\} = \{ \pm (\bfa_1 - \bfa_2), \pm \bfa_2, \mp \bfa_1\}$, where the sign depends on the specific pair $(i,j)$ considered, but does not matter up to second order. We peform the summation over nearest-neighbor vectors $\boldsymbol{\delta}^k_{ij}$ to find $\sum_{k=1}^3 (\boldsymbol{\delta}^k_{ij} \cdot \nabla) = 0$ and $\sum_{k=1}^3 (\boldsymbol{\delta}^k_{ij} \cdot \nabla)^2 = \frac32 a_0^2 (\partial_x^2 + \partial_y^2)$. The Hamiltonian thus takes the form
\begin{align}
  \label{eq:214}
  &H_{tt} = \frac{J_{t}t}{6} \sum_{\bfr_m} \sum_{i=1}^3 \sum_{j \neq i} \bfss_i \Bigl[ 3 \bfss_j + \frac12 \frac{3 a_0^2}{2} (\partial_x^2 + \partial_y^2) \bfss_j \Bigr] \\
&= \frac{J_{tt}}{2} \sum_{\bfr_m} \Bigl[ (\bfss_1 + \bfss_2 + \bfss_3)^2 - \sum_{i=1}^3 \bfss_i^2 \nonumber \\ &+ (\bfss_1 + \bfss_2 + \bfss_3) \frac{a_0^2}{4} \partial_\mu^2 (\bfss_1 + \bfss_2 + \bfss_3) \nonumber \\ & - \sum_{i=1}^2 \bfss_i  \frac{a_0^2}{4} \partial_\mu^2 \bfss_i \Bigr]\,.
\end{align}
We observe that $\sum_{i=1}^3 \bfss_i^2$ is just a constant and the third term is of fourth order in $a_0 |\bfll| \partial_\mu$ since $(\bfss_1 + \bfss_2 + \bfss_3) = 3 a_0 S R (T \bfll)$ is of the order of $\bfll$ already. In the last term, it is sufficient to expand to lowest order and use $\bfss_i = S R \bfn_i$. Keeping only the first and the last term, we arrive at $H_{tt} = \frac{J_{tt}}{2} \sum_{\bfr_m} \Bigl[ (\bfss_1 + \bfss_2 + \bfss_3)^2 - \sum_{i=1}^2 \bfss_i  \frac{a_0^2}{4} \partial_\mu^2 \bfss_i \Bigr]$. We now extremize with respect to $\bfll$ which yields $\bfll = 0$ for classical spins. In case of quantum spins, there would be a Berry phase term linear in $\bfll$. It is absent in the classical limit, where the action reads
\begin{align}
  \label{eq:215}
  S_{tt} = \frac{H_{tt}}{T} &= - \frac{J_{tt}}{2 T} \sum_{\bfr_m} \sum_{i=1}^3 S R \bfn_i \frac{a_0^2}{4} \partial_\mu^2 S R \bfn_i \,.
\end{align}
We write this expression as $\sum_{i, \alpha, \beta, \gamma} R_{\alpha \beta} \bfn_i^\beta \partial_\mu^2 R_{\alpha \gamma} \bfn_i^\gamma = \sum_{\alpha, \beta, \gamma} (R^{-1})_{\beta \alpha} \partial_\mu^2  R_{\alpha \gamma} \sum_{i} n^\beta_i n_i^\gamma $. Using that we can write the sum $\sum_{i} n^\beta_i n_i^\gamma = \frac32 P_{\gamma \beta} $ in terms of the projector matrix $P_{\gamma \beta} = \left(\begin{smallmatrix} 1 & 0 & 0 \\ 0 & 1 & 0 \\ 0 & 0 & 0 \end{smallmatrix}\right) $, that $(R^{-1} \partial_\mu R)^2 = - (\partial_\mu R^{-1}) (\partial_\mu R)$ and taking the continuum limit $\sum_{\bfr_m} = \int d^2x/V_{\text{unit cell}}$ with $V_{\text{unit cell}} = \sqrt{3} a_0^2/2$, the action can be written as
\begin{align}
  \label{eq:216}
  S_{tt} &= - \frac{3 J_{tt} S^2}{16 T} \sum_{\bfr_m} \sum_{\alpha, \beta, \gamma}  P_{\gamma \beta} (R^{-1})_{\beta \alpha} (\partial_\mu^2 R_{\alpha \gamma}) \nonumber \\ &= \frac12 K_1 \int d^2x \, \text{Tr} \bigl[ P (R^{-1} \partial_\mu R)^2 \Bigr] 
\end{align}
with $K_1 = \sqrt{3} J_{tt} S^2/4 T$. 

\subsection{Honeycomb lattice}
\label{sec:honeycomb-lattice}
The part of the Hamiltonian describing spins on the honeycomb lattice reads
\begin{align}
  \label{eq:217}
  H_{hh} &= J_{hh} \sum_{\bfr_m} \frac12 \sum_{i=1 \atop j\neq i}^2 \sum_{\boldsymbol{\delta}_\alpha = 1}^3 \bfss^A_i(\bfr_m) \bfss^B_j(\bfr_m + \boldsymbol{\delta}_\alpha) \,.
\end{align}
Here, $\{\boldsymbol{\delta}_\alpha\} = \{\boldsymbol{0}, - \bfa_1, - \bfa_2\}$ denote the Bravais lattice vectors to nearest-neighbor sites. We parametrize the spin directions as $\bfss_1 = \frac{S (\bfn + a_0 \bfll)}{\sqrt{1 + a_0^2 \bfll^2}}$ and $\bfss_2 = \frac{S (-\bfn + a_0 \bfll)}{\sqrt{1 + a_0^2 \bfll^2}}$ with $\bfn(x) \cdot \bfll(x) = 0$ and $\bfss_1 + \bfss_2 = 2 S a_0 \bfll + \mathcal{O}(\bfll^4)$. We now perform the gradient expansion $ \bfss^B_j(\bfr_m + \boldsymbol{\delta}_\alpha) = \bfss_j + \frac12 (\tilde{\boldsymbol{\delta}}_\alpha \cdot \nabla)^2 \bfss_j$, where $\{\tilde{\boldsymbol{\delta}}_\alpha \} = \{\boldsymbol{\delta}_\alpha = \boldsymbol{b}_B - \boldsymbol{b}_A\}$ with $\boldsymbol{b}_A = \boldsymbol{0}$ and $\boldsymbol{b}_B = (1, 1/\sqrt{3})$. We also use that $\sum_{\tilde{\boldsymbol{\delta}}_\alpha = 1}^3 (\tilde{\boldsymbol{\delta}}_\alpha \cdot \nabla)^2 = \frac{a_0^2}{2} (\partial_x^2 + \partial_y^2)$. Extremizing with respect to $\bfll$ yields $\bfll =0 $ in the classical limit, and the action 
\begin{align}
  \label{eq:218}
  S_{hh} = \frac{H_{hh}}{T} &= - \frac{J_{hh} a_0^2}{8 T} \sum_{\bfr_m} \sum_{i=1}^2 \bfss_i  \partial_\mu^2 \bfss_i\,,
\end{align}
where we have used that $\bfss_1 \partial_\mu^2 \bfss_2 + \bfss_2 \partial_\mu^2 \bfss_1 = (\bfss_1 + \bfss_2) \partial_\mu^2 (\bfss_1 + \bfss_2) - \bfss_1 \partial_\mu^2 \bfss_1 - \bfss_2 \partial_\mu^2 \bfss_2$ and have neglected the first term which is $\mathcal{O}(\bfll^2 \partial_\mu^2)$. To the order we consider it is sufficient to replace $\bfss_1 = S \bfn$ and $\bfss_2 = - S \bfn$. Taking the continuum limit $\sum_{\bfr_m} = \int d^2x/V_{\text{unit cell}}$ with $V_{\text{unit cell}} = \sqrt{3} a_0^2/2$, we find
\begin{align}
  \label{eq:219}
  S_{hh} &= \frac12 \int d^2 x \, K (\partial_\mu \bfn)^2 \,,
\end{align}
with $K = J_{hh} S^2/\sqrt{3} T$. 

\subsection{Windmill coupling term} 
\label{sec:coupling-term}
We now demonstrate that the inter-sublattice coupling term between triangular and honeycomb lattice does not contribute to second order to the gradient part of the long-wavelength action $S_0$. It does contribute to the long-wavelength action $S$ via the potential term $S_c$. This is shown in detail in Sec.~\ref{sec:spin-wave-expansion}. The coupling term in the windmill lattice Heisenberg model reads
\begin{align}
  \label{eq:220}
  H_{th} &= \sum_{\bfr_m} \sum_{\alpha=A,B} \sum_{\delta^{t\alpha}_k = 1}^3 S^t(\bfr_m) \bfss^\alpha(\bfr_m + \boldsymbol{\delta}^{t\alpha}_k) \,.
\end{align}
where $\boldsymbol{\delta}^{tA}_k = \{ (0,1), (-1/2, \sqrt{3}/2), (-1/2, -\sqrt{3}/2) \}$ and $\boldsymbol{\delta}^{tB}_k = \{ (0,-1), (1/2, -\sqrt{3}/2), (1/2, \sqrt{3}/2) \}  = - \boldsymbol{\delta}^{tA}_k $. We perform a long-wavelength approximation around a certain ground state configuration. We choose the spins on the honeycomb lattice to (almost) point along direction $\bfn$ on basis sites $A$, \emph{i.e.}, $\bfss^A \propto S (\bfn + a_0 \bfll)$ and the spins on basis sites $B$ to point along direction $-\bfn$, \emph{i.e.}, $\bfss^B \propto S (- \bfn + a_0 \bfll)$. For the triangular lattice, we choose to sum over the three sublattice magnetizations $\bfss_1 = \bfm_1 = (0,1,0)$, $\bfss_2= \bfm_2 = (\frac{\sqrt{3}}{2}, -\frac12, 0 )$ and $\bfss_3 = \bfm_3 = (-\frac{\sqrt{3}}{2}, -\frac12, 0)$ and divide by a factor of three. In other words, we average over the configuration where the directions of the triangular spin at each triangular plaquette are interchanged with each other. Apart from the central triangular spin, each honeycomb plaquette looks identical. Its contribution will thus be identical in all three cases such that the Hamiltonian can thus be written as
\begin{align}
  \label{eq:222}
  H_{th} &= J_{th} \sum_{\bfr_m} \frac13 \sum_{i=1}^3 \Bigl[ \sum_{\boldsymbol{\delta}^{tA}_k = \boldsymbol{\delta}^{tA}_1}^{\boldsymbol{\delta}^{tA}_3} \bfss^t_i(\bfr_m) \cdot \bfss^A_1(\bfr_m + \boldsymbol{\delta}^{tA}_k)\nonumber \\ &  + \sum_{\boldsymbol{\delta}^{tB}_k = \boldsymbol{\delta}^{tB}_1}^{\boldsymbol{\delta}^{tB}_3} \bfss^t_i(\bfr_m) \cdot \bfss^B_2(\bfr_m + \boldsymbol{\delta}^{tB}_k) \Bigr] \,,
\end{align}
We now perform the gradient expansion and use $\sum_{\boldsymbol{\delta}^{tA}_k = 1}^3 (\boldsymbol{\delta}^{tA}_k \cdot \nabla)^2 = \frac{a_0^2}{2} (\partial_x^2 + \partial_y^2)$ and $\sum_{\boldsymbol{\delta}^{tB}_k = 1}^3 (\boldsymbol{\delta}^{tB}_k \cdot \nabla)^2 = \frac{a_0^2}{2} (\partial_x^2 + \partial_y^2)$. We keep only the second order term, since the zeroeth order is proportional to $\bfll^2$, which can be neglected in the classical limit, to obtain 
\begin{align}
  \label{eq:224}
  H_{th} &= \frac{J_{th} a_0^2}{12} \sum_{\bfr_m} \sum_{i=1}^3 \Bigl[\bfss^t_i(\bfr_m) \cdot (\partial_x^2 + \partial_y^2) \nonumber \\ & \quad \times \bigl\{  \bfss^A_1(\bfr_m) + \bfss^B_2(\bfr_m ) \bigr\} \Bigr] \,.
\end{align}
Finally, we employ that $\bfss^A_1(\bfr_m) + \bfss^B_2(\bfr_m) = 2 S a_0 \bfll_h$ and $ \bfss^t_1 (\bfr_m) + \bfss^t_2 (\bfr_m) + \bfss^t_3 (\bfr_m) = 3 a_0 S R T_{\alpha \gamma} L^t_\gamma \sim \bfll_t$ (see discussion below Eq.~\eqref{eq:210}) to conclude that Eq.~\eqref{eq:224} is already of fourth order in the small quantities $\mathcal{O}(\bfll_t, \bfll_h, \partial_\mu^2)$. To second order, the coupling term $H_{th}$ thus does not contribute to $S_0$.


\section{Spin-wave theory on windmill lattice}
\label{sec:spin-wave-expansion}
In this section, we provide details to the calculation of the spin-wave spectrum on the windmill lattice. We perform a Holstein-Primakov spin-wave analysis of the Heisenberg Hamiltonian~\eqref{eq:1} around the classical ground state for $J_{th} \ll J_{tt}, J_{hh}$ shown in Fig.~\ref{fig:2}(a). 

\begin{figure}[t]
  \centering
  \includegraphics[width=.9\linewidth]{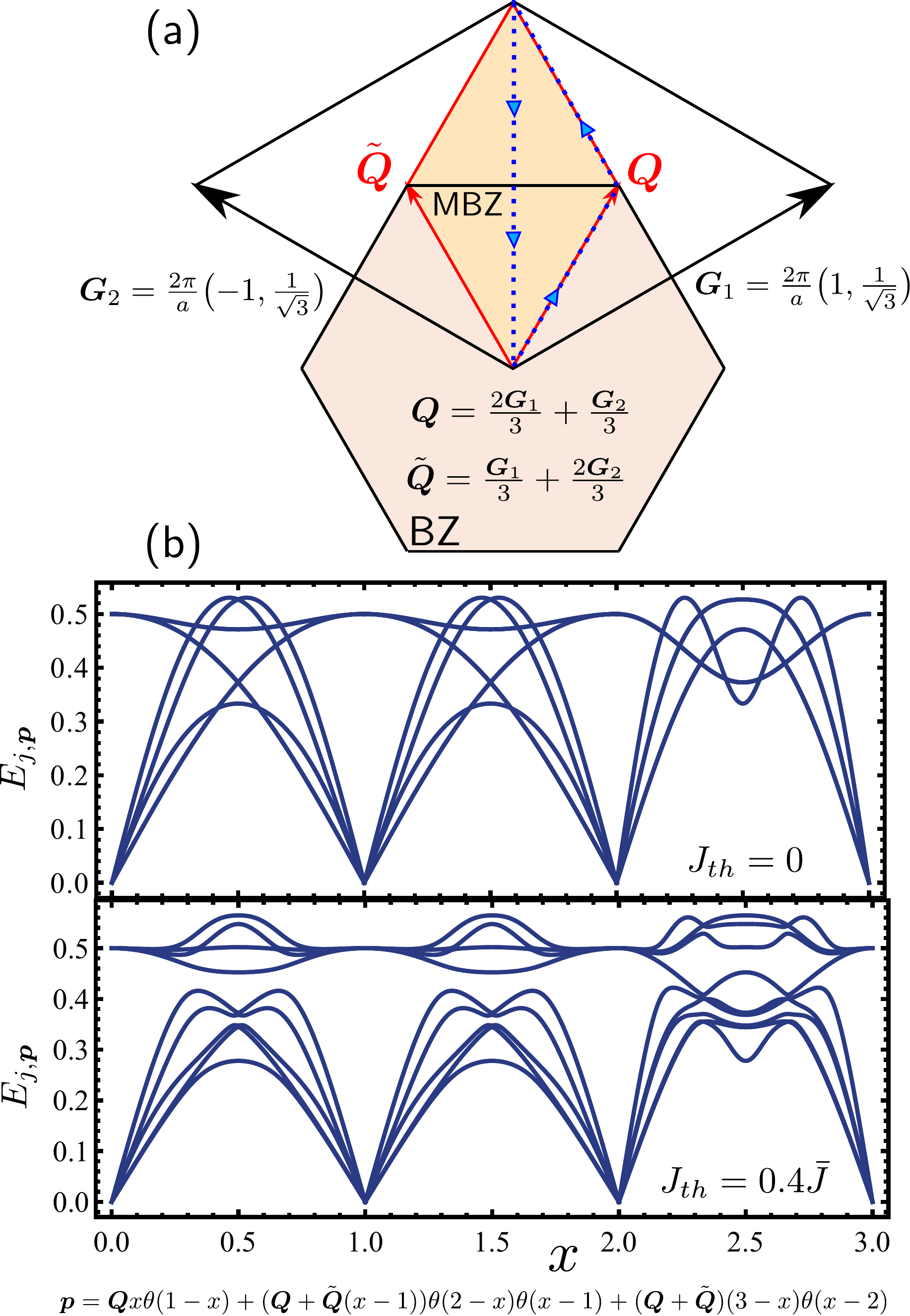}
  \caption{(color online). (a) First Brillouin zone (BZ) and magnetic Brillouin zone (MBZ) with location of ordering vectors $\bfqq = \frac{2 \pi}{\sqrt{3}} (1,1)$ and $\tilde{\bfqq} = \frac{4 \pi}{\sqrt{3}} (1,1)$. (b) Spin-wave spectrum $E_{j,\bfp}$ for the path in the MBZ shown in panel (a) (blue dashed with arrows) }
  \label{fig:8}
\end{figure}

The biaxial magnetic order in the classical ground state of the triangular lattice is described by the ordering wavevector $\bfqq = \frac{2 \pi}{\sqrt{3}} (1,1)$ (see Fig.~\ref{fig:8}(a), which allows us to write 
\begin{align}
  \label{eq:84}
  \bft_1(\bfr_m) &= \bigl( \cos (\bfqq \cdot \bfr_m) , \sin (\bfqq \cdot \bfr_m), 0 \bigr) \\
\label{eq:85}
  \bft_2(\bfr_m) &= \bigl(-\sin (\bfqq \cdot \bfr_m), \cos( \bfqq \cdot \bfr_m), 0\bigr) \\
\label{eq:86}
  \bft_3(\bfr_m) &= \bigl(0,0,1 \bigr) \,.
\end{align}
Here, $\bfr_m = m_1 \bfa_1 + m_2 \bfa_2$ with $\bfa_1 = \frac12 ( 1, \sqrt{3})$ and $\bfa_2 = \frac12 (-1, \sqrt{3})$ is a Bravais lattice vector, and we set the lattice constant $a_0 = 1$. There also exists a magnetically ordered state with opposite chirality. This state is described by the wavevector $\tilde{\bfqq} = \frac{4 \pi}{\sqrt{3}} (1,1)$, and triangular spins have an opposite sense of rotation around one lattice plaquette. Since low energy excitations do not induce transitions between states with different chirality, in the following we assume the magnetic order described by $\bfqq = \frac{2 \pi}{\sqrt{3}} (1,1)$.  

The uniaxial magnetic order on the bipartite honeycomb lattice is described by a normalized unit vector $\bfn \equiv \bfh_1$, which is parallel (anti-parallel) to the direction of the spins on the $A$ ($B$) basis sites.  

We define local triads of orthonormal vectors $\bft_j(\bfr_m)$ and $\bfh_j$ with $j = 1,2,3$ on both sublattices. Here, $\bft_1(\bfr_m)$ points along the direction of the spin $\bfss_t(\bfr_m)$ at the triangular basis site in the unit cell at site $\bfr_m$. In case of quantum spins, it defines a local quantization axis for triangular spins. Together with $\bft_2(\bfr_m)$ it spans the plane of the triangular magnetization. The unit vector $\bfh_1$ points along the direction of the spins on the honeycomb sublattice $A$ in the classical ground state. It defines a local quantization axis for quantum spins on the $A$ sites of the honeycomb lattice. Spins on the honeycomb $B$ sites point along $(-\bfh_1)$ in the classical ground state.

The honeycomb triad $\bfh_j = R(\alpha, \beta) \bft_j(0)$ is rotated with respect to the triangular triad at the origin with rotation matrix
\begin{align}
  \label{eq:87}
  R(\alpha, \beta) = \bmx  \sin \beta \cos \alpha & -\sin \alpha & -\cos \beta \cos \alpha \\ \sin \beta \sin \alpha & \cos \alpha & -\cos \beta \sin \alpha \\ \cos \beta & 0 & \sin \beta \emx \,.
\end{align}
We define local quantization axis' for the spins via the triad $t_j(\bfr_m)$ and $\bfh_j$ as 
\begin{align}
  \label{eq:88}
  \bfss_t(\bfr_m) &= \tilde{S}^z_t  \bft_1 + \tilde{S}^x_t \bft_2 + \tilde{S}^y_t \bft_3 \\
\label{eq:89}
  \bfss_A(\bfr_m) &= \tilde{S}^z_A \bfh_1 + \tilde{S}^x_A \bfh_2 + \tilde{S}^y_A \bfh_3 \\
\label{eq:90}
  \bfss_B(\bfr_m) &= - \tilde{S}^z_B \bfh_1 + \tilde{S}^x_B \bfh_2 - \tilde{S}^y_B \bfh_3 \,.
\end{align}
Both the spin operators $\tilde{S}^j_a(\bfr_m)$ and the local triad $\bft_j(\bfr_m)$ depend on the unit cell vector $\bfr_m$. The vectors $\bfh_j$ are independent of $\bfr_m$. 
Note that we have defined the $1$-axis of the local triad, $\bft_1$ and $\bfh_1$, as the spin quantization axis $\tilde{S}^z_a$. 
The last equation~\eqref{eq:90} follows from the fact that the rotation matrix $R$ on the $B$ sites is given by $R(\alpha, \beta + \pi)$ (see Eq.~\eqref{eq:87}). We then perform a Fourier expansion 
\begin{align}
  \label{eq:91}
  \tilde{S}_a^j(\bfr_m) = \frac{1}{\sqrt{N_L}} \sum_{\bfp \in \text{BZ}} e^{i \bfp (\bfr_m + \bfb_a)} \tilde{S}_a^j(\bfp)
\end{align}
and introduce three different types of Holstein-Primakov (HP) bosons $a_\bfp, b_\bfp$ and $c_\bfp$, one on each basis site $a \in\{ t, A, B\}$. For the triangular lattice, we write
\begin{align}
  \label{eq:92}
  \tilde{S}^x_t(\bfp) &= \sqrt{\frac{S}{2}} (c^\dag_{-\bfp} + c_{\bfp}) \\
\label{eq:93}
\tilde{S}^y_t(\bfp) &= i\sqrt{\frac{S}{2}} (c^\dag_{-\bfp} - c_{\bfp}) \\
\label{eq:94}
\tilde{S}^z_t(\bfp) &= \sqrt{N_L} S \delta_{\bfp,0} - \frac{1}{\sqrt{N_L}} \sum_{\bfk} c^\dag_{\bfk - \bfp} c_{\bfk} \,.
\end{align}
In the same way we define HP bosons $a_\bfp$ and $b_\bfp$ for the honeycomb $A$ and $B$ sites. Up to terms of order $S^0$, the part $H_{tt} + H_{AB}$ of the Heisenberg Hamiltonian (see Eq.~\eqref{eq:1}), takes the form
\begin{align}
  \label{eq:95}
  H_{tt} &= -\frac{3}{2} J_{tt} S^2 N_L + 3 J_{tt} S \sum_{\bfp \in \text{BZ}} \Bigl[ c^\dag_{\bfp} c_\bfp \bigl( 1 + \frac{\nu_\bfp}{2} \bigr) \nonumber \\ & - \frac34 \nu_\bfp c^\dag_{-\bfp} c^\dag_{\bfp} - \frac34 \nu_\bfp c_{-\bfp} c_{\bfp} \Bigr] \\
\label{eq:96}
  H_{AB} &= - 3 J_{hh} S^2 N_L + J_{hh} S \sum_{\bfp \in \text{BZ}} \Bigl[ \eta_{\boldsymbol{0}} \bigl( a^\dag_\bfp a_\bfp + b^\dag_\bfp b_\bfp \bigr) \nonumber \\ & + \eta_\bfp \bigl( a^\dag_\bfp b^\dag_{-\bfp} + a_{-\bfp} b_\bfp \bigr) \Bigr] \,,
\end{align}
where we have defined the lattice functions
\begin{align}
  \label{eq:97}
  \nu_\bfp &= \frac13 \biggl( \cos p_1 + 2 \cos \frac{p_1}{2} \cos \frac{\sqrt{3} p_2}{2} \biggr) \\
\label{eq:98}
 \eta_\bfp &= e^{i p_2 /\sqrt{3}} + 2 e^{i p_2/(2 \sqrt{3})} \cos\frac{p_1}{2} \,. 
\end{align}
The part of the Hamiltonian describing exchange between the two sublattices $H_{th} = H_{tA} + H_{tB} $ depends on the relative orientation of spins on the two sublattices and reads in real-space as 
\begin{widetext}
\begin{align}
  \label{eq:99}
  H_{th} &= J_{th} \sum_{\bfr_m} \sum_{k=1}^3 \Bigl\{ \sum_{\{ \delta^{tA}_n\}} \tilde{S}^k_A(\bfr_m + \delta^{tA}_n) \Bigl[  \tilde{S}^3_t(\bfr_m) R_{3k}(\alpha,\beta)  + \cos (\bfqq \cdot \bfr_m) \Bigl( \tilde{S}^1_t(\bfr_m) R_{1k}(\alpha,\beta) + \tilde{S}^2_t(\bfr_m) R_{2k}(\alpha,\beta)  \Bigr) \nonumber \\ 
& + \sin(\bfqq \cdot \bfr_m) \Bigl( \tilde{S}^1_t(\bfr_m) R_{2k}(\alpha,\beta) - \tilde{S}^2_t(\bfr_m) R_{1k}(\alpha,\beta) \Bigr)  \Bigr]  + \sum_{\{ \delta^{tB}_n\}} \tilde{S}^k_B(\bfr_m + \delta^{tB}_n) \Bigl[ R(\alpha, \beta) \rightarrow R(\alpha, \beta + \pi)  \Bigr] \Bigr\}
\end{align}
Note that we are using the convention $\tilde{S}^1_a = \tilde{S}^z_a$, $\tilde{S}^2_a = \tilde{S}^x_a$, $\tilde{S}^3_a = \tilde{S}^y_a$ (see Eq.~\eqref{eq:88}). Next, we make the transformation to momentum space using Eq.~\eqref{eq:91}, which yields
\begin{align}
  \label{eq:100}
  &H_{th} = J_{th} \sum_{\bfp \in \text{BZ}} \sum_{k=1}^3 \Bigl\{ f_A(\bfp) \tilde{S}^k_A(\bfp) \Bigl[ \tilde{S}^3_t(-\bfp) R_{3k} + \frac{R_{1k}}{2} \Bigl( \tilde{S}^1_t(-\bfp-\bfqq) + \tilde{S}^1_t (-\bfp + \bfqq) + i \tilde{S}^2_t(-\bfp - \bfqq) - i\tilde{S}^2_t(-\bfp + \bfqq) \Bigr) \nonumber \\
&+ \frac{R_{2k}}{2} \Bigl( \tilde{S}^2_t(-\bfp - \bfqq) + \tilde{S}^2_t(-\bfp + \bfqq) + i \tilde{S}^1_t(-\bfp + \bfqq) - i \tilde{S}^1_t(-\bfp - \bfqq) \Bigr)  \Bigr] + f_B(\bfp) \tilde{S}^k_B(\bfp) \Bigl[ R(\alpha, \beta) \rightarrow R(\alpha, \beta + \pi)  \Bigr] \Bigr\} \,,
\end{align}
\end{widetext}
where the lattice functions are given by $f_A(\bfp) = \eta_\bfp$ and $f_B(\bfp) = \eta_\bfp^*$. Note that the lattice functions vanish at the ordering wavevector $\eta_{\pm \bfqq} =0$. 
It is now straightforward to insert the HP bosonic representation of the spins, noting that we have to use separate bosonic operators for wavevectors $\bfp, \bfp\pm \bfqq$, and thus work with the following bosonic operators $\{b^\dag_\bfp, b_{-\bfp}, b^\dag_{\bfp- \bfqq}, b_{-\bfp + \bfqq}, b^\dag_{\bfp + \bfqq}, b_{-\bfp -\bfqq}, c^\dag_{-\bfp}, c_{\bfp}, d^\dag_{-\bfp}, d_{\bfp}  \}  $. Later we will symmetrize the expression with respect to adding a wavevector $\pm \bfqq$ and work with a total number of $18$ bosonic operators. 

We notice a few simplifications: the contributation of order $\mathcal{O}(S^2)$ vanishes, which is due to the fact that the two sublattices are only coupled via fluctuations. To $\mathcal{O}(S)$ only terms of the form $\tilde{S}^{x,y}_t \tilde{S}^{x,y}_{A,B}$ and $\tilde{S}^{z}_t \tilde{S}^{z}_{A,B}$ occur. The terms containing $\tilde{S}^{z}_t \tilde{S}^{z}_{A,B}$ in fact all vanish, either since they are multiplied by a factor of $\eta_{\pm \bfqq} = 0$ or because they cancel when the sum over the $A$ and $B$ sublattices is performed because $R_{j1}(\alpha, \beta + \pi) = - R_{j1}(\alpha, \beta)$. Note that in our notation $\tilde{S}^1_t = \tilde{S}^z_t$. 

In terms of the HP bosons the expression in Eq.~\eqref{eq:100} to $\mathcal{O}(S)$ can be cast into the form
\begin{align}
  \label{eq:101}
  H_{th} = \sum_{\bfp \in \text{BZ}} \beta^\dag_\bfp \bmx \mathcal{F}_\bfp & \mathcal{G}_\bfp \\ \mathcal{G}_{-\bfp}^* & \mathcal{F}_{-\bfp}^* \emx \beta_\bfp 
\end{align}
with bosonic (Nambu) vector
\begin{align}
  \label{eq:102}
  \beta_\bfp &= \bigl( b_\bfp, b_{\bfp - \bfqq}, b_{\bfp + \bfqq}, c_{\bfp}, d_{\bfp}, \nonumber \\ & \qquad \qquad b^\dag_{-\bfp}, b^\dag_{- \bfp + \bfqq}, b^\dag_{- \bfp - \bfqq}, c^\dag_{-\bfp}, d^\dag_{-\bfp} \bigr)^T 
\end{align}
and block matrices
\begin{align}
  \label{eq:103}
  \mathcal{F}_\bfp &= \bmx X_\bfp & 0 & 0 & A_\bfp & -A^*_\bfp \\ 0 & X_{\bfp-\bfqq} & 0 & B_\bfp & C^*_\bfp \\ 0 & 0 & X_{\bfp+\bfqq } & C_\bfp & B^*_\bfp \\ A^*_\bfp & B^*_\bfp & C^*_\bfp & V_\bfp & 0 \\ -A_\bfp & C_\bfp & B_\bfp & 0 & V_\bfp   \emx
\end{align}
and 
\begin{align}
\label{eq:104}
  \mathcal{G}_\bfp &= \bmx Y_\bfp & 0 & 0 & -A_\bfp & A^*_\bfp \\ 0 & Y_{\bfp-\bfqq} & 0 & C^*_{-\bfp} & B_{-\bfp} \\ 0 & 0 & Y_{\bfp+\bfqq} & B^*_{-\bfp} & C_{-\bfp} \\ -A_{-\bfp} & B^*_{\bfp} & C^*_{\bfp} & 0 & W_\bfp \\ A^*_{-\bfp} & C_{\bfp} & B_{\bfp} & W_{-\bfp} & 0    \emx \,.
\end{align}
Note the reversed order of $C^*_{-\bfp}$, $B^*_{-\bfp}$ and $C^*_{\bfp}$, $B^*_{\bfp}$ along column and row, because we must only reverse the sign of $\bfp$ (and not the sign of $\bfqq$) when going from column to row. The same applies to the pair $(B_{\pm\bfp}$, $C_{\pm\bfp})$. We have defined the (triangular lattice) functions~\cite{chubukov_large-s_1994}
\begin{align}
  \label{eq:105}
  X_\bfp &= \frac12 J_{tt} S \bigl( 1 + \frac{\nu_\bfp}{2} \bigr) \\
 \label{eq:106}
  Y_\bfp &= - \frac34 J_{tt} S \nu_\bfp 
\end{align}
and the (honeycomb lattice) functions
\begin{align}
  \label{eq:107}
  V_\bfp &= \frac32 J_{hh} S \\
\label{eq:108} 
  W_\bfp &= \frac12 J_{hh} S \eta_\bfp 
\end{align}
The functions that appear in the off-diagonal entries coupling different sublattices depend on the relative angles $\alpha$ and $\beta$ and are given by
\begin{align}
  \label{eq:109}
  A_\bfp &= \frac14 J_{th} S \eta_\bfp \sin \beta \\
\label{eq:110}
  B_\bfp &= \frac18 J_{th} S \eta_\bfp e^{i \alpha} ( 1 + \cos \beta) \\
\label{eq:111}
  C_\bfp &= \frac18 J_{th} S \eta_\bfp e^{- i \alpha} ( 1 - \cos \beta) \,.
\end{align}
To obtain the spin-wave spectrum via Bogoliubov transformation, we first have to symmetrize the matrices $\mathcal{F}_\bfp$ and $\mathcal{G}_\bfp$ with respect to adding a wavevector $\pm \bfqq$. We thus work with a matrix of dimension $d = 18$. To avoid double counting, the sum over wavevectors $\bfp$ is restricted to the magnetic Brillouin zone (MBZ), which is spanned by the vectors $\bfqq$ and $\tilde{\bfqq} = \frac{4 \pi}{\sqrt{3}} (1,1)$ and covers $1/3$ of the first Brillouin zone. 

Per wavevector $\bfp$ there are three degrees of freedom $\{b_\bfp, c_\bfp, d_\bfp \}$. As usual for Bogoliubov transformations, we symmetrize with respect to $\bfp \rightarrow -\bfp$ and we obtain twice as many bands. Thus, we should only sample over one half of the magnetic Brillouin zone, \emph{i.e.}, $1/6$ of the first Brillouin zone. 

We find the spin-wave spectrum by diagonalizing the matrix~\cite{0953-8984-16-20-005} 
\begin{align}
  \label{eq:117}
  L = g H = \bmx \mathcal{F}_\bfp & \mathcal{G}_\bfp \\ - \mathcal{G}_{-\bfp} & - \mathcal{F}_{-\bfp} \emx \,,
\end{align}
where $g = \left( \begin{smallmatrix} 1 & 0 \\ 0 & 1  \end{smallmatrix} \right)$. In diagonal form, the complete Hamiltonian $H = H_{tt} + H_{hh} + H_{th}$ takes the form
\begin{widetext}
\begin{align}
  \label{eq:118}
  H &= \Bigl( - \frac32 J_{tt} N_L - 3 J_{hh} N_L \Bigr) S^2 - \Bigl( \frac32 J_{tt} N_L - 3 J_{hh} N_L \Bigr) S + \sum_{\bfp \in \text{MBZ}} \sum_{j = 1}^9 \frac{E_{j,\bfp}(\alpha, \beta)}{2} (1 + B^\dag_{j, \bfp} B_{j, \bfp} ) \,,
\end{align}
\end{widetext}
where $B_{j, \bfp}$ denote the Bogoliubov mode operators. The spin-wave spectrum for $J_{th} = 0$ and $J_{th} = 0.4 \bar{J}$ is shown in Fig.~\ref{fig:8}(b). It exhibits different degeneracies: for $J_{th} = 0$ there is a degeneracy between $c_\bfp$ and $d_\bfp$ as well as $c_{\bfp \pm \bfqq}$ and $d_{\bfp \pm \bfqq}$. As a result, from a total of $9$ bands there are at most $6$ different energies at a given wavevector $\bfp$ for $J_{th} = 0$. Some of the degeneracies are lifted for $J_{th} \neq 0$; we observe band mixing that leads to avoided crossings. There are always five zero modes around $\bfp = 0$; the three zero modes stemming from the triangular lattice are also zero modes at $\bfp = \pm \bfqq$. The additional zero modes at $\bfp = \bfqq$ occur since we have symmetrized the Hamiltonian with respect to $c_\bfp \rightarrow c_{\bfp \pm \bfqq}$ and $d_\bfp \rightarrow d_{\bfp \pm \bfqq}$. For $J_{th} \neq 0$, 

The angle dependent fluctuation correction to the free energy $\delta F(J_{th}, \alpha, \beta) = F(J_{th}) - F(J_{th} = 0)$ follows from the diagonal Hamiltonian in Eq.~\eqref{eq:118} as 
\begin{align}
  \label{eq:119}
  \delta F(J_{th}, \alpha, \beta) = T  \sum_{\bfp \in \text{MBZ}} \sum_{j = 1}^9 \ln \biggl[ \frac{\sinh( E_{j,\bfp}(J_{th})/2 T)}{\sinh (E_{j, \bfp}(0)/2T) } \biggr] 
\end{align}
This result is given in Eq.~\eqref{eq:9} and shown in Fig.~\ref{fig:3} of the main text.

\section{Derivation of RG equations in uncoupled regime}
\label{sec:derivation-rg-flow}
In this section we explicitly derive the RG equations for the spin stiffnesses $K(l)$ and $K_i(l)$ in the uncoupled sublattice regime at high temperatures $T_{cp} < T < \bar{J}$. 

In Sec.~\ref{sec:matrix-formalism} we present the derivation of the one-loop result using the method of Wilson-Polyakov scaling for uniaxial $\text{O(3)/O(2)}$ order on the honeycomb lattice. In Sec.~\ref{sec:poly-scal-triang} we use the same method to derive the flow for the biaxial SO(3) order on the triangular lattice. In Sec.~\ref{sec:rg-equations-as} we use Friedan scaling to calculate the RG equations for both cases up to two loop order. In the electronic Supplementary Materials we provide a \emph{Mathematia} file that includes the calculation of the RG equations via Friedan scaling~\cite{SupplMat-FriedanScaling}. 
 
\subsection{Wilson-Polyakov scaling for honeycomb lattice}
\label{sec:matrix-formalism}
The $\text{O(3)/O(2)}$ NLSM action in $d = 2 + \epsilon$ dimensions is given by
\begin{align}
  \label{eq:120}
  S &= \frac12 \int_x K (\partial_\mu \bfn)^2 \,,
\end{align}
where $\int_x = \int d^dx$ and the normalized field $\bfn(x)$ with $|\bfn(x)| = 1$ is the fluctuating order parameter of an uniaxial magnet. An order parameter configuration $\bfn(x)$ is uniquely decribed by two Euler angle fields $\theta(x)$ and $\phi(x)$. Instead of the vectorial expression in Eq.~\eqref{eq:120} we use a matrix form of the action in the following. To this end we introduce a local orthonormal triad, which we combine into a matrix $h = \bigl( \bfn, \bfh_2, \bfh_3 \bigr) \in \text{SO(3)}$. Defining the NLSM matrix field 
\begin{align}
  \label{eq:121}
  Q_h = h K_h h^{-1} 
\end{align}
with coupling matrix $K_h = \sqrt{K} \text{diag}(1,0,0)$, the action~\eqref{eq:120} becomes
\begin{align}
  \label{eq:122}
  S &= \frac14 \int_x \text{Tr} \Bigl[ \bigl( \partial_\mu Q_h\bigr)^T \bigl( \partial_\mu Q_h \bigr) \Bigr] \,.
\end{align}
It is useful to define the matrix $Q_h$, because it is constant $Q = K_h $, if the order parameter field $h(x)$ commutes with the coupling matrix $[K_h, h(x) ] = 0$. A functional integral over the field $Q_h$ thus automatically runs over elements of the coset space $\text{O(3)/O(2)}$. Physically, this corresponds to the fact that two configurations of the order parameter $h$ (or $\bfn$) that only differ by a local rotation around the $\bfn$ axis are identical. Note that Eq.~\eqref{eq:122} is in fact equal to $S = \frac12 \int_x K \Bigl[ (\partial_\mu \bfn)^2 - (\bfn \cdot (\partial_\mu \bfn))^2 \Bigr]$. The additional second term vanishes due to the constraint that $\bfn(x)$ is a normalized field $|\bfn(x)| = 1$. 

We now express the matrix in terms of Euler angles as 
\begin{align}
  \label{eq:123}
  h &= e^{ - i \phi \tau_2} e^{- i \theta \tau_3} e^{- i \psi \tau_1}\,,
\end{align}
where the matrices $(\tau_a)_{bc} = i \epsilon_{bac}$ fulfill the SU(2) algebra $[\tau_a, \tau_b] = i \epsilon_{abc} \tau_c$. Note that the angle $\psi$ is purely a gauge degree of freedom as it describes rotations around the local $\bfn$ axis and drops out in Eq.~\eqref{eq:122}.

We then decompose the Euler angle fields in Fourier space into slow and fast modes, and write 
\begin{align}
  \label{eq:206}
  h = h_< h_>
\end{align}
as a product of matrices $h_<$ and $h_>$, where $h_<$ contains only slow Fourier components $|bfq| \in [0, \Lambda/b]$ and $h_>$ only fast ones ($|\bfq| \in [\Lambda/b, \Lambda]$. Here, $\Lambda \sim 1/a_0$ is a momentum cutoff due to the lattice, and $b > 1$. That this can be done follows most clearly by writing the rotation matrix as $h = \exp( - i \boldsymbol{\phi} \cdot \boldsymbol{\tau})$, where $\boldsymbol{\tau} = (\tau_x, \tau_y, \tau_z)$ and $\boldsymbol{\phi}$ defines the axis and angle of rotation. 

The decoupling into slow and fast modes can be performed in two ways: either as $h = h_> h_<$, which corresponds to decoupling in the body frame of the magnet, or as $h = h_< h_>$, which corresponds to decoupling in the lab frame (see Ref.~\onlinecite{ChandraColeman-LesHouches} for details). The form of $Q_h = h K_h h^{-1}$ suggests to use the lab frame decoupling $h = h_< h_>$, because the Euler angle $\psi$ then drops out immediately because $[K_h, \tau_1] = 0$. Inserting this decoupling
\begin{align}
  \label{eq:124}
  h = h_< h_> 
\end{align}
into the action in Eq.~\eqref{eq:122} yields
\begin{align}
  \label{eq:125}
  S &= \frac12 \int_x \text{Tr} \Bigl[ \frac12 (\partial_\mu Q_>)^2 - Q_>^2 \Omega_\mu^2 + Q_> \Omega_\mu Q_> \Omega_\mu \nonumber \\ &\qquad + Q_> (\partial_\mu Q_>) \Omega_\mu - (\partial_\mu Q_>) Q_> \Omega_\mu   \Bigr] 
\end{align}
where the Einstein summation for $\mu = x,y$ is used. We have defined $Q_> = h_> K_h h_>^{-1}$ and 
\begin{align}
  \label{eq:126}
  \Omega_\mu &= s^{-1} (\partial_\mu s) \,,
\end{align}
which is the slow angular velocity (matrix) along direction $\mu$ in the lab frame.  

In terms of Euler angles $h_< = e^{ - i \phi_< \tau_2} e^{- i \theta_< \tau_3} e^{- i \psi_< \tau_1}$ the different components of the angular velocity
\begin{align}
  \label{eq:207}
  \Omega_\mu = - i \sum_{j=1}^3 \Omega_\mu^j \tau_j
\end{align}
are given by 
\begin{align}
  \label{eq:127}
  \Omega^1_\mu &= \partial_\mu\psi_< + \partial_\mu\phi_< \sin \theta_< \\
\label{eq:128}
  \Omega^2_\mu &= \partial_\mu\phi_< \cos \theta_< \cos \psi_< + \partial_\mu\theta_< \sin \psi_< \\
\label{eq:129}
  \Omega^3_\mu &= \partial_\mu\theta_< \cos \psi_< - \partial_\mu\phi_< \cos \theta_< \sin \psi_< \,.
\end{align}
Next, we expand the action to second order in the fast fields
\begin{align}
  \label{eq:130}
  h_> &= e^{ - i \phi_> \tau_2} e^{- i \theta_> \tau_3} e^{- i \psi_> \tau_1} \,.
\end{align}
which yields 
\begin{align}
  \label{eq:131}
  Q_> = h_> K_h h_>^{-1} = \left( \begin{smallmatrix} 1 - \theta_>^2 - \phi_>^2 & \theta_> & - \phi_> \\ \theta_> & \theta_>^2 & - \theta_> \phi_> \\ - \phi_> & - \phi_> \theta_> & \phi_>^2   \end{smallmatrix}   \right) \,.
\end{align}
Since we express slow fields in terms of the angular velocity $\Omega_\mu$, let us drop the subscript $(>)$ from the Euler angles in what follows. The action then reads
\begin{widetext}
\begin{align}
  \label{eq:132}
  S &= \frac{K}{2} \int_x \Bigl\{ \bigl[ (\partial_\mu \theta)^2 + (\partial_\mu \phi)^2 \bigr]  + \bigl[ \bigl(\Omega_\mu^2\bigr)^2 + \bigl(\Omega_\mu^3\bigr)^2 \bigr] + 2 \bigl[ (\partial_\mu \theta) \Omega_\mu^3  + (\partial_\mu \phi) \Omega_\mu^2 - \Omega_\mu^1 \bigl( \theta \Omega_\mu^2 - \phi \Omega_\mu^3\bigr) \bigr]\nonumber \\ & \qquad + \bigl[  2\Omega_\mu^1 \bigl(  \phi (\partial_\mu \theta) - \theta (\partial_\mu \phi) \bigr) + (\Omega_\mu^1)^2 (\theta^2 + \phi^2) - \theta^2 (\Omega_\mu^2)^2 - \phi^2 (\Omega_\mu^3)^2 + 2 \theta \phi \Omega_\mu^2 \Omega_\mu^3  \bigr] \Bigr\}\,.
\end{align}
\end{widetext}
The combination $(\Omega_\mu^2)^2 + (\Omega_\mu^3)^2$ is independent of the angle $\psi_<$ as required. The component $(\Omega_\mu^1)$, which depends on $\partial_\mu \psi_>$, however, appears in the term coupling slow and fast and slow fields. This dependence will cancel after the integration over the fast variables. The integration over the fast variables yields
\begin{align}
  \label{eq:133}
  Z &= \int \mathcal{D}[\Omega_\mu^j, \phi_>, \theta_>] e^{- S_< - S_{0>} - S_c} 
\nonumber \\ &
= \int \mathcal{D}[\Omega_\mu^j] e^{- S_< - \delta S_<} =  \int \mathcal{D}[\Omega'^j_\mu] e^{- S'} \,. 
\end{align}
In the last step the slow fields have been rescaled $\Omega_\mu^j \rightarrow \Omega'^j_\mu$. Calculating $\delta S_<$ gives
\begin{widetext}
\begin{align}
  \label{eq:134}
  e^{-\delta S_<} &=  \int \mathcal{D}[\phi_>, \theta_>] \exp \biggl[ - \frac12 \int_{p, p'}^>  \Phi_p^T  M_{pp'} \Phi_{p'} - \int_p^> B_p^T \Phi_p \biggr] = \exp \biggl[ - \frac12 \int_{p,p'}^> B_p^T (M^{-1})_{pp'} B_p - \frac12 \text{Tr} \bigl( \ln M \bigr)  \biggr] 
\end{align}
where $\Phi_p^T = (\theta_p, \phi_p)$ and $\int_p^> = \frac{1}{( 2 \pi)^2} \int_{\Lambda/b}^\Lambda p dp \int_0^{2 \pi} d \varphi$ is an integral over the fast variables. We have defined $M_{pp'} = (G^{-1})_{pp'} + C_{pp'}$ with
\begin{align}
  \label{eq:135} 
  (G^{-1})_{pp'} &= K p^2 \delta_{p,-p'} \left( \begin{smallmatrix} 1 & 0 \\ 0 & 1 \end{smallmatrix} \right) \\
\label{eq:136}
  C_{pp'} &= K \bmx \bigl[(\Omega_\mu^1)^2 - (\Omega_\mu^2)^2 \bigr]_{-p-p'} & i (p'_\mu - p_\mu) (\Omega^1_\mu)_{-p-p'} + (\Omega^2_\mu \Omega^3_\mu)_{-p-p'} \\
            - i (p'_\mu - p_\mu) (\Omega^1_\mu)_{-p-p'} + (\Omega^2_\mu \Omega^3_\mu)_{-p-p'}  & \bigl[(\Omega_\mu^1)^2 - (\Omega_\mu^3)^2 \bigr]_{-p-p'} \emx
\end{align}
\end{widetext}
and the linear coupling term
\begin{align}
\label{eq:137}
  B_p^T &= K \bmx - (\Omega^1_\mu \Omega^2_\mu)_{-p}, - (\Omega^1_\mu \Omega^3_\mu)_{-p} \emx  \,.
\end{align}
In Eq.~\eqref{eq:137} we have dropped terms by using that $p_\mu (\Omega^a_\mu)_{-p} = 0$ for fast momenta $|p| \in [\Lambda/b,\Lambda]$, since $\Omega^a_\mu$ only contains slow Fourier components. 

Let us analyze the first term in the final expression of Eq.~\eqref{eq:134}. Since $B_p \sim (\Omega^a_\mu)^2$ it follows that all terms in $B_p M^{-1} B_p$ contain at least four derivatives in the slow fields, which makes them irrelevant in the RG sense. We expand the second term in the final expression of Eq.~\eqref{eq:134} as $\ln M = \ln G^{-1} (G C -\frac12 G C G C + \ldots)$. All higher order terms contain more than two derivatives of slow fields. The first order contribution reads
\begin{widetext} 
\begin{align}
  \label{eq:138}
  \text{Tr} ( G C) &= \sum_{i,j=1}^2 \int_{p,p'}^> G^{ij}_{pp'} C^{ji}_{p'p} = \sum_{j} \int_{p}^> G^{jj}_{p,-p} C^{jj}_{-p,p} = \frac{\ln b}{2 \pi} \Bigl[ 2 (\Omega^1_\mu)^2 - (\Omega^2_\mu)^2 - (\Omega^3_\mu)^2 \Bigr] \,, 
\end{align}
where we have used that $\int_p p^{-2} = \frac{1}{2 \pi} \int_{\Lambda/b}^\Lambda \frac{dp}{p} = \frac{\ln b}{2 \pi}$. For brevity we write $(\Omega^j_\mu)^2 \equiv \int_q^< (\Omega^j_\mu)_q (\Omega^j_\mu)_{-q} = \int_x (\Omega^j_\mu)_x (\Omega^j_\mu)_x$. In real space, the action remains local. In the second order contribution we only need to keep the part in $C_{pp'}$ that contains a single slow derivative, \emph{i.e.}, that is linear in $\Omega_\mu^j$. We find
\begin{align}
  \label{eq:139}
  -\frac12 \text{Tr} G C G C &= -\int_{p,p'}^> \frac{(p_\mu + p'_\mu) (p_\nu + p'_\nu)}{p^2 p'^2} (\Omega^1_\mu)_{p'-p} (\Omega^1_{\nu})_{p-p'} = - \int_q^< \frac{\ln b}{2 \pi} 2 (\Omega^1_\mu)_q (\Omega^1_\mu)_{-q} \,.
\end{align}
Here, we have used that $p'-p = q$ is a slow momentum variable, and we can approximate $p \pm q \approx p$. We can evaluate the momentum integrals in $d=2$ dimensions, because we only keep terms up to $\mathcal{O}(\epsilon)$. This yields $\int_p \frac{p_\mu p_\nu}{p^4} = \delta_{\mu \nu} \frac{\ln b}{2 \pi d} = \delta_{\mu \nu} \frac{\ln b}{4 \pi}$. \end{widetext}
To obtain the final result we add $S_<$ to $\delta S_<$ (see Eqs.~\eqref{eq:138} and~\eqref{eq:139}). Then, we rescale length $x' = x/b$ and fields $(\Omega'^j_\mu)_{x'} = b (\Omega^j_\mu)_{x'}$, which yields the renormalized action after one RG step
\begin{align}
  \label{eq:140}
  S' = \frac12 \int_{x'} b^{d-2} \Bigl( K - \frac{\ln b}{2 \pi} \Bigr) \Bigl( (\Omega'^2_\mu)_{x'}^2 + (\Omega'^3_\mu)_{x'}^2 \Bigr) \,.
\end{align}
With $b= e ^l$ and running cutoff $\Lambda(l) = \Lambda_0 e^{-l}$, we arrive at the flow equation of the $\text{O(3)/O(2)}$ NLSM in $d= 2 + \epsilon$ dimensions
\begin{align}
  \label{eq:141}
  \frac{d}{dl} K &= - \frac{1}{2 \pi} + \epsilon K \,.
\end{align}
In terms of the small expansion parameter $g = 1/K$, it becomes 
\begin{align}
  \label{eq:208}
  \frac{d}{dl} g = - \epsilon g + \frac{g^2}{2 \pi} \,.
\end{align}
\subsection{Wilson-Polyakov scaling for triangular lattice}
\label{sec:poly-scal-triang}
In this section we derive the RG flow equations for the SO(3) NLSM using Wilson-Polyakov scaling. The action of the SO(3) NLSM reads
\begin{align}
  \label{eq:142}
  S &= \frac12 \int_x \text{Tr} \Bigl[ K_t \bigl( \partial_\mu t^{-1} \bigr) \bigl( \partial_\mu t \bigr) \Bigr] 
\end{align}
with matrix field $t(x) \in $ SO(3) and spin stiffness matrix $K_t = \text{diag} \bigl( K_1, K_2, K_3 \bigr)$. We decompose $t$ into slow and fast modes using the lab frame decoupling $t = t_< t_>$ which yields
\begin{align}
  \label{eq:143}
  S &= - \frac12 \int_\bfx  \text{Tr} \Bigl[ K_t \bigl( \Omega^>_\mu + t_>^{-1} \Omega^<_\mu t_> \bigr)^2 \Bigr]\,.
\end{align}
Here, $\Omega^<_\mu = t_<^{-1} (\partial_\mu t_<)$ and $\Omega^>_\mu = t_>^{-1} (\partial_\mu t_>)$ denote slow and fast angular velocities, respectively. We parametrize the fast fluctuations using Euler angles as
\begin{align}
  \label{eq:144}
  t_> &= e^{-i \phi_> \tau_2} e^{-i \theta_> \tau_3} e^{-i \psi_> \tau_1} \,,
\end{align}
and expand the action to quadratic order in the fast fields 
\begin{widetext}
\begin{align}
  \label{eq:145}
  e^{- \delta S_<} &= \int \mathcal{D}[ \theta_>, \phi_>, \psi_> ] e^{- S_{0>} - S_c} = \int \mathcal{D}[ \theta_>, \phi_>, \psi_> ] \exp \biggl[ -\frac12 \int_{p, p'} \Phi_{p}^T M_{p, p'} \Phi_{p'} - \int_p B_p^T \Phi_{p}    \biggr] \,.
\end{align}
We combine the three Euler angles into the vector $\Phi_p^T = \bigl( \theta_{>p}, \phi_{>p}, \psi_{>p} \bigr)$. The quadratic part $M = (G)^{-1} + C$ consists of the inverse free propagator 
\begin{align}
  \label{eq:146}
  \bigl( G^{-1} \bigr)_{p, p'} &= \delta_{p, -p'} \; p^2 \bmx K_1 + K_2 & 0 & 0 \\ 0 & K_1 + K_3 & 0  \\ 0 & 0 & K_2 + K_3 \emx  
\end{align}
and the coupling matrix
\begin{align}
\label{eq:147}
  C_{p, p'} &= \bmx K_{12}^- \bigl[ (\Omega_\mu^1)^2 - (\Omega_\mu^2)^2 \bigr]  & -i  \Omega_{\mu}^1 [ K_{21}^+ p_{1,\mu} + K_{21}^- p_{2, \mu} ]  & -i \Omega_\mu^2 [ K_{32}^+ p_{2, \mu} + K_{32}^- p_{1, \mu} ] \\ 
-i  \Omega_{\mu}^1 [ K_{21}^+ p_{2,\mu} + K_{21}^- p_{1, \mu} ] & K_{13}^- \bigl[ (\Omega_\mu^1)^2 - (\Omega_\mu^3)^2 \bigr] & - i \Omega_\mu^3 [ - K_{32}^+ p_{2,\mu} + K_{32}^- p_{1,\mu} ] \\
-i \Omega_\mu^2 [ K_{32}^+ p_{1, \mu} + K_{32}^- p_{2, \mu} ] & - i \Omega_\mu^3 [ - K_{32}^+ p_{1,\mu} + K_{32}^- p_{2,\mu} ] & K_{23}^- \bigl[ (\Omega_\mu^2)^2 - (\Omega_\mu^3)^2 \bigr]    \emx \,.
\end{align}
\end{widetext}
We have dropped the subscript $(<)$ on the $\Omega_\mu^j$, and have defined $K^\pm_{ij} = K_i \pm K_j$. The slow fields are evaluated at momentum $(-p-p')$, \emph{i.e.}, $(\Omega^j_\mu)^2_{-p-p'} = \int_q^< (\Omega^j_\mu)_{-p-p'+q} (\Omega^j_\mu)_-q$. The coupling matrix fulfills the relation $C^{ji}_{p, p'} = C^{ij}_{p', p }$. In the off-diagonals of $C^{ij}_{p,p'}$ we kept only first order terms in $\Omega_\mu^j$, because higher order terms become higher order gradient terms in the slow action and such terms are RG irrelevant. The terms $B_p^T \Phi_p$ that are linear in the fast fields do not contribute to the flow equation: they either vanish since a slow field is evaluated at a fast momentum or are of quadratic order in the slow fields. This is analogous to the uniaxial case (see discussion below Eq.~\eqref{eq:137}).

As in Sec.~\ref{sec:matrix-formalism}, we now integrate over the fast modes and perform a (derivative) expansion in the slow fields of the resulting $\text{Tr}[ \ln (1 + G C)]$ term. The first order result is given by
\begin{widetext}  
\begin{align}
  \label{eq:148}
    \text{Tr} (G C) &= \sum_{j=1}^3 \int_p^> \bigl( G \bigr)^{jj}_{p} C^{jj}_{-p, p} = \frac{\ln b}{2 \pi} \biggl[ \Bigl\{ \frac{K_{12}^-}{K_{12}^+} + \frac{K_{13}^-}{K_{13}^+} \Bigr\} \bigl(\Omega_\mu^1 \bigr)^2 + \Bigl\{ \frac{K_{23}^-}{K_{23}^+} - \frac{K_{12}^-}{K_{12}^+} \Bigr\} \bigl(\Omega_\mu^2 \bigr)^2 - \Bigl\{  \frac{K_{23}^-}{K_{23}^+} + \frac{K_{13}^-}{K_{13}^+} \Bigr\} \bigl(\Omega_\mu^3 \bigr)^2 \biggr] \,.
\end{align}
The second order contribution reads $- \frac12 \text{Tr} G C G C = - \frac12 \int_{p, p'}^> \sum_{i,j = 1}^3 G_{p}^{ii} C_{-p, p'}^{ij} G_{p'}^{jj} C_{-p', p}^{ji}$. Let us evaluate one of the three terms explicitly ($i=1,j=2$):
\begin{align}
  \label{eq:149}
  - \frac12 \int_{p, p'}^> 2 G_{p}^{11} C_{-p, p'}^{12} G_{p'}^{22} C_{-p', p}^{21} &= - \int_q^< \frac{(\Omega_{\mu}^1)_{q} (\Omega_\nu^1)_{-q} }{K_{12}^+ K^+_{13}} \bigl(K^+_{21} - K^-_{21} \bigr)^2 \int_{p}^> \frac{p_\mu p_\nu}{p^4 }  = - \int_q^< \frac{\ln b}{2 \pi} \frac{2 K_1^2 (\Omega_{\mu}^1)^2}{K_{12}^+ K^+_{13}} \,.
\end{align}
The other two terms $(i=1,j=3)$ and $(i=2,j=3)$ are obtained in the same way. We combine first and second order terms, rescale fields and momenta to arrive at the renormalized action in $d = 2 + \epsilon$ dimension
\begin{align}
  \label{eq:150}
    S' &= \frac12 \int_{x'} b^{d-2}  \biggl\{ \bigl( \Omega_\mu^1\bigr)^2 \Bigl[ K_{23}^+ + \frac{\ln b}{2 \pi} \Bigl\{ \frac{K_{12}^-}{K_{12}^+} + \frac{K_{13}^-}{K_{13}^+} - \frac{2 K_1^2}{K_{12}^+ K_{13}^+} \Bigr\} \Bigr] 
+ \bigl( \Omega_\mu^2 \bigr)^2  \Bigl[ K_{13}^+ + \frac{\ln b}{2 \pi} \Bigl\{ \frac{K_{23}^-}{K_{23}^+} - \frac{K_{12}^-}{K_{12}^+} - \frac{2 K_2^2 }{K_{12}^+ K_{23}^+} \Bigr\} \Bigr] \\
& \qquad + \bigl( \Omega_\mu^3 \bigr)^2  \Bigl[ K_{12}^+ + \frac{\ln b}{2 \pi} \Bigl\{ - \frac{K_{23}^-}{K_{23}^+} - \frac{K_{13}^-}{K_{13}^+} - \frac{2 K_3^2}{K_{13}^+ K_{23}^+} \Bigr\} \Bigr] \biggr\} \,.
\end{align}
\end{widetext}
We can now read-off the RG flow equations. First, we note that in case of $K_1 = K$ and $K_2 = K_3 = 0$, we recover the previous $\text{O(3)/O(2)}$ result of Eq.~\eqref{eq:141}. If at least two stiffnesses are non-zero, we find 
\begin{align}
  \label{eq:151}
  \frac{d}{dl} I_a = - \frac{2 K_b K_c }{2 \pi I_b I_c} = - \frac{ - I_a^2 + (I_b - I_c)^2 }{ 4 \pi I_b I_c} \,,
\end{align}
where we have defined 
\begin{align}
  \label{eq:209}
  I_a = K_b + K_c
\end{align}
with $a \neq b \neq c$. On the triangular lattice, two of the stiffnesses are initially identical $K_1 = K_2$. This equality is preserved during the RG flow, and we find
\begin{align}
  \label{eq:152}
  \frac{d K_1}{dl} &= - \frac{1}{2 \pi} \frac{K_1^2}{(K_1 + K_3)^2} \\
\label{eq:153}
  \frac{d K_3}{dl} &= \frac{1}{2 \pi} \frac{K_1^2 - K_3 (K_1 + K_3)}{(K_1 + K_3)^2} \,.
\end{align}
Clearly, a non-zero value of $K_3$ is generated during the flow. If we define the stiffness anisotropy  
\begin{align}
  \label{eq:154}
  \eta = \frac{K_1 - K_3}{K_1 + K_3} \,,
\end{align}
the flow equations take the form (see Eqs.~\eqref{eq:38} and \eqref{eq:39})
\begin{align}
  \label{eq:155}
  \frac{d K_1}{dl} &= - \frac{(1 + \eta)^2}{8 \pi} \\
\label{eq:156}
  \frac{d \eta}{dl} &= - \frac{\eta ( 1 + \eta)^2}{4 \pi K_1}\,.
\end{align}
The anisotropy flows to zero and the system approaches an isotropic fixed point with all stiffnesses being equal $K_1 = K_2 = K_3$.

\subsection{O(3)/O(2) and SO(3) scaling from Ricci flow}
\label{sec:rg-equations-as}
The RG equations for the O(3)/O(2) and the SO(3) NLSM can also be derived using Friedan scaling via the Ricci flow of the stiffness metric tensor. We provide as electronic Supplementary Material a \emph{Mathematica} file that includes this calculation~\cite{SupplMat-FriedanScaling}. 
To find the metric tensor $g_{ij}$, we write the NLSM action in Eq.~\eqref{eq:4} in covariant form. First, we write
\begin{align}
  \label{eq:157}
  S_0 &=\frac12 \int_x \Bigl\{ K [ (\Omega_\mu^{2})^2 + (\Omega_\mu^{3})^2 ]  + \sum_{a = 1}^3 I_a  (\tilde{\Omega}_\mu^{a})^2 \Bigr\}
\end{align}
where $\Omega_\mu^j$ and $\tilde{\Omega}_\mu^j$ denote components of the angular velocities $\Omega_\mu = h^{-1} (\partial_\mu h) = -i \Omega_\mu^a \tau_a$ and $\tilde{\Omega}_\mu = t^{-1} (\partial_\mu t) = -i \tilde{\Omega}_\mu^a \tau_a$, and the $I_a$ are defined in Eq.~\eqref{eq:209}. Note the analogy of Eq.~\eqref{eq:157} to the Hamiltonian of a spinning top with moments of inertia $I_a$ along the three principal axes. For the honeycomb lattice, only rotations around the $\bfh_2$ and $\bfh_3$ axes have a finite moment of inertia and thus $I_1^h = 0$, $I_2^h = I_3^h = K$. 

In covariant form Eq.~\eqref{eq:157} reads 
\begin{align}
  \label{eq:159}
  S_0 &= \frac12 \int_x \sum_{i,j=1}^2 g^h_{ij} (\partial_\mu X^i) (\partial_\mu X^j) \nonumber \\ & \quad + \frac12 \int_x \sum_{i,j=1}^3 g^t_{ij} (\partial_\mu Y^i) (\partial_\mu  Y^j) \,,
\end{align}
where the coordinate vectors $X^i, Y^i$ contain the Euler angles for the spins on the honeycomb lattice $X = (\phi_h,\theta_h )$ and on the triangular lattice $Y = (\phi_t, \theta_t, \psi_t)$.
We use the Euler angle convention 
\begin{align}
  \label{eq:160}
  h &=  e^{-i \phi_h \tau_2} e^{-i \theta_h \tau_3} e^{-i \psi_h \tau_1} \\
  t &=  e^{-i \phi_t \tau_3} e^{-i \theta_t \tau_1} e^{-i \psi_t \tau_3} \,.
\end{align}
This is the Euler angle convention for $t$, that we also employ in the coplanar regime. The resulting RG equations are of course independent of the choice of Euler angles. The covariant metric tensors read
\begin{widetext}
\begin{align}
  \label{eq:161}
  g^{(h)}_{ij} &= K \bmx \cos^2\theta_h & 0 \\ 0 & 1 \emx \\
\label{eq:162}
  g^{(t)}_{ij} &= \bmx I_3 \cos^2 \theta_t + \sin^2\theta_t ( I_1 \sin^2 \psi_t + I_2 \cos^2 \psi_t) & (I_1 - I_2) \sin \theta_t \sin \psi_t \cos \psi_t & I_3 \cos \theta_t \\ 
(I_1 - I_2) \sin \theta_t \sin \psi_t \cos \psi_t & I_1 \cos^2 \psi_t + I_2 \sin^2 \psi_t & 0 \\
I_3 \cos \theta_t & 0 & I_3 \emx \,.
\end{align}
\end{widetext}
The contravariant tensors $g^{(h);ij}$ and $g^{(t);ij}$ are given by the inverse of $g^h_{ij}$ and $g^t_{ij}$ due to $g^{h;ij}g^h_{jk} = \delta^i_k$ and $g^{t;ij}g^t_{jk} = \delta^i_k$. 

Following Friedan~\cite{PhysRevLett.45.1057,Friedan-AnnPhys-1985}, the renormalization group flow of the spin stiffnesses up to the order of two loops is given by the Ricci flow of the metric tensor
\begin{align}
  \label{eq:163}
  \frac{d g_{ij}}{dl} &= - \frac{1}{2 \pi} R_{ij} - \frac{1}{8 \pi^2} R_i{}^{klm} R_{jklm} \,.
\end{align}
The Riemann tensor $R^k{}_{lij}$ is determined by the Christoffel symbols 
\begin{align}
  \label{eq:164}
  \Gamma^i_{jk} &= \frac12 g^{il} \bigl( g_{jl,k} + g_{kl,j} - g_{jk,l} \bigr)
\end{align}
as
\begin{align}
  \label{eq:165}
  R^k{}_{lij} &= \Gamma^k_{lj,i} - \Gamma^k_{li,j} + \Gamma^k_{ni} \Gamma^n_{lj} - \Gamma^k_{nj} \Gamma^n_{li}\,.
\end{align}
We use the common notation $g_{ij,k} = \frac{\partial g_{ij}}{\partial X^k}$. The first loop contribution of the RG flow is determined by the Ricci tensor $R_{ij}$, which is a contraction of the Riemann tensor
\begin{align}
  \label{eq:166}
  R_{ij} &= R^k{}_{ikj} \,.
\end{align}
The Ricci tensor for the honeycomb metric reads explicitly
\begin{align}
  \label{eq:167}
  R^{(h)}_{ij} &= \bmx \cos^2\theta_h & 0 \\ 0 & 1 \emx \,.
\end{align}
In the most general case of $I_1 \neq I_2 \neq I_3$, the Ricci tensor is rather lengthy. For the triangular lattice magnet, where $K_1 = K_2$ and thus $I_1 = I_2$, it reads  
\begin{align}
  \label{eq:168}
  R^{(t)}_{ij} &= \bmx \frac{I_3}{2 I_1^2}  c^2_{\theta_t} + (1 - \frac{I_3}{2 I_1})  s^2_{\theta_t} & 0 & \frac{I_3^2}{2 I_1^2} c^2_{\theta_t} \\ 
0 & 1 - \frac{I_3}{2 I_1} & 0 \\
\frac{I_3^2}{2 I_1^2} c^2_{\theta_t} & 0 & \frac{I_3^2}{2 I_1^2} \emx \,. 
\end{align}
where $c_{Y^j} = \cos Y^j$ and $s_{Y^j} = \sin Y^j$. Note that all entries appear with the same prefactor in the corresponding entries of the covariant metric tensor $g_{ij}^t$ in Eq.~\eqref{eq:162}. The contraction of the Riemann tensor that appears at two loop order is given for the honeycomb lattice by
\begin{align}
  \label{eq:169}
  R^{(h)}_i{}^{klm} R^{(h)}_{jklm} &= \frac{2}{K_h} \bmx \cos^2\theta_h & 0 \\ 0 & 1 \emx \,.
\end{align}
and for the triangular lattice by
\begin{widetext}
\begin{align}
\label{eq:170}
  R^{(t)}_i{}^{klm} R^{(t)}_{jklm} &= \frac{1}{4 I_1^4} \bmx I_1 (8 I_1^2 - 12 I_1 I_3 + 5 I_3^2)  \sin \theta_t +  I_3^3 \cos \theta_t & 0 & I_3^3 \cos \theta_t \\ 
0 & I_1( 8 I_1^2 - 12 I_1 I_3 + 5 I_3^2) & 0 \\
I_3^2 \cos \theta_t & 0 & I_3^3 \emx \,.
\end{align}
According to Eq.~\eqref{eq:163}, a comparison with the covariant metric tensor allows to read-off the RG flow equation up to two loops as 
\begin{align}
  \label{eq:171}
  \frac{d K}{dl} &= - \frac{1}{2 \pi} - \frac{1}{4 \pi^2 K} \\
\label{eq:172}
  \frac{d I_1}{dl} &= \frac{- I_1^2 + (I_2 - I_3)^2}{4 \pi I_2 I_3}  - \frac{1}{32 \pi^2 I_1 I_2^2 I_3^2} \bigl[ I_1^4 + (I_2 - I_3)^2 \bigl\{   5 (I_2^2 + I_3^2) + 2 I_1^2 + 6 I_2 I_3 - 8 (I_2 + I_3) I_1 \bigr\} \bigr]\\ 
\label{eq:173}
  \frac{d I_2}{dl} &= \frac{- I_2^2 + (I_1 - I_3)^2}{4 \pi I_1 I_3}  - \frac{1}{32 \pi^2 I_1^2 I_2 I_3^2} \bigl[ I_2^4 + (I_1 - I_3)^2 \bigl\{ 5 (I_1^2 + I_3^2) + 2 I_2^2 + 6 I_1 I_3 - 8 (I_1 + I_3) I_2 \bigr\} \bigr] \\ 
\label{eq:174}
  \frac{d I_3}{dl} &= \frac{- I_3^2 + (I_1 - I_2)^2}{4 \pi I_1 I_2}  - \frac{1}{32 \pi^2 I_1^2 I_2^2 I_3} \bigl[  I_3^4 + (I_1 - I_2)^2 \bigl\{  5 (I_1^2 + I_2^2) + 2 I_3^2 + 6 I_1 I_2 - 8 (I_1 + I_2) I_3 \bigr\} \bigr]  \,.
\end{align}
\end{widetext}
The one loop result agrees with Wilson-Polyakov scaling [see Eqs.~\eqref{eq:141},~\eqref{eq:152} and~\eqref{eq:153}].

\section{Derivation of RG equations in coplanar regime}
\label{sec:derivation-rg-flow-1}
In this section, we compute the renormalization group flow of the coplanar action (see Eq.~\eqref{eq:50})
\begin{align}
  \label{eq:175}
  &S = \frac12 \int_x \biggl\{ I_1 (\Omega_\mu^1)^2 + I_2 (\Omega_\mu^2)^2 + I_3 (\Omega_\mu^3)^2 + I_\alpha (\partial_\mu \alpha)^2 \nonumber \\ 
& \qquad + \frac{\kappa}{2} (\partial_\mu \alpha) \Omega_\mu^3+ \frac{\lambda}{4} \cos (6 \alpha) \biggr\} \,,
\end{align}
where $I_1 = K_2 + K_3$, $I_2 = K + K_1 + K_3$, $I_3 = K + K_1 + K_2$, $I_\alpha = K_1 + K_2$ and $\kappa = 2 (K_1 + K_2)$. The angular velocity $\Omega_\mu = h^{-1} (\partial_\mu h) = - i \Omega_\mu^a \tau_a$ with SU(2) matrices $(\tau_a)_{bc} = i \epsilon_{bac}$ describes the locally fluctuating SO(3) magnetic order parameter. The U(1) phase angle $\alpha$ is coupled only to the component $\Omega_\mu^3$, since we choose the local axis that is perpendicular to the common plane of triangular and honeycomb spins to be the $\tau_3$ direction. It holds that $t = h U$ with $U = \exp(- i \alpha \tau_3)$. 

We derive the scaling of the spin stiffnesses $I_{j}$ and $I_\alpha$ as well as the SO(3)$\times$ U(1) coupling constant $\kappa$ first using Wilson-Polyakov scaling in Sec.~\ref{sec:poly-scal-copl} and then using the Friedan approach in Sec.~\ref{sec:ricci-flow-rg}. The flow of the six-fold potential $\lambda$ is calculated in Sec.~\ref{sec:flow-six-fold}. The effect of the potential term $\lambda$ on the scaling of $I_j$ and $\kappa$ is small and thus neglected in the following. In contrast, the flow of $\lambda$ strongly affects that of $I_\alpha$ and vice versa.

\subsection{Coplanar flow from Wilson-Polyakov scaling}
\label{sec:poly-scal-copl}
In this section, we derive the flow equations for $\{I_j, I_\alpha, \kappa\}$ using Wilson-Polyakov scaling. We employ the Euler angle parametrization 
\begin{align}
  \label{eq:176}
  h = e^{- i \phi \tau_2} e^{- i \theta \tau_3} e^{- i \psi \tau_1}\,.
\end{align}
This choice ensures that the Euler angle $\psi$ drops out of the O(3)/O(2) NLSM action immediately since $[K, \tau_1] = 0$, and the fast propagator $G^{-1}$ (defined below) can be inverted. We will use a different Euler angle parametrization in the Friedan approach in Sec.~\ref{sec:ricci-flow-rg}, since it allows for an easier identification of decoupling criteria of the SO(3) and U(1) sectors. The result for the RG equations is, of course, independent of the choice of Euler angle parametrization.    

We first separate $h$ and $U = \exp(- i \alpha \tau_3)$ into slow and fast modes $h = h_< h_>$ and $  U = U_< U_>$. This yields 
\begin{align}
  \label{eq:211}
  \Omega_\mu = h^{-1} (\partial_\mu h) = h_>^{-1} \Omega^<_\mu h_> + \Omega_\mu^>
\end{align}
with $\Omega^<_\mu = h_<^{-1} (\partial_\mu h_<)$ and $\Omega^>_\mu = h_>^{-1} (\partial_\mu h_>)$, as well as $U_\mu = - i (\partial_\mu \alpha) \tau_3 = - i \tau_3 \bigl[ (\partial_\mu \alpha_<) + (\partial_\mu \alpha_>) \bigr]$. Expanding to quadratic order in the fast fields and performing the functional integration gives the correction to the slow action
\begin{widetext}
\begin{align}
  \label{eq:177}
  e^{- \delta S_<} &= \int \mathcal{D}[ \phi_>, \theta_>, \psi_>, \alpha_> ] e^{- S_{0>} - S_c} = \int \mathcal{D}[ \phi_>, \theta_>, \psi_>, \alpha_> ] \exp \biggl[ -\frac12 \int_{p, p'} \Phi_{p}^T M_{p, p'} \Phi_{p'} - \int_p B_p^T \Phi_{p}    \biggr]
\end{align}
\end{widetext}
with $\Phi_p^T = \bigl( \phi_>, \theta_>, \psi_>, \alpha_> \bigr)_p$. The quadratic part $M = G^{-1} + C$ contains the inverse free propagator 
 \begin{align}
\label{eq:178}
   \bigl( G^{-1} \bigr)_{p, p'} &= \delta_{p, -p'} \; p^2 \bmx I_2 & 0 & 0 & 0 \\ 0 & I_3 & 0 & \kappa/2 \\ 0 & 0 & I_1 & 0 \\ 0 & \kappa/2 & 0 & I_\alpha \emx   \,.
 \end{align}
For the propagator one finds
 \begin{align}
   \label{eq:179}
   G_{p, p'} &= \frac{\delta_{p, -p'}}{p^{2}} \bmx I_2^{-1} & 0 & 0 & 0 \\ 0 & \frac{1}{I_3 - \kappa^2/4 I_\alpha}& 0 & \frac{2 \kappa}{\kappa^2 - 4 I_\alpha I_3} \\ 0 & 0 & I_1^{-1} & 0 \\ 0 & \frac{2 \kappa}{\kappa^2 - 4 I_\alpha I_3} & 0 & \frac{1}{I_\alpha - \kappa^2/4 I_3} \emx \,.
 \end{align}
It also contains the coupling matrix
 \begin{widetext}
 \begin{align}
   \label{eq:180}
   C_{p, p'} &= \bmx  I_{31} \bigl[ (\Omega_\mu^1)^2 - (\Omega_\mu^3)^2 \bigr] - \frac{\kappa}{2} \Omega_\mu^3 (\partial_\mu \alpha_<) & - i \Omega_\mu^1 [ I_{12} p + I_3 p']  & i \Omega_\mu^3 [ I_{32} p + I_1 p'] & - \frac{i \kappa}{2} p' \Omega_\mu^1 \\
 -i \Omega_\mu^1 [ I_{12} p' + I_3 p] & - I_{12} [ (\Omega_\mu^1)^2 - (\Omega_\mu^2)^2 ] & - i \Omega_\mu^2 [ I_{23} p + I_1 p'] & 0 \\
i \Omega_\mu^3 [ I_{32}  p' + I_1 p] & - i \Omega_\mu^2 [ I_{23} p' + I_1 p] &  I_{32} [ (\Omega_\mu^2)^2 - (\Omega_\mu^3)^2] - \frac{\kappa}{2} \Omega_\mu^3 (\partial_\mu \alpha_<) & \frac{i \kappa}{2} p' \Omega_\mu^2 \\
- \frac{i \kappa}{2} p \Omega_\mu^1 & 0 & \frac{i \kappa}{2} p \Omega_\mu^2 & 0 
\emx
 \end{align}
 \end{widetext}
where $I_{ij} = I_i - I_j$. We have dropped the index $(<)$ on $\Omega_\mu$. The linear coupling term $B_p^T$ contains terms that are linear and quadratic in the slow fields: the linear terms vanish in Fourier space because they involve evaluating a slow fields at a fast momentum. The terms quadratic in the slow fields lead after functional integration to irrelevant operators (see discussion below Eq.~\eqref{eq:137}). We thus do not give $B_p^T$ explicitly here. 

To find the renormalization of $I_j, I_\alpha, \kappa$, we integrate over the fast fields (see Eq.~\eqref{eq:134}) and expand $\text{Tr} [ \ln (1 + G C)]$ to second order in $C$. We then rescale momenta and fields. From the expressions of the renormalized parameters, we extract the one-loop RG flow equations as
\begin{align}
  \label{eq:181}
  \frac{d}{dl} I_1 &= \frac{-I_1^2 + (I_2 - I_3)^2}{4 \pi I_2 I_3} - \frac{(I_1^2 - I_2^2) \kappa^2}{16 \pi I_2 I_3^2 (I_\alpha - \kappa^2/4 I_3)}\\
\label{eq:182}
  \frac{d}{dl} I_2 &= \frac{-I_2^2 + (I_1 - I_3)^2}{4 \pi I_1 I_3} + \frac{(I_1^2 - I_2^2) \kappa^2}{16 \pi I_1 I_3^2 (I_\alpha - \kappa^2/4 I_3)} \\
\label{eq:183}
  \frac{d}{dl} I_3 &= \frac{-I_3^2 +(I_1 - I_2)^2}{4 \pi I_1 I_2}  \\
\label{eq:184}
  \frac{d}{dl} \kappa &= - \frac{I_3 \kappa}{4 \pi I_1 I_2} \\
\label{eq:185}
  \frac{d}{dl} I_\alpha &= - \frac{\kappa^2}{16 \pi I_1 I_2} \,.
\end{align}
Changing variables to $r = \kappa/2 I_3$ and $I'_\alpha = I_\alpha - \kappa^2/4 I_3$ yields the RG equations~\eqref{eq:65}-\eqref{eq:70} given in the main text. 

\subsection{Coplanar RG equations from Ricci flow}
\label{sec:ricci-flow-rg} 
In this section, we derive the flow equations using the Ricci flow. In the electronic Supplementary Material we provide a \emph{Mathemtica} file that includes this calculation~\cite{SupplMat-FriedanScaling}. Besides being technically more straightforward to implement, the main advantage of this approach is that it provides us with clear decoupling criteria of the SO(3) and U(1) sectors. 

In the coplanar regime, the order parameter triad on the triangular lattice is related to the one on the honeycomb lattice by a simple rotation around the common $\bft_3$ axis as
\begin{align}
  \label{eq:186}
  t = h U = h \exp( - i \alpha \tau_3)
\end{align}
Here we employ the Euler angle parametrization 
\begin{align}
  \label{eq:187}
  h = e^{- i \phi \tau_3} e^{- i \theta \tau_1} e^{- i \psi \tau_3}\,,
\end{align}
since it allows for a transparent derivation of decoupling criteria between the SO(3) and U(1) sectors. This arises from the fact that for this choice of Euler angles, the relative angle $\alpha$ adds to the Euler angle $\psi$ via $h U = e^{- i \phi \tau_3} e^{- i \theta \tau_1} e^{- i (\psi + \alpha) \tau_3}$. 

To derive the RG equations as the Ricci flow of a metric tensor $g_{ij}(X)$, we first need to write the gradient part of the action in Eq.~\eqref{eq:175} in covariant form as
\begin{align}
  \label{eq:188}
  S_0 &= \frac12 \int_x \sum_{i,j = 1}^4 g_{ij} (\partial_\mu X^i) (\partial_\mu X^j)\,,
\end{align}
where we have combined the Euler angles and the relative angle $\alpha$ into the coordinate vector $X = (\phi, \theta, \psi, \alpha)$. The covariant metric tensor is given by
\begin{widetext}
\begin{align}
  \label{eq:189}
  g_{ij} &= \bmx g^{\text{SO(3)}} & \mathcal{K}^T \\ \mathcal{K} & I_\alpha \emx = \bmx I_3 \cos^2 \theta + \sin^2 \theta (I_1 \sin^2 \psi + I_2 \cos^2 \psi) & (I_1 - I_2) \sin \theta \sin \psi \cos \psi & I_3 \cos \theta & \frac{\kappa}{2} \cos \theta \\
(I_1 - I_2) \sin \theta \sin \psi \cos \psi & I_1 \cos^2 \psi + I_2 \sin^2 \psi & 0 & 0 \\
I_3 \cos \theta & 0 & I_3 & \frac{\kappa}{2} \\
\frac{\kappa}{2} \cos \theta & 0 & \frac{\kappa}{2} & I_\alpha \emx \,.
\end{align}
\end{widetext}
Here, the block matrix $g_{ij}^{\text{SO(3)}}$ is identical to the metric tensor of the isolated SO(3) magnet in Eq.~\eqref{eq:162}. Note, however, that $I_1 \neq I_2$ in the coplanar case. The SO(3) and U(1) sectors are coupled via the off-diagonal elements
\begin{align}
  \label{eq:190}
  \mathcal{K} = \frac{\kappa}{2} \bigl( \cos \theta, 0, 1 \bigr) \,. 
\end{align}
As described in the main text below Eq.~\eqref{eq:61}, these off-diagonal elements can be formally eliminated by a shift of the Euler angle $\psi \rightarrow \psi' = \psi + r \alpha$ with $r = \kappa/2I_3$. While $\mathcal{K}$ now vanishes, the SO(3) metric $g_{ij}^{\text{SO(3)}}$ implicitly depends on the relative angle $\alpha$ via $\psi'(\alpha)$. The U(1) stiffness changes to $I_\alpha \rightarrow I'_\alpha = I_\alpha - \kappa^2/2I_3$. This transformation provides us with two transparent decoupling criteria: either $|I_1 - I_2| \ll \sqrt{I_1 I_2}$ such that $g_{ij}^{\text{SO(3)}}$ becomes independent of Euler angle $\psi'$ (and thus of $\alpha$) or $r \ll 1$ such that the shift of $\psi$ is negligible. 

Following Frieds, the flow of $I_j$, $I'_\alpha$ and $r$ is given by the Ricci flow of the metric tensor (see Eq..~\eqref{eq:18}). This allows us to confirm that the system flows towards a decoupled regime at longer lengthscales. 
In the coplanar regime, the Ricci tensor takes the form
\begin{widetext}
\begin{align}
  \label{eq:191}
  R_{ij} &= \bmx R_{11}  & \frac{(I_1 - I_2) \{ 4 I_\alpha [(I_1 + I_2)^2 - I_3^2] + I_3 \kappa^2 \} \sin \theta \sin 2 \psi}{4 I_1 I_2 (4 I_3 I_\alpha - \kappa^2)}  &  \frac{ (I_3^2 - (I_1 - I_2)^2 )\cos \theta}{2 I_1 I_2} & \frac{\kappa I_3 \cos \theta}{4 I_1 I_2} \\
\frac{(I_1 - I_2) \{ 4 I_\alpha [(I_1 + I_2)^2 - I_3^2] + I_3 \kappa^2 \} \sin \theta \sin 2 \psi}{4 I_1 I_2 (4 I_3 I_\alpha - \kappa^2)} & R_{22}  & 0 & 0 \\
\frac{(I_3^2 - (I_1 - I_2)^2)\cos \theta}{2 I_1 I_2}  & 0 & \frac{I_3^2 - (I_1 - I_2)^2}{2 I_1 I_2} & \frac{\kappa I_3}{4 I_1 I_2} \\
\frac{I_3 \kappa \cos \theta}{ 4 I_1 I_2 } & 0 & \frac{I_3 \kappa}{4 I_1 I_2} & \frac{\kappa^2}{8 I_1 I_2} \emx \,,
\end{align}
where 
\begin{align}
\label{eq:192}
R_{11} &=  \frac{I_3^2 - (I_1 - I_2)^2}{2 I_1 I_2} \cos^2 \theta + \sin^2 \theta \biggl\{ \cos^2 \psi \frac{I_2 \bigl( 4 I_\alpha ( I_2^2 - (I_1 - I_3)^2 + \kappa^2 (I_3 - 2 I_1)  \bigr)}{8 I_1 I_2 I_3 (I_\alpha - \kappa^2/4 I_3)} \nonumber \\ & \quad  + \sin^2 \psi \frac{I_1 \bigl( 4 I_\alpha ( I_1^2 - (I_2 - I_3)^2 + \kappa^2 (I_3 - 2 I_2)  \bigr)}{8 I_1 I_2 I_3 (I_\alpha - \kappa^2/4 I_3)}  \biggr\} \\
  \label{eq:193}
  R_{22} &= \frac{\cos^2\psi \Bigl[ - 8 I_1 I_\alpha ( - I_1^2 + (I_2 - I_3)^2) + 2 I_1 \kappa^2 (I_3 - 2 I_2)\Bigr]}{16 I_1 I_2 I_3 (I_\alpha - \kappa^2/4 I3)} \nonumber \\ & \quad  + \frac{\sin^2 \psi \Bigl[ - 8 I_2 I_\alpha ( - I_2^2 + (I_1 - I_3)^2) + 2 I_2 \kappa^2 (I_3 - 2 I_1)\Bigr]}{16 I_1 I_2 I_3 (I_\alpha - \kappa^2/4 I3)} \,.
\end{align}
The contraction of the Riemann tensor that corresponds to the two loop result is a lengthy expression that is straightforwardly computed.
It allows to extract the two-loop RG equations
\begin{align}
  \label{eq:194}
  \frac{d}{dl} I_1 &= \frac{- I_1^2 + (I_2 - I_3)^2}{4 \pi I_2 I_3} - \frac{(I_1^2 - I_2^2) \kappa^2}{ 16 \pi I_2 I_3^2 (I_\alpha - \kappa^2/4 I_3)} - \frac{1}{4 I_1^2 I_2} \Bigl[ 4 I_1 I_2 + 4 I_2^2 - 4 I_1 I_3 - 8 I_2 I_3 + 5 I_3^2 \nonumber \\ & \qquad + \frac{(I_1 - I_2)^2}{I_\alpha - \kappa^2/4 I_3} \Bigl\{ \frac{(4 I_1^2 + (I_1 + I_2)^2) I_\alpha^2}{I_3} - 2 (2 I_2 + I_3) I_\alpha \Bigr\} \Bigr] \\
\label{eq:195}
\frac{d}{dl} I_2 &= \frac{- I_2^2 + (I_1 - I_3)^2}{4 \pi I_1 I_3} - \frac{(I_2^2 - I_1^2) \kappa^2}{ 16 \pi I_1 I_3^2 (I_\alpha - \kappa^2/4 I_3)} - \frac{1}{4 I_1 I_2^2} \Bigl[ 4 I_1 I_2 + 4 I_1^2 - 4 I_2 I_3 - 8 I_1 I_3 + 5 I_3^2 \nonumber \\ & \qquad + \frac{(I_1 - I_2)^2}{I_\alpha - \kappa^2/4 I_3} \Bigl\{ \frac{(4 I_2^2 + (I_1 + I_2)^2) I_\alpha^2}{I_3} - 2 (2 I_1 + I_3) I_\alpha \Bigr\} \Bigr]  \\
\label{eq:196}
\frac{d}{dl} I_3 &= -\frac{I_3^2 + (I_1 - I_2)^2}{4 \pi I_1 I_2} - \frac{1}{32 \pi^2 I_1^2 I_2^2 I_3 (I_\alpha - \kappa^2/4 I_3)} \Bigl[ I_\alpha \Bigl\{ (I_1 - I_2)^2 (5 I_1^2 + 6 I_1 I_2 + 5 I_2^2) \nonumber \\ & \qquad - 8 (I_1 - I_2)^2 (I_1 + I_2) I_3 + 2 (I_1 - I_2)^2 I_3^2 + I_3^4 \Bigr\} + \kappa^2 \Bigl\{ (I_1 - I_2)^2 (I_1 + I_2 + I_3/4) + I_3^2/4  \Bigr\}   \Bigr]
\\
\label{eq:197}
\frac{d}{dl} \kappa &= - \frac{\kappa I_3}{4 \pi I_1 I_2} - \frac{I_3^3 + I_3 (I_1 - I_2)^2 - 2 (I_1 + I_2) (I_1 - I_2)^2 }{16 \pi^2 I_1^2 I_2^2 I_3 (I_\alpha - \kappa^2/4 I_3)} \\
\label{eq:198}
\frac{d}{dl} I_\alpha &= - \frac{\kappa^2}{16 \pi I_1 I_2} - \frac{ \bigl[ ( I_1 - I_2)^2 + I_3^2) I_\alpha \kappa^2 - I_3 \kappa^4/4}{ 128 \pi^2 I_1^2 I_2^2 I_3 (I_\alpha - \kappa^2/4 I_3)} \,.
\end{align}
\end{widetext}
The one-loop contribution agrees with the result in Eqs.~\eqref{eq:181}-\eqref{eq:185} obtained from Wilson-Polyakov scaling. 

\subsection{Flow of six-fold potential $\lambda$}
\label{sec:flow-six-fold}
Let us derive the flow of the six-fold potential $\lambda$ in the coplanar regime. Rewriting the action in Eq.~\eqref{eq:175} in terms of $r = \kappa/2 I_3$ and $I'_\alpha = I_\alpha - \kappa^2/4 I_3$, one finds
 \begin{align}
   \label{eq:199}
   &S = \frac12 \int_x \biggl\{ I_1 \Bigl[ (\Omega_\mu^1)^2 + (\Omega_\mu^2)^2 \Bigr] + I_3 (\Omega_\mu^3)^2 + (I_2 - I_1) \nonumber \\
 & \times \Bigl[ \sin(\psi - r \alpha) (\partial_\mu \theta) + \cos \theta \cos(\psi - r \alpha) (\partial_\mu \phi)^2 \Bigr]^2 \nonumber \\ & + I_\alpha' (\partial_\mu \alpha)^2 + \frac{\lambda}{4} \cos (6 \alpha) \,.
 \end{align}
The potential $\lambda$ is renormalized by spin waves in the phase angle $\alpha$. We decompose $\alpha = \alpha_< + \alpha_>$ into fast modes $\alpha_>$ and slow modes $\alpha_<$, and keep only those parts of the action that are relevant to the renormalization of $\lambda$ to arrive at 
\begin{align}
  \label{eq:200}
  S_{\mathbb{Z}_6} &= \int_x \Bigl\{ \frac{I'_\alpha}{2} (\partial_\mu \alpha_<)^2 + \frac{I'_\alpha}{2} (\partial_\mu \alpha_>)^2 \nonumber \\ & \quad + \frac{\lambda}{4} \cos(p \alpha_< + p \alpha_>) \,.
\end{align}
Here, we have generalized to a potential with $\mathbb{Z}_p$ symmetry, where $p=6$ in our case. For the renormalization of $\lambda$ we can focus on the derivative terms $(\partial_\mu \alpha)^2$, and neglect the terms in the second line of Eq.~\eqref{eq:199}. 
Expanding to quadratic order in the fast fields, we find
\begin{align}
  \label{eq:201}
  &S_{\mathbb{Z}_6} = \int_x \biggl\{ \frac{I'_\alpha}{2} (\partial_\mu \alpha_<)^2 + \lambda \cos (p \alpha_<)  \\
& + \frac{I'_\alpha}{2} (\partial_\mu \alpha_>)^2 - \frac{\lambda p^2}{2} \alpha^2_> \cos (p \alpha_<) - p \alpha_> \sin (p \alpha_< ) \biggr\}  \nonumber 
\end{align}
In momentum space this becomes
\begin{align}
  \label{eq:202}
  S_{\mathbb{Z}_6} &= S^<_{\mathbb{Z}_6} + \int^>_{k, k'} \alpha_>(k) \alpha_>(k') \biggl\{ \frac{I'_\alpha}{2} k^2 \delta(k+k') \nonumber \\ &  - \frac{\lambda p^2}{2} \cos(p \alpha_<)_{k+k'} \biggr\} \,,
\end{align}
where $S^<_{\mathbb{Z}_6} = \int_x \{ \frac{I'_\alpha}{2} (\partial_\mu \alpha_<)^2 + \lambda \cos (p \alpha_<) \}$ contains only slow modes and $\int_{k}^> = \frac{1}{(2 \pi)^2} \int_{\Lambda/b}^\Lambda dk k \int_0^{2 \pi} d \phi$. We have disregarded the last term in Eq.~\eqref{eq:201} because it involves a function of slow modes $f(\alpha_<)$ evaluated at a fast momentum $|k| \in [\Lambda/b, \Lambda]$, where this function vanishes. 

The next step is to perform the functional integration over the fast modes $\alpha_>$, which yields
\begin{align}
  \label{eq:203}
  & S_{\mathbb{Z}_6} = S^<_{\mathbb{Z}_6} + \frac12 \text{Tr} \ln ( G_{k,k'}^{-1} - C_{k,k'}) \\
&= S^<_{\mathbb{Z}_6} + \frac12 \text{Tr} \ln [ G_{k,k}^{-1}]  - \frac12 \text{Tr} [G_{k,k'} C_{k',k} ] + \mathcal{O}(C^2) \nonumber 
\end{align}
with inverse propagator $G_{k,k'}^{-1} = I'_\alpha k^2 \delta(k + k')$, propagator $G_{k,k'} = [I'_\alpha k^2]^{-1} \delta(k+k')$ and potential term $C_{k,k'} = \lambda p^2 \cos(p \alpha_<)_{k+k'} $. Evaluating the trace in Eq.~\eqref{eq:203} gives 
\begin{align}
  \label{eq:204}
  \frac12 \text{Tr} [G_{k,k'} C_{k',k} ] &= \frac12 \int_k^> \frac{1}{I'_\alpha k^2} \lambda p^2 \cos(p \alpha_<)_0  \nonumber \\ 
&= \frac{\ln b}{4 \pi I'_\alpha}\lambda p^2 \cos(p \alpha_<)_0 \,.
\end{align}
Finally, we rescale momenta $k' = b k$ and fields $\alpha'(k') = \alpha_<(b k)$ to obtain the renormalized value $\lambda' = b^2 \lambda - \frac{\ln b}{4 \pi I'_\alpha} \lambda p^2$. The resulting flow equation for the $p$-fold potential is thus given by 
\begin{align}
  \label{eq:205}
  \frac{d}{dl} \lambda &= \Bigl( 2 - \frac{\lambda p^2}{4 \pi I'_\alpha} \Bigr) \lambda \,,
\end{align}
which for $p=6$ results in Eq.~\eqref{eq:70}.

%

\end{document}